\documentclass[a4paper,11pt]{article}
\pdfoutput=1
\usepackage{jheppub}
\usepackage[T1]{fontenc}
\usepackage{lmodern}
\usepackage{float}
\usepackage{orcidlink}
\usepackage{tabularx}
\usepackage{array}
\usepackage{xcolor}
\hypersetup{hypertexnames=false}
\newcommand{\paperpart}[1]{
\bigskip \phantomsection \addcontentsline{toc}{section}{#1} \noindent{\large\bfseries #1} \par\medskip}

\usepackage[normalem]{ulem}

\newcommand{\nocontentsline}[3]{}
\let\origcontentsline\addcontentsline
\newcommand\stoptoc{\let\addcontentsline\nocontentsline}
\newcommand\resumetoc{\let\addcontentsline\origcontentsline}

\title{Dynamical tidal response of regular black holes: Perturbative analysis and shell EFT interpretation}

\author[a]{Arpan Bhattacharyya \orcidlink{0000-0002-7933-6441},}
\author[a]{Naman Kumar \orcidlink{0000-0001-8593-1282},}
\author[a,b]{Shailesh Kumar,\orcidlink{0000-0001-7072-9452}}

\affiliation[a]{Department of Physics, Indian Institute of Technology Gandhinagar, Palaj, Gujarat 382355, India}
\affiliation[b]{Department of Physics, Indian Institute of Technology, Kharagpur 721302, India}

\emailAdd{abhattacharyya@iitgn.ac.in}
\emailAdd{naman.kumar@iitgn.ac.in}
\emailAdd{shaileshkumar.1770@gmail.com}
\date{\today}

\abstract{We compute the frequency-dependent quadrupolar tidal response of Bardeen, Hayward, and Fan-Wang regular black holes in the polar and axial sectors by solving the coupled gravitational-electromagnetic perturbation equations numerically. Our analysis independently recovers the static Love numbers and their scaling at small regularization, while differences can occur at finite regularization due to higher-order corrections. The ratios of metric source-response coefficients at low frequencies ($\omega$) have smooth corrections starting at $\mathcal{O}(\omega^{2})$. 
Furthermore, we compare the peaks in the response coefficients with the real parts of the quasinormal mode (QNM) frequencies and find that in the polar sector for Bardeen, Hayward and Fan-Wang, peaks for small values of the regularization parameter, align with the corresponding QNM frequencies  within the damping width provided by the imaginary part of the respective QNM. On the other hand, in the axial sector, the Bardeen and Hayward maxima of the response coefficients are not aligned with the real part of the QNM, whereas all the Fan-Wang peaks are. 
Moreover, we perform a shell EFT calculation using a scalar field as a simpler probe. The shell EFT construction expresses the response in terms of renormalised Wilson coefficients and helps isolate the scheme-dependent finite terms from the scheme-independent part. We also show that it yields the same source–response ratio as the direct calculation when the source is subtracted in the same background. This agreement, obtained in the simpler probe case, further supports the broader interpretation of the dynamical tidal response as a well-defined gauge-invariant observable.}

\begin{document}
\maketitle
\flushbottom

\section{Introduction}
The direct detection of gravitational waves (GWs) has opened a new observational window on the universe, particularly the strong-field and dynamical regime of gravity, offering high-precision means to probe the nature of compact objects, from binary black holes to neutron stars \cite{LIGOScientific:2016aoc, LIGOScientific:2017bnn, LIGOScientific:2016sjg, LIGOScientific:2021qlt, LIGOScientific:2017vwq, Lasky:2015uia}. These observations from compact binaries confirm the existence of black holes and enable us to investigate their distinct properties through GW signals. GWs from such binaries provide direct access to finite-size effects associated with the internal structure of the spacetime geometry \cite{Cardoso:2019rvt, LISA:2022kgy, Hinderer:2007mb}. Among these effects, tidal interactions are an important factor during the inspiral phase of binary dynamics, as they alter the GW phase evolution. In other words, in a binary system, each component perturbs its companion during its inspiral dynamics, leading to deformations that influence the orbital motion and accumulate over many cycles. Tidal effects or deformability describe how the multipole structure of a compact object is altered by an external tidal field, the Love number encoding the real part of this response \cite{Zhang:1986cpa, Binnington:2009bb, Damour:2009vw, Hinderer:2007mb, Poisson:2014gka, Poisson:2020vap}. Although these effects may be small, as the binary separation decreases, their impact on the GW phase can become significant. Consequently, the inspiral waveform carries imprints of finite-size physics that are sensitive to the nature of the compact objects and can be isolated through accurate modelling of the phase evolution \cite{Flanagan:2007ix,Yagi:2013sva,Chakraborty:2025wvs}.

\textcolor{black}{The static tidal response of four-dimensional vacuum Ricci-flat black hole solutions of classical general relativity (GR), namely Schwarzschild and Kerr, has been studied extensively using different definitions and matching prescriptions, finding vanishing conservative static tidal Love numbers (TLNs) \cite{Landry:2015zfa, Charalambous:2021mea, Chia:2020yla, Hui:2021vcv, BenAchour:2022uqo, Ivanov:2022qqt,LeTiec:2020bos, LeTiec:2020spy}. 
The vanishing of static TLNs has also been extended beyond linear order for both Schwarzschild and Kerr black holes \cite{Gounis:2024hcm, Kehagias:2024rtz, Combaluzier-Szteinsznaider:2024sgb, Riva:2023rcm}. This vanishing conservative response has also been found across all bosonic degrees of freedom, including scalar (spin-0), electromagnetic (spin-1), and gravitational perturbations (spin-2) \cite{Chia:2020yla, Charalambous:2021mea, Ivanov:2022qqt, Kol:2011vg, Parra-Martinez:2025bcu, Berens:2025okm}. Neutron stars, in contrast, generally exhibit a nonzero tidal response, making tidal deformability a useful probe for distinguishing them from black holes and other compact objects with similar macroscopic properties \cite{Hinderer:2007mb, Damour:2009vw, Yagi:2013bca, Pani:2015hfa, Pani:2015nua, Cardoso:2019rvt, Steinhoff:2021dsn, Creci:2023cfx}. A complementary description is provided by the effective field theory (EFT) approach, in which a compact object is treated as a point particle with additional worldline degrees of freedom that encode its internal dynamics \cite{Goldberger:2004jt, Goldberger:2005cd, Porto:2016pyg, Porto:2016zng, Ivanov:2025ozg, Charalambous:2021mea, Hadad:2024lsf, Bhattacharyya:2023kbh}.}

\textcolor{black}{In the worldline EFT description, static tidal responses are encoded in Wilson coefficients multiplying finite-size operators \cite{Goldberger:2004jt,Porto:2016zng}.
For four-dimensional Schwarzschild black holes, matching gives vanishing static tidal Wilson coefficients \cite{Kol:2011vg,Ivanov:2022hlo,Charalambous:2022rre}.
For Kerr black holes, a vanishing conservative static tidal response has likewise been obtained in Refs.~\cite{Charalambous:2021mea,Chia:2020yla,Ivanov:2022qqt}.} The absence of static TLN of black holes in vacuum GR in four dimensions implies an unusual cancellation or the fine-tuning problem analogous to the cosmological constant problem \cite{Porto:2016zng}. However, in vacuum GR beyond four dimensions, as well as in four-dimensional theories with modified field equations and additional degrees of freedom, black holes can exhibit a nonzero static tidal response \cite{Hui:2020xxx,DeLuca:2023mio, DeLuca:2022tkm,Cardoso:2017cfl,Cardoso:2018ptl,Pereniguez:2021xcj,Katagiri:2023umb, Katagiri:2024fpn,Charalambous:2025ekl,Coviello:2025pla, Kobayashi:2025swn,Pang:2025myy,Bhattacharyya:2025slf, Barbosa:2026qcv,Rodriguez:2026iot, Chakraborty:2026qru, Cohen:2026ovm}. This distinction is useful for GW observations, as it provides a way of testing the robustness of GR and constraining possible deviations from it \cite{Flanagan:2007ix,Yagi:2013sva,Chakraborty:2025wvs, Zi:2025lio, Kumar:2024our, AbhishekChowdhuri:2023gvu, Kumar:2024utz,Rodriguez:2026iot, Chakraborty:2026qru}.


\textcolor{black}{On the other hand, the singularity theorems of classical GR show that, under suitable physical and geometric assumptions, gravitational collapse can lead to geodesically incomplete spacetimes \cite{Hawking:1970zqf, Penrose:1964wq}. This indicates a limitation or breakdown of the classical spacetime description and motivates the possibility that additional physics, potentially of quantum-gravitational origin, becomes relevant in the black hole interior \cite{Ashtekar:2005cj, 1968qtr..conf...87B}.}
Motivated by this idea, a class of phenomenological models has been developed in which the central singularity is replaced by a regular core characterized by a short-distance length scale, where the resultant geometry is known as regular black holes \cite{1968qtr..conf...87B, Fan:2016hvf, Hayward:2005gi, Lan:2023cvz, Ansoldi:2008jw, Carballo-Rubio:2018pmi, Simpson:2019mud, Simpson:2018tsi, Carballo-Rubio:2022kad, Bueno:2024eig, Bueno:2025gjg, Johannsen:2013szh, Carballo-Rubio:2021bpr}. These models offer a well-motivated and consistent framework to address intriguing notions that alter black hole interiors and singularities \cite{Dymnikova:1992ux, Frolov:2016pav, Zi:2024jla}. 

\textcolor{black}{Regular black holes retain an event horizon while replacing the central curvature singularity with a regular core supported by effective matter sources. As a result, their background geometry differs from that of the corresponding vacuum black hole.} These crucial properties make them suitable candidates for describing non-singular gravitational collapse as well as modifications to GR \cite{Cardoso:2019rvt, Ansoldi:2008jw, Johannsen:2013szh, Kumar:2024our, Carballo-Rubio:2019fnb}. Apart from this, we also know that regular black holes asymptotically approach the Schwarzschild and Kerr solutions, and distinguishing them observationally is intricate; however, they still exhibit differences in their finite-size effects. These regular geometries, depending on the regularization length-scale, admit configurations with or without horizons that allow one to study how short distances impact black hole physics, whose tidal responses have \textcolor{black}{recently been studied in} 
\cite{Coviello:2025pla, Wang:2025oek}.

When the external field or tidal field varies in time, static TLNs provide an incomplete description of tidal interactions; however, in such scenarios, 
\textcolor{black}{the relevant quantity is the dynamical TLN, which characterizes the conservative, i.e., real part of the frequency-dependent tidal response of a compact object} \cite{Steinhoff:2016rfi,Chia:2020yla,Creci:2021rkz, Saketh:2023bul, Perry:2023wmm, Caron-Huot:2025tlq, Kosmopoulos:2025rfj,Combaluzier--Szteinsznaider:2025eoc}. \textcolor{black}{For black holes beyond vacuum GR, these quantities can be computed using metric perturbation theory. In this approach, the compact object is treated as a background spacetime and the tidal field as a linear perturbation, with the relevant theory corrections derived consistently from the action.} Solving the corresponding perturbation equations with appropriate boundary conditions allows the extraction of the tidal response directly from the asymptotic behavior of the perturbed metric. We know that for black holes in 4-dimensional vacuum GR, the Teukolsky formalism is a natural and efficient approach to compute TLNs that gives rise to zero static TLNs at all spin orders \cite{PhysRevLett.29.1114, 1973ApJ...185..635T, Chia:2020yla}. Further, the modified Teukolsky approach has recently been developed, opening new avenues to explore tidal effects for theories beyond GR \cite{Cano:2025zyk, Guo:2024bqe, Li:2022pcy, Kumar:2025jsi}. Similarly, EFT-based methods provide a complementary approach to this problem using techniques from quantum field theory \cite{Kol:2011vg, Ivanov:2024sds, Caron-Huot:2025tlq, Kosmopoulos:2025rfj}. Together, these approaches offer consistent, physically transparent frameworks for connecting the tidal responses of compact objects to observable imprints in GW signals.

In the present study, we compute the quadrupolar frequency-dependent tidal response of the Bardeen, Hayward, and Fan-Wang regular black holes, which are solutions of Einstein's equations sourced by nonlinear electrodynamics \cite{1968qtr..conf...87B,Hayward:2005gi,Fan:2016hvf,Coviello:2025pla}. 
\textcolor{black}{In the full gravitational calculation, the polar and axial sectors are treated as coupled gravitational--electromagnetic systems. We solve the coupled perturbation equations in terms of gauge-invariant master variables and then reconstruct the corresponding Regge--Wheeler-gauge metric amplitudes }\cite{Regge, Zerelli, Katagiri:2023yzm}. \textcolor{black}{The dynamical tidal response is then extracted numerically in the intermediate region}\footnote{When determining the dynamical Love numbers by solving the perturbation equation numerically, we analyze the source and response in the intermediate or buffer region $M\ll r\ll\omega^{-1}$. For a finite real frequency, the solution from the perturbation equations in the far zone is of an oscillatory nature. Hence, it is numerically less stable to fit it directly to a form that has the growing and decaying series to extract the response coefficients. In the buffer region, both $M/r$ and $\omega r$ are small, and the source and response powers can be clearly separated numerically \cite{Chia:2020yla, Katagiri:2024fpn}. On the other hand, effective-field-theory methods instead separate the long-distance solution directly into source and response parts \cite{Ivanov:2022hlo}. Recent work has matched the horizon-ingoing Mano-Suzuki-Takasugi (MST) solution \cite{Mano:1996vt,Mano:1996mf} to worldline EFT in the common buffer region \cite{Kobayashi:2025vgl}.} 
The source is chosen to be purely gravitational, while the induced electromagnetic response is retained. The low-frequency calculation also gives the static Love numbers and their scaling in the small-regularization limit, reduces to the Regge-Wheeler and Zerilli equations in the Schwarzschild limit \cite{Regge, Zerelli, Katagiri:2023yzm}, and gives regular $O(\omega^2)$ conservative corrections whose signs depend on the geometry and parity. Therefore, these frequency-dependent corrections provide information beyond the static coefficients and distinguish the three regular black hole models considered here within the range where near-zone matching is reliable. Further, wider-frequency scans are also used to test possible resonant structures. \textcolor{black}{Using the damping-width criterion defined below, all sampled Bardeen and Hayward polar maxima are QNM aligned. For Fan--Wang, the maximum is aligned only at the lowest sampled regularization, while the two larger regularization values are not.} In the axial sector, the Bardeen and Hayward maxima are not aligned, while the sampled Fan-Wang maxima are QNM aligned. At this point, note that the frequency-dependent tidal response may, in principle, leave observable imprints inside the gravitational-wave phase \cite{Chakraborty:2025wvs}. A detailed study in that direction, i.e, waveform modeling by incorporating these effects and their observational consequences, is, however, beyond the scope of the present work. In this work, we mainly focus on the structure of the dynamical TLNs themselves and study their frequency-dependent features, which provide the response coefficients needed for future waveform analysis for these theories.

Further, as a complementary EFT calculation, we study a minimally coupled scalar field using a finite-shell matching construction \cite{Kosmopoulos:2025rfj}. We use the same scalar equation for both the direct source--response extraction and the shell calculation. The finite-radius shell quantity obtained from the derivative jump is the Wilson coefficient of the regulated shell EFT. In the original shell-EFT construction, the corresponding Love coefficient is obtained by expanding this quantity in the point-particle limit. Here we use the same renormalization logic with a different finite prescription: we subtract the canonical source solution in the same background and then extrapolate the remaining inverse-radius terms. This places the shell response in the same source convention as the direct calculation. The logarithmic coefficients and their running are unchanged by this finite subtraction, whereas the finite part depends on the renormalization prescription. The resulting matched response is independent of the arbitrary shell radius and agrees with the direct scalar source--response ratio. Although the scalar field is only a simpler probe, this agreement provides a gauge-invariant EFT description of its dynamical response and illustrates how the shell construction can separate the finite response from regulator-dependent terms.

\textcolor{black}{To the best of our knowledge, this is the first application of the shell EFT approach to a non-vacuum black hole solution in GR.} This scalar response should be distinguished from the coupled gravitational Love number in the second part of the paper, which is obtained independently from the metric perturbations. For comparison, Appendix \ref{app:tensor_probe} gives a separate fixed-background tensor test-field calculation that isolates ordinary spin-2 scattering from the coupled gravitational response. 


The paper is organized as follows. In Section~\ref{regular BH}, we start by discussing the regular black hole backgrounds and then detail the perturbation equations required for solving the dynamical TLNs. The entire paper is essentially divided into two parts. Part I is devoted entirely to shell-EFT computations, in which we use a simpler probe in the form of a minimally coupled scalar field. In this part, Section~\ref{sec:methods_eft} discusses the shell-EFT construction, Section~\ref{sec:pert-shellEFT} derives the analytic structure of scalar response in a small-frequency asymptotic expansion, and separates the logarithmic terms from the finite terms. Section~\ref{subsec:shell-probe-matching} then determines the remaining global response numerically by evolving the full scalar equation with the ingoing horizon condition and directly compares the resulting shell response with the source-response ratio obtained from the same scalar
solution. Then, in Part II, we address the computation of the dynamical gravitational TLNs. 
In this part, Section~\ref{sec:methods} discusses how the response coefficients are extracted after solving the coupled gravitational--electromagnetic perturbation equations. Then Secs.~\ref{sec:results_polar} and~\ref{sec:results_axial} present the polar and axial results, and Section~\ref{sec:conclusion} summarizes the conclusions with future prospects. In addition, Appendix~\ref{app:dynLove} provides details on finite-frequency perturbation equations and a consistency check by showing that these equations reduce to the Schwarzschild case when the regularization parameter vanishes in the GR limit. Appendix~\ref{app:numerical_asymptotics} gives the numerical asymptotic solutions of the intermediate-near-zone used for the extraction of dynamical response. Finally, Appendix~\ref{app:tensor_probe} presents a probe tensor spin-2 field computation for comparison with the response obtained from solving the full coupled gravitational and electromagnetic perturbation.


\section{Regular Black Hole Background and Perturbation Setup}\label{regular BH}
\textcolor{black}{We consider Einstein gravity coupled to nonlinear electrodynamics (NLED) and a scalar field.
The Einstein--NLED sector follows the standard formulation used for regular black hole solutions \cite{Ayon-Beato:2000mjt,Bronnikov:2000vy}, while the scalar field is included for
the more general perturbative setup considered below \cite{Coviello:2025pla}.
The action is}

\begin{equation}
   \textcolor{black}{
    S=\int d^4x\,\sqrt{-g}\left[
    \frac{1}{16\pi}R
    -\frac{1}{4\pi}\mathcal{L}(F)
    +\frac{1}{2}(\partial\Phi)^2
    -V(\Phi)
    \right],
    }
\end{equation}
\textcolor{black}{where $R$ is the Ricci scalar, $g$ is the determinant of the spacetime metric,
$\mathcal{L}(F)$ is the NLED Lagrangian, and
$F\equiv F_{\mu\nu}F^{\mu\nu}/4$, with $F_{\mu\nu}$ the electromagnetic
field-strength tensor. The scalar field is denoted by $\Phi$, with potential
$V(\Phi)$, and $(\partial\Phi)^2\equiv
g^{\mu\nu}\partial_\mu\Phi\,\partial_\nu\Phi$.
For the backgrounds considered here, the electromagnetic field is purely magnetic,
and the magnetic charge is related to the corresponding regularization scale.}

\textcolor{black}{Variation of the action with respect to the metric gives}
\begin{align}
    G_{\mu\nu}
    =&\,2\left(
    \mathcal{L}_{,F}F_{\mu}{}^{\sigma}F_{\nu\sigma}
    -g_{\mu\nu}\mathcal{L}
    \right)-8\pi\left[
    \partial_\mu\Phi\,\partial_\nu\Phi
    +g_{\mu\nu}\left(
    -\frac{1}{2}(\partial\Phi)^2+V(\Phi)
    \right)
    \right],
    \label{eq:Einstein}
\end{align}
\textcolor{black}{where $\mathcal{L}_{,F}\equiv\partial\mathcal{L}/\partial F$.}

\textcolor{black}{Variation with respect to the gauge field gives the NLED field equation}
\begin{equation}
    \textcolor{black}{
    \nabla_\mu\left(\mathcal{L}_{,F}F^{\nu\mu}\right)=0
    }
    \label{eq:modMaxwell}
\end{equation}
\textcolor{black}{while the scalar field satisfies}
\begin{equation}
   \textcolor{black}{
    \Box\Phi-\frac{\partial V}{\partial\Phi}=0.
    }
    \label{eq:KG}
\end{equation}

For the three backgrounds, the scalar field and its potential vanish, while the corresponding NLED Lagrangians take the forms
\begin{equation}
\begin{split}
    &\textup{Bardeen:}\qquad \,\,\,\,\,\Phi_{B}=0\qquad \,\,\,\,\,\,\mathcal{L}_{B}=\frac{3M\ell_B^2}{(r^2+\ell_B^2)^{5/2}}\qquad \,\,\,\,\,V_{B}=0,\\&
    \textup{Hayward:}\qquad\,\,\, \Phi_{H}=0\qquad \,\,\,\,\,\mathcal{L}_{H}=\frac{6M^2\ell_H^2}{(r^3+2M\ell_H^2)^2}\qquad V_{H}=0,\\&
\textup{Fan-Wang:}\qquad \Phi_{FW}=0\qquad \mathcal{L}_{FW}=\frac{3M\ell_{FW}}{(r+\ell_{FW})^4}\qquad \,\,\,\,V_{FW}=0.\label{eq:matt_content}
\end{split}
\end{equation} 


The three regular black hole geometries can be written in the common static, spherically symmetric form
\begin{align}
ds^{2} =& -f(r)dt^{2}+f(r)^{-1}dr^{2}
+r^{2}\left(d\theta^{2}+\sin^{2}\theta\,d\varphi^{2}\right),
\label{metric}
\end{align}
with
\begin{equation}
f(r)=1-\frac{2m(r)}{r}.
\end{equation}
The corresponding mass functions for the Bardeen \cite{1968qtr..conf...87B},
Hayward \cite{Hayward:2005gi}, and Fan--Wang \cite{Fan:2016hvf}
solutions are
\begin{equation}
\begin{split}
\textup{Bardeen:}\qquad
&m_B(r)=\frac{Mr^{3}}{(r^{2}+\ell_B^{2})^{3/2}},
\qquad
\ell_{\textup{ext}}=\frac{4M}{3\sqrt{3}},\\
\textup{Hayward:}\qquad
&m_H(r)=\frac{Mr^{3}}{r^{3}+2M\ell_H^{2}},
\qquad
\ell_{\textup{ext}}=\frac{4M}{3\sqrt{3}},\\
\textup{Fan--Wang:}\qquad
&m_{FW}(r)=\frac{Mr^{3}}{(r+\ell_{FW})^{3}},
\qquad
\ell_{\textup{ext}}=\frac{8M}{27}.
\end{split}
\label{met}
\end{equation}

For each model, the regularization parameter controls the horizon structure. For $\ell < \ell_{\rm ext}$, the spacetime has two horizons, which merge at $\ell = \ell_{\rm ext}$. All three metrics are asymptotically flat, possess a regular de Sitter-like core, and approach the Schwarzschild solution as the regularization scale tends to zero. Since the three models have different mass profiles, they allow us to distinguish features of the tidal response that are common to all models from those that depend on the specific model. 
\textcolor{black}{The TLNs of these regular black hole backgrounds have been studied in Refs.~\cite{Wang:2025oek,Coviello:2025pla}.}

\textcolor{black}{We now consider linear perturbations of the metric and matter fields around these static, purely magnetic backgrounds. Owing to spherical symmetry, the perturbations can be decomposed into polar (even-parity) and axial (odd-parity) sectors. For the magnetic backgrounds considered here, even-parity gravitational perturbations couple to odd-parity electromagnetic perturbations and, when present, to the even-parity scalar perturbation. Odd-parity gravitational perturbations instead couple to even-parity electromagnetic perturbations, while the scalar field has no independent odd-parity sector. We work in Regge--Wheeler gauge for the metric \cite{Regge,Zerelli} and fix the electromagnetic gauge. Therefore, the perturbations can be written as}
\begin{align}
\delta g_{\mu\nu}
&=
\delta g^{(\mathrm{polar})}_{\mu\nu}
+
\delta g^{(\mathrm{axial})}_{\mu\nu}\,,
\\[1mm]
\delta A_\mu
&=
\delta A^{(\mathrm{polar})}_\mu
+
\delta A^{(\mathrm{axial})}_\mu\,,
\\[1mm]
\delta\Phi
&=
\delta\Phi^{(\mathrm{polar})}\,,
\end{align}
where, for each \((\ell,m)\)-mode, the even-parity metric perturbation takes the form
\begin{align}\label{perteqs_gen_pol}
\delta g^{(\mathrm{polar})}_{\mu\nu}
=
\begin{pmatrix}
-f(r)e^{-2\phi(r)}\,H^{\ell m}_{0}(t,r) &
-i\omega H^{\ell m}_{1}(t,r) & 0 & 0 \\
-i\omega H^{\ell m}_{1}(t,r) &
\dfrac{H^{\ell m}_{2}(t,r)}{f(r)} & 0 & 0 \\
0 & 0 & r^{2}K^{\ell m}(t,r) & 0 \\
0 & 0 & 0 & r^{2}\sin^{2}\theta\,K^{\ell m}(t,r)
\end{pmatrix}
Y^{\ell m}(\theta,\varphi)\,,
\end{align}
while the odd-parity metric perturbation is
\begin{align}\label{perteqs_gen_ax}
\delta g^{(\mathrm{axial})}_{\mu\nu}
=&
-\frac{h^{\ell m}_{0}(t,r)}{\sin\theta}
\frac{\partial Y^{\ell m}(\theta,\varphi)}{\partial\varphi}
\left(\delta^{0}_{\mu}\delta^{2}_{\nu}+\delta^{2}_{\mu}\delta^{0}_{\nu}\right)
+h^{\ell m}_{0}(t,r)\sin\theta
\frac{\partial Y^{\ell m}(\theta,\varphi)}{\partial\theta}
\left(\delta^{0}_{\mu}\delta^{3}_{\nu}+\delta^{3}_{\mu}\delta^{0}_{\nu}\right)
\nonumber\\
&\quad
-\frac{h^{\ell m}_{1}(t,r)}{\sin\theta}
\frac{\partial Y^{\ell m}(\theta,\varphi)}{\partial\varphi}
\left(\delta^{1}_{\mu}\delta^{2}_{\nu}+\delta^{2}_{\mu}\delta^{1}_{\nu}\right)
+h^{\ell m}_{1}(t,r)\sin\theta
\frac{\partial Y^{\ell m}(\theta,\varphi)}{\partial\theta}
\left(\delta^{1}_{\mu}\delta^{3}_{\nu}+\delta^{3}_{\mu}\delta^{1}_{\nu}\right)\,.
\end{align}
The general polar electromagnetic perturbation is
\begin{align}\label{pertApol_gen}
\delta A^{(\mathrm{polar})}_{\mu}
=
\left(
a^{\ell m}_{0}(t,r)\,Y^{\ell m},
\,a^{\ell m}_{1}(t,r)\,Y^{\ell m},
\,a^{\ell m}_{2}(t,r)\,\partial_{\theta}Y^{\ell m},
\,a^{\ell m}_{2}(t,r)\,\partial_{\varphi}Y^{\ell m}
\right)\,,
\end{align}
After fixing the electromagnetic gauge, this reduces to
\begin{align}\label{pertApol_gf}
\delta A^{(\mathrm{polar})}_{\mu}
=
\left(
\frac{u^{\ell m}_{1}(t,r)}{r},\,0,\,0,\,0
\right)Y^{\ell m}(\theta,\varphi)\,.
\end{align}
The axial electromagnetic perturbation is
\begin{align}\label{pertAax_gen}
\delta A^{(\mathrm{axial})}_{\mu}
=
\left(
0,\,
0,\,
-\frac{u^{\ell m}_{4}(t,r)}{\sin\theta}\frac{\partial Y^{\ell m}(\theta,\varphi)}{\partial\varphi},\,
u^{\ell m}_{4}(t,r)\sin\theta\frac{\partial Y^{\ell m}(\theta,\varphi)}{\partial\theta}
\right),
\end{align}
This is the standard axial vector-harmonic decomposition. Lastly, the scalar perturbation is purely polar:
\begin{align}\label{pertPhi_gen}
\delta\Phi^{(\mathrm{polar})}
=
\frac{\delta\Phi^{\ell m}(t,r)}{r}\,
Y^{\ell m}(\theta,\varphi)\,.
\end{align}
Each radial amplitude has harmonic time dependence,
\begin{align}
\gamma^{\ell m}(t,r)=\gamma^{\ell m}(r)e^{-i\omega t},
\end{align}
for any of the radial perturbation functions
\(
\gamma^{\ell m}\in
\{H_0,H_1,H_2,K,h_0,h_1,u_1,u_4,\delta\Phi\}.
\)
Setting $\omega=0$ gives the static sector, for which the field equations imply $H_1^{\ell m}=h_1^{\ell m}=0$. Summation over $(\ell,m)$ is understood. Spherical symmetry allows us to set $m=0$ without loss of generality. Appendix~\ref{app:dynLove} gives the finite-frequency equations and their static and GR limit.

\textcolor{black}{For all three backgrounds, the scalar field vanishes, $\Phi=0$, and $V=0$. The black hole solutions are therefore supported entirely by the NLED sector. A scalar fluctuation $\delta\Phi$ can nevertheless be introduced as a separate probe. At linear order, it does not couple to the gravitational--electromagnetic perturbations and hence does not contribute to the gravitational multipole response.} 

\paperpart{PART I: Scalar probe response and shell EFT}
This part uses a minimally coupled scalar field equation for both the direct and the shell EFT calculation. We use this setup to describe the finite-frequency response in terms of renormalized EFT coefficients and to compare it with the usual source--response calculation. Since this calculation involves only the scalar field, it does not mix the spin sectors. 
The resulting scalar coefficient is a probe quantity and is not the coupled gravitational Love number calculated in Part II. \textcolor{black}{Although this is a probe calculation, the shell EFT provides analytic control over the low-frequency asymptotic structure of the response, allowing the universal logarithmic running to be separated from the finite, matching-dependent terms.}

\section{Shell EFT construction and renormalized response}
\label{sec:methods_eft}

We apply the scalar shell EFT of \cite{Kosmopoulos:2025rfj} to a minimally coupled massless scalar. The purpose of this part is to compare two descriptions of the same scalar response: a direct near-zone source--response decomposition and a finite-shell EFT matching. With $\Phi=e^{-i\omega t}Y_{\ell m}\psi(r)/r$, the exterior equation is
\begin{equation} \label{mster}
\begin{split}
&\frac{d^2\psi}{dr_*^2}+\Big[\omega^2-V(r)\Big]\psi=0\,,
\qquad
\frac{dr_*}{dr}=f^{-1}(r),\\&
\qquad
V(r)=f(r)\left[\frac{\ell(\ell+1)}{r^2}+\frac{f'(r)}{r}\right]\,.
\end{split}
\end{equation}
\begingroup
At a finite radius $r=R$, the exterior solution is matched to a regular
interior solution.  Write $f_R\equiv f(R)$.  Continuity of the induced metric
on the shell requires the flat-interior coordinates
\begin{equation*}
 \widetilde t=\sqrt{f_R}\,t,
 \qquad
 \widetilde r=R+\frac{r-R}{\sqrt{f_R}},
 \qquad
 \widetilde\omega=\frac{\omega}{\sqrt{f_R}}.
\end{equation*}
The relation $\widetilde r(R)=R$ ensures that the shell has the same areal radius on both sides. The radius $R$ is both the matching scale and the finite regulator of the shell description.

In terms of the reduced variable $\psi$ defined above, the regular interior solution is
\begin{equation}\label{psi_in}
 \psi_{\rm in}(r)
 =\widetilde A\,r\,
 j_\ell\!\left(\widetilde\omega\widetilde r(r)\right).
\end{equation}
At the shell, with $x_R=\omega R/\sqrt{f_R}$,
\begin{align*}
 \psi_{\rm in}(R)&=\widetilde A Rj_\ell(x_R),\\
 \psi'_{\rm in}(R)&=\widetilde A\left[
 j_\ell(x_R)+\frac{\omega R}{f_R}j_\ell'(x_R)\right].
\end{align*}
The amplitude $\widetilde A$ is fixed by continuity,
\begin{equation}
 \psi_{\ell}^{\rm out}(R)=\psi_\ell^{\rm in}(R)\equiv\psi_\ell(R).
\end{equation}
After rescaling the exterior solution to enforce this equality, the exact
shell response follows from the derivative jump,
\begin{align}
 F_\ell(\omega;R)
 &=\mathcal{N}_\ell(R)
 \frac{\psi'_{\rm in}(R)-\psi'_{\rm out}(R)}{\psi(R)},
 \label{eq:shell_derivative_jump}\\
 \mathcal{N}_\ell(R)
 &=\frac{4\pi\,2^\ell\ell!}{(2\ell+1)!}
 R^{2\ell+2}f_R^{(2\ell+1)/2}.
 \label{eq:shell_exact_normalization}
\end{align}
The numerical calculation evaluates this derivative jump with the factor  $\mathcal N_\ell$.
\endgroup

\textcolor{black}{
Following Ref.~\cite{Kosmopoulos:2025rfj}, the derivative jump given by Eq.(\ref{eq:shell_derivative_jump}) defines the finite-radius Shell EFT Wilson coefficient $F_\ell(\omega;R)$. Since the shell radius $R$ acts as a regulator, a prescription is required to extract the finite point-particle response. In the original Shell EFT construction this is done by expanding the finite-radius coefficient in the compact-object scale and subsequently taking the regulated point-particle limit. For the numerical matching performed here, we instead adopt a finite subtraction prescription that is more convenient to implement directly.}

\textcolor{black}{For each exterior geometry $X$, we compute the Wilson coefficient
$F_{\ell,\rm full}^{X}$ from the physical solution and the corresponding coefficient $F_{\ell,\rm src}^{X}$ from a canonical source solution in the same geometry.\footnote{Using the same exterior geometry is important because the background itself fixes the subleading structure of the applied-field branch, including the terms and logarithms forced by the radial equation. The reference solution therefore differs from the physical solution only by the freely addable homogeneous response, rather than by a change of background.}
The canonical source is normalized to unit growing amplitude and contains all terms required by the radial equation. Whenever the asymptotic recurrence reaches the radial power of the independent decaying homogeneous solution, the recurrence becomes resonant and a logarithmic term is required. At such an order, an arbitrary multiple of the decaying homogeneous solution may also be added. We set this free response coefficient to zero in defining the canonical
source.
}

We then define
\begin{equation}
F_{\ell,\rm ren}^{X}(\omega;R)
=
F_{\ell,\rm full}^{X}(\omega;R)
-
F_{\ell,\rm src}^{X}(\omega;R).
\end{equation}
\textcolor{black}{This subtraction fixes the canonical source configuration as the zero-response reference. It is therefore a finite renormalization prescription for the Shell EFT Wilson coefficient, chosen so that the shell result can be matched directly to the independently extracted source--response ratio. The resulting quantity is fitted as}
\begin{equation}\label{ren_F}
F_{\ell,\rm ren}^{X}(\omega;R)
=
B^{X}(\omega)
+\sum_{n\ge1}C_n^{X}(\omega)
\left(\frac{R_S}{R}\right)^n
+\mathcal O\!\left[(\omega R)^{2N+2}\right],
\end{equation}
\textcolor{black}{where $N$ denotes the truncation order of the low-frequency basis. As such, retaining the basis through $\omega^{2N}$ leaves the first omitted frequency term at order $(\omega R)^{2N+2}$. In the numerical calculation below we use $N=4$, corresponding to a basis through $\omega^8$. The finite coefficient $B^{X}(\omega)$ is the scalar response in this subtraction convention and is the quantity compared with the independently extracted source--response ratio.
}

\textcolor{black}{The logarithm associated with a resonant order requires a separate
qualification. It belongs to the canonical source branch\footnote{The logarithm appears in the canonical source branch because it is generated when the large-$r$ expansion of the unit-normalized growing solution reaches a resonant order. At that order, the radial power coincides with that of the independent decaying homogeneous solution, so the ordinary power-series recurrence degenerates and the radial equation requires a logarithmic particular term. Its coefficient is therefore fixed once the growing source amplitude is fixed. By contrast, the accompanying non-logarithmic decaying homogeneous solution may be added with an arbitrary coefficient and is identified with the free response.} and is therefore removed explicitly when $F_{\ell,\rm src}^{X}$ is subtracted from $F_{\ell,\rm full}^{X}$.} \textcolor{black}{This does not mean that the logarithmic information is lost. Rather, the coefficient multiplying the logarithm controls how the finite response changes under a change of logarithmic convention. To see this, write}
\begin{equation}
    L(r)\rightarrow L(r)+Q,
    \qquad
    L(r)\equiv \log\!\left(\frac{2M}{r}\right),
    \label{eq:shift}
\end{equation}
\textcolor{black}{where $Q$ is a constant. If the canonical source contains a term
$\lambda_{\log}^{X}(\omega)\,L(r)\,\psi_{\ell,\rm resp}(r)$ at the resonant order, the shift in Eq.~\eqref{eq:shift} generates the additional term $Q\,\lambda_{\log}^{X}(\omega)\,\psi_{\ell,\rm resp}(r)$. Since this has the same radial form as the homogeneous response, it can be absorbed into a finite redefinition of $B^{X}(\omega)$. Thus, the explicit logarithm does not appear separately in Eq.~\eqref{ren_F}, because it is part of the subtracted canonical
source, whereas its coefficient $\lambda_{\log}^{X}(\omega)$ remains relevant because it determines the scheme dependence, or running, of the finite response $B^{X}(\omega)$.}

Within the EFT description, these logarithmic and finite terms get multiplied by the local scalar tidal operators. Their coefficients are obtained by matching the full scalar source
and response solutions, without referring to any particular metric component. The shell EFT-based calculation, therefore, provides a gauge-invariant description of the probe response. The finite Wilson coefficients depend on the renormalization prescription and change if one shifts the  
logarithmic term as shown in Eq.~(\ref{eq:shift}). In contrast, the coefficients of the logarithmic terms remain unchanged, and subsequently the fully matched response is independent
of the arbitrary shell radius. 

\section{Perturbative shell EFT computation of the scalar probe response}
\label{sec:pert-shellEFT}
\begingroup
We now use the low-frequency expansion of the minimally coupled scalar equation
to determine the asymptotic EFT structure of the response. This analysis based on a large-$r$ and low-frequency expansion, shows how the EFT response is organized and identifies the terms that are fixed by the large-$r$ asymptotic equation before the global boundary condition and full matching are imposed to extract the Love numbers. We start with
\begin{equation}
 \Phi(t,r,\theta,\varphi)=e^{-i\omega t}Y_{\ell m}(\theta,\varphi)
 \frac{\psi(r)}{r}.
\end{equation}
The general spin-$s$ test-field equation of \cite{Coviello:2025pla}, specialized to $s=0$, is
\begin{align}
 \frac{d^2\psi}{dr_*^2}+[\omega^2-V_0(r)]\psi=0, \qquad
 &\frac{dr_*}{dr}=\frac{e^{\phi}}{f},
 \label{eq:scalar_master_cov}\\
 V_0=fe^{-2\phi}\left[
 \frac{\ell(\ell+1)}{r^2}+\frac{2m}{r^3}
 -\frac{2m'}{r^2}-\frac{f\phi'}{r}\right].
 \label{eq:scalar_cov_potential}
\end{align}
For the three geometries studied here, $\phi=0$ and
$f=1-2m/r$.  Therefore
\begin{equation}
 V_0(r)=f(r)\left[\frac{\ell(\ell+1)}{r^2}
 +\frac{f'(r)}{r}\right].
 \label{eq:scalar_potential_simple}
\end{equation}
Equation~\eqref{eq:scalar_potential_simple} is used in both the direct and shell calculations. 

\subsection{Static scalar exterior solution}

At $\omega=0$, Eq.~\eqref{eq:scalar_master_cov} becomes
\begin{equation}
 f\psi_0''+f'\psi_0'
 -\left[\frac{\ell(\ell+1)}{r^2}+\frac{f'}{r}\right]\psi_0=0.
 \label{eq:scalar_static_radial}
\end{equation}
We set $\ell=2$ and normalize the applied-field branch so that the coefficient of the leading term at $\mathcal{O}(r^3)$ is always unity.  The two asymptotic branches then scale as $r^3$ and $r^{-2}$.
Solving Eq.~\eqref{eq:scalar_static_radial} recursively in $1/r$ gives
\begin{equation}
 \psi_{\rm out}^{X}(r,0)=P_X(r)
 +\frac{\lambda_{\rm fin}^{X}(\mu)
 +\lambda_{\log}^{X}\ln(\mu r)}{r^2}
 +\mathcal O\!\left(\frac{\ln r}{r^3}\right),
 \label{eq:scalar_psi_out_static}
\end{equation}
where $X=(B,H,FW)$ corresponds to Bardeen, Hayward, and Fan-Wang regular geometries, respectively.
\textcolor{black}{The source-side polynomials, kept through the order where the source expansion first reaches the radial scaling of the decaying homogeneous response, are}
\begin{equation}
\begin{aligned}
 P_B(r)&=r^3-2Mr^2+\frac{2M^2}{3}r
 +\frac{3M^2\ell_B^2}{2r},\\
 P_H(r)&=r^3-2Mr^2+\frac{2M^2}{3}r
 -\frac{2M^2\ell_H^2}{r},\\
 P_{FW}(r)&=r^3-2Mr^2+\frac{2M(M+3\ell_{FW})}{3}r
 -\frac{2M\ell_{FW}^2(3M+5\ell_{FW})}{r}.
\end{aligned}
\label{eq:scalar_source_polynomials}
\end{equation}
\textcolor{black}{We truncate the source-side polynomials at this order because, for $\ell=2$, the growing source branch starts as $r^3$, while the independent response branch scales as $r^{-2}$. At the order where the source recurrence reaches the same $r^{-2}$ scaling, the two branches can mix, and the Frobenius solution develops a logarithmic term. Therefore, terms beyond this order are not needed to identify the leading response coefficient and its associated logarithmic running.}

The corresponding logarithmic coefficients are
\begin{equation}
\begin{aligned}
 \lambda_{\log}^{B}
 &=\frac{3M\ell_B^2}{5}(5\ell_B^2-4M^2),\\
 \lambda_{\log}^{H}
 &=\frac{32}{5}M^3\ell_H^2,\\
 \lambda_{\log}^{FW}
 &=\frac{8M\ell_{FW}^2}{5}
 (6M^2+20M\ell_{FW}+15\ell_{FW}^2).
\end{aligned}
\label{eq:scalar_lambda_logs}
\end{equation}

The large-$r$ asymptotic equation fixes the logarithmic coefficient. The associated rational constant changes if a different particular solution is used, so it has no separate invariant meaning. Also, both finite coefficients ($\lambda_{\rm fin}^{X}, \lambda_2^{X}$) are affected by the horizon condition and the choice of the finite subtraction convention. These are presented in Table~\ref{tab:scalar_scheme_summary}. The logarithmic entries in each column follow from the large-$r$ asymptotic equation; the finite terms depend on the global information and the subtraction convention. This separates universal running from scheme-dependent finite coefficients in the shell description.

\begin{table}[t]
\centering

\small
\renewcommand{\arraystretch}{1.45}
\begin{tabularx}{\textwidth}{@{}l X X X@{}}
\hline
Geometry & Asymptotic running coefficient
& Scheme-independent logarithmic  coefficients
& Finite and matching-dependent coefficients \\
\hline
Bardeen
& $\displaystyle \lambda_{\log}^{B}
=\frac{3M\ell_B^2}{5}(5\ell_B^2-4M^2)$
& $\displaystyle [\ln(\mu R)]\, \Delta F_2^{(0)}
=\frac{4\pi}{3}\lambda_{\log}^{B}$\newline
  $\displaystyle [\ln(\mu r)]\,\psi_{2,\rm resp}^{B}
=\frac{\lambda_{\log}^{B}}6$
& $\displaystyle \lambda_{\rm fin}^{B}(\mu),\ \lambda_2^{B}(\mu)$\newline
  finite constants accompanying the logarithms\newline
  $\displaystyle \sum_n C_n^{B}(2M/R)^n$ \\
\hline
Hayward
& $\displaystyle \lambda_{\log}^{H}=\frac{32}{5}M^3\ell_H^2$
& $\displaystyle [\ln(\mu R)]\,\Delta F_2^{(0)}
=\frac{4\pi}{3}\lambda_{\log}^{H}$\newline
  $\displaystyle [\ln(\mu r)]\,\psi_{2,\rm resp}^{H}
=\frac{\lambda_{\log}^{H}}6$
& $\displaystyle \lambda_{\rm fin}^{H}(\mu),\ \lambda_2^{H}(\mu)$\newline
  finite constants accompanying the logarithms\newline
  $\displaystyle \sum_n C_n^{H}(2M/R)^n$ \\
\hline
Fan--Wang
& $\displaystyle \lambda_{\log}^{FW}
=\frac{8M\ell_{FW}^2}{5}$\newline
  $\displaystyle \times(6M^2+20M\ell_{FW}+15\ell_{FW}^2)$
& $\displaystyle [\ln(\mu R)]\,\Delta F_2^{(0)}
=\frac{4\pi}{3}\lambda_{\log}^{FW}$\newline
  $\displaystyle [\ln(\mu r)]\,\psi_{2,\rm resp}^{FW}
=\frac{\lambda_{\log}^{FW}}6$
& $\displaystyle \lambda_{\rm fin}^{FW}(\mu),\ \lambda_2^{FW}(\mu)$\newline
  finite constants accompanying the logarithms\newline
  $\displaystyle \sum_n C_n^{FW}(2M/R)^n$ \\
\hline
\end{tabularx}
\caption{Structure of the renormalized scalar quadrupolar response for the three regular
black hole backgrounds are shown in this table. The second and third columns show the logarithmic terms
determined by the large-$r$ expansion. These coefficients control the logarithmic running and remain
unchanged under a finite redefinition of the source solution (Eq.~(\ref{eq:scalar_running}). The last column
lists the finite coefficients, whose values depend on the horizon boundary condition and the chosen subtraction convention. After evolving the finite coefficients according to Eq.~\eqref{eq:scalar_running}, the complete response is independent of the arbitrary shell radius.}
\label{tab:scalar_scheme_summary}
\end{table}

The three values of $\lambda_{\log}^{X}$, together with their logarithmic
coefficients, are determined locally from the large-$r$ recurrence. In contrast,
$\lambda_{\rm fin}^{X}$ multiplies the independent homogeneous response, so its
value is fixed only after imposing the horizon boundary condition and choosing
a finite subtraction prescription. The rational constant appearing separately
in Eq.~\eqref{eq:scalar_scheme_split} is therefore not an observable by itself.
Only its combination with the running finite coefficient is independent of the
scale.
Changing the scale as
$\mu\rightarrow e^{\sigma}\mu$ leaves
Eq.~\eqref{eq:scalar_psi_out_static} unchanged when
\begin{equation}
 \lambda_{\rm fin}^{X}(e^\sigma\mu)
 =\lambda_{\rm fin}^{X}(\mu)-\sigma\lambda_{\log}^{X}.
 \label{eq:scalar_running}
\end{equation}
The logarithmic running is fixed, whereas an isolated finite constant depends on the convention.

\textcolor{black}{The coefficients in Eqs.~\eqref{eq:scalar_source_polynomials} and \eqref{eq:scalar_lambda_logs}
were obtained symbolically from the large-$r$ recurrence of the exact radial equation. As an independent check, we reproduced the same coefficients in a separate \texttt{Python} implementation and verified that substituting the resulting expansions into the exact radial equation gives a vanishing residual through the retained order}
\cite{BhattacharyyaKumarKumar_DynamicTLN_RBH_Code}.

\subsection{Low-frequency scalar solution}

In the intermediate near zone, $M\ll r\ll\omega^{-1}$, we expand the scalar field as
\begin{equation}
\psi=\psi_0+\omega^2\psi_2+\mathcal O(\omega^4).
\label{eq:scalar_low_frequency_expansion}
\end{equation}
The source part of $\psi_2$ contains the growing terms that arise from the
small-$\omega r$ expansion of the applied field. We denote these terms by
$\psi_{2,{\rm src}}^X$. They are retained in the unit-normalized source branch but are separated from the response calculation. At leading large-$r$ order, the part generated
by the static response obeys
\begin{equation}
 \left(\frac{d^2}{dr^2}-\frac{6}{r^2}\right)
 \psi_{2,{\rm resp}}^X
 =-\frac{\lambda_{\rm fin}^X
 +\lambda_{\log}^X\ln(\mu r)}{r^2}.
 \label{eq:scalar_omega2_inhomogeneous}
\end{equation}
A convenient particular solution is
\begin{equation}
 \psi_{2,{\rm resp}}^X=
 \frac{\lambda_{\log}^X}{6}\ln(\mu r)
 +\frac{\lambda_{\rm fin}^X-\lambda_{\log}^X/6}{6}.
 \label{eq:scalar_omega2_particular}
\end{equation}

\textcolor{black}{Combining the static response with this $O(\omega^2)$ correction, the response branch can be written as}
\begin{align}
 \psi_{{\rm resp}}^{X}(r,\omega)={}&
 \frac{\lambda_{\log}^{X}\ln(\mu r)+\lambda_{\rm fin}^{X}(\mu)}{r^2}
 \nonumber\\
 &+\omega^2\left\{
 \frac{\lambda_{\log}^{X}}{6}
 \left[\ln(\mu r)-\frac{1}{6}\right]
 +\frac{\lambda_{\rm fin}^{X}(\mu)}{6}
 +\frac{\lambda_2^{X}(\mu)}{r^2}\right\}
 +\mathcal O(\omega^4).
 \label{eq:scalar_scheme_split}
\end{align}

\textcolor{black}{Using Eq.(\ref{eq:scalar_low_frequency_expansion})}, the corresponding low-frequency asymptotic exterior expansion through this order may therefore be written as
\begin{align}
 \psi_{\rm out}^{X}(r,\omega)={}&\psi_{\rm out}^{X}(r,0)
 +\omega^2\bigg[\psi_{2,{\rm src}}^X(r)
 +\frac{\lambda_{\log}^X}{6}\ln(\mu r)
 +\frac{\lambda_{\rm fin}^X-\lambda_{\log}^X/6}{6}
 +\frac{\lambda_2^X(\mu)}{r^2}\bigg]
 +\mathcal O(\omega^4).
 \label{eq:scalar_psi_out_dynamic}
\end{align}
Equation~\eqref{eq:scalar_omega2_particular} is a convenient particular
solution generated by the static response.  Its logarithmic coefficient
$\lambda_{\log}^{X}/6$ follows from the far-zone equation. However, the finite constant has to be considered together with the running of $\lambda_{\rm fin}^{X}$. The homogeneous coefficient $\lambda_2^X$ cannot be determined from the far-zone equation alone. It must be fixed from the global horizon or by shell matching, as for the static finite coefficient. This has been explicitly done in the numerical shell matching of Section~\ref{subsec:shell-probe-matching}. Thus, the present analytic calculation determines the asymptotic EFT structure and its running, while the complete dynamical response is obtained only after the global matching condition is imposed.

\subsection{Shell response, source subtraction, and convention dependence}
With the asymptotic source and response structure identified, we now
return to the shell construction and apply the renormalization prescription defined above (in Section~\ref{sec:methods_eft}).

Let $F_{2,\rm full}^{X}(\omega;R)$ denote the finite-radius Shell-EFT Wilson coefficient obtained from the full ingoing black hole solution, and let $F_{2,\rm src}^{X}(\omega;R)$ denote the corresponding coefficient obtained from the canonical source solution in the same geometry. Using Eq.~(\ref{eq:shell_derivative_jump}), we have
\begin{align}
 F_{2,\rm full}^{X}(\omega;R)
 &=\mathcal N_2(R)\left[L_{\rm in}(R)-L_{\rm full}^{X}(R)\right],\\
 F_{2,\rm src}^{X}(\omega;R)
 &=\mathcal N_2(R)\left[L_{\rm in}(R)-L_{\rm src}^{X}(R)\right],
 \qquad L\equiv\frac{\psi'}{\psi}.
\end{align}
where the quadrupolar normalization entering the derivative jump is given as (using Eq.(\ref{eq:shell_exact_normalization}))
\begin{equation}
 \mathcal N_2(R)=\frac{4\pi}{15}R^6 f_R^{5/2}.
 \label{eq:scalar_exact_quadrupole_normalization}
\end{equation}

Their difference is independent of the interior solution,
\begin{equation}
 F_{2,\rm ren}^{X}(\omega;R)
 \equiv F_{2,\rm full}^{X}-F_{2,\rm src}^{X}
 =\mathcal N_2(R)\left[L_{\rm src}^{X}(R)-L_{\rm full}^{X}(R)\right].
 \label{eq:scalar_same_background_subtraction}
\end{equation}
This is the same-background finite renormalization used in the numerical
matching.


For direct comparison with the source--response coefficient extracted in the intermediate near-zone, we define 
\begin{equation}
 \Lambda_{2,\rm shell}^{X}(\omega;R)
 =\frac{3}{4\pi}F_{2,\rm ren}^{X}(\omega;R).
 \label{eq:scalar_renormalized_response}
\end{equation}
In the overlap region, it is fitted as
\begin{equation}
 \Lambda_{2,\rm shell}^{X}(\omega;R)
 =\Lambda_0^{X}(\omega)
 +\sum_{n\geq1}D_n^{X}(\omega)\left(\frac{2M}{R}\right)^n
 +\mathcal O\!\left[(\omega R)^{2N+2}\right],
 \label{eq:scalar_shell_dynamic_fit}
\end{equation}
\textcolor{black}{where $N$ is the truncation index of the low-frequency basis. In our calculation $N=4$, corresponding to a basis through $\omega^8$.
The constant $\Lambda_0^{X}(\omega)$ is the finite renormalized scalar response and is compared directly with the source--response ratio in the next section. The coefficients $D_n^{X}(\omega)$ describe finite-$R$ corrections that vanish under extrapolation and therefore do not represent additional Love coefficients. As discussed above, changing the logarithmic convention shifts the finite coefficient, while the logarithmic coefficient listed in Table~\ref{tab:scalar_scheme_summary} remains unchanged.}
\endgroup

\section{Numerical scalar matching between shell EFT and the direct scalar response}
\label{subsec:shell-probe-matching}
\begingroup
We now implement the global matching numerically. Unlike the asymptotic
analysis in Section~\ref{sec:pert-shellEFT}, the physical black hole solution is
obtained by integrating the full scalar radial equation with the ingoing
boundary condition at the horizon. The shell and direct calculations therefore
use the same globally evolved scalar solution, while the low-frequency
Frobenius expansion is used only to define the source--response basis in the
overlap region (large-$r$, low frequency expansion). At finite frequency, the reference source
and response solutions are constructed through $\omega^8$, including the
logarithmic terms required by the Frobenius recurrence. These solutions are
initialized at $r=120M$, within the overlap region, and then integrated inward
using the full scalar equation.

For the direct calculation, the ingoing horizon solution is decomposed as
\begin{equation}
 \psi_{\rm full}^{X}
 =C_{\rm src}^{X}\psi_{\rm src}^{X}
 +C_{\rm resp}^{X}\psi_{\rm resp}^{X},
\end{equation}
from which we define
\begin{equation}
 \Lambda_0^{X}(\omega)
 =
 \left(\frac{C_{\rm resp}}{C_{\rm src}}\right)_{X}.
 \label{eq:direct_scalar_response}
\end{equation}
For the shell calculation, Eq.~\eqref{eq:shell_derivative_jump} is evaluated
for both the full solution and the reference source solution in the same
geometry, and the same-background subtraction in
Eq.~\eqref{eq:scalar_same_background_subtraction} is then applied. The results
at
\begin{equation}
 R/M=12,16,20,24,30,40,50,60,80
\end{equation}
are extrapolated using Eq.~\eqref{eq:scalar_shell_dynamic_fit}.

Figure~\ref{fig:scalar_shell_all} therefore provides a direct comparison
between the shell calculation and the usual near-zone source--response
decomposition. The two results agree once the local source contributions are
treated consistently in both calculations.

\begin{figure}[t]
 \centering
 \includegraphics[width=0.96\textwidth]{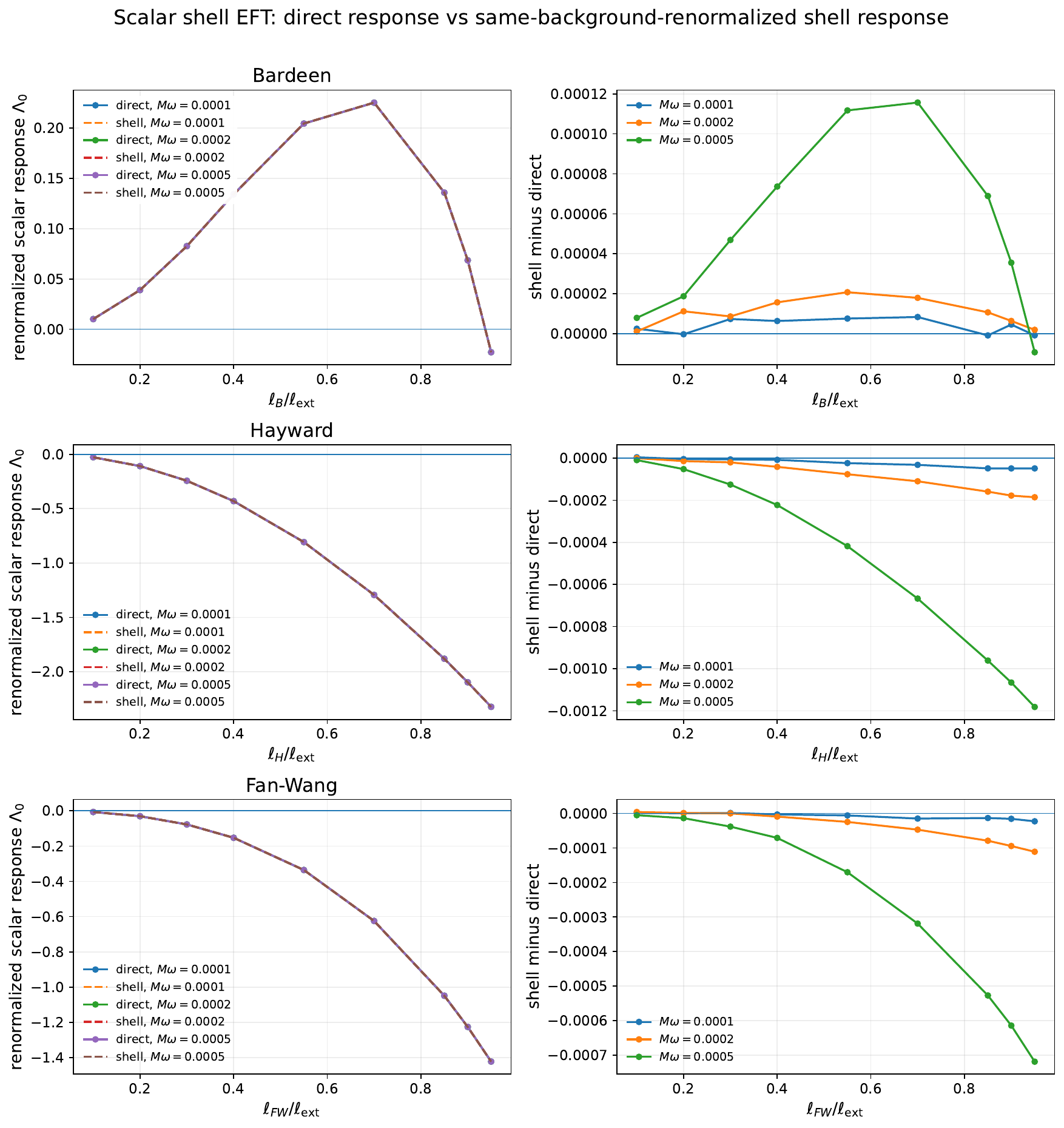}
 \caption{A comparison between the direct scalar response and the scalar shell-EFT matching for the cases of Bardeen, Hayward, and Fan--Wang. The solid curves are obtained by decomposing the ingoing
black hole solution into its scalar source and response components in the
intermediate near zone. The dashed curves are obtained independently from the
shell derivative jump after subtracting the canonical source solution in the
same background, as explained in the main text. These two almost overlap. The right panels show the residual
$\Lambda_0^{\rm shell}-\Lambda_0^{\rm direct}$, which indicates the precision of the agreement between the two calculations.}
 \label{fig:scalar_shell_all}
\end{figure}

At $M\omega=10^{-4}$, the relative root-mean-square (RMS) differences between
the renormalized shell response and the direct unsubtracted response are
$4.20\times10^{-5}$, $2.36\times10^{-5}$, and $1.56\times10^{-5}$ for
Bardeen, Hayward, and Fan--Wang, respectively. At the largest frequency
considered, $M\omega=5\times10^{-4}$, these differences are $5.30\times10^{-4}$, $5.10\times10^{-4}$, and $5.04\times10^{-4}$,
respectively. The largest absolute residuals in the real part over the \textcolor{black}{full scan in the regularization parameter} also occur at this frequency, with values
$1.16\times10^{-4}$, $1.18\times10^{-3}$, and $7.18\times10^{-4}$ for
Bardeen, Hayward, and Fan--Wang, respectively.

The importance of same-background subtraction can be clearly illustrated as follows. For example, for the Bardeen geometry at $\ell_B/\ell_{\rm ext}=0.70$ and $M\omega=10^{-4}$. As a
check, if one subtracts instead the Schwarzschild shell jump, the result grows from about $-18$ at
$R=12M$ to $-1.06\times10^3$ at $R=80M$. This growth comes from local source terms that scale with positive powers of $R$ and points to the absence of numerical convergence. In contrast, after subtracting the reference Bardeen source solution, the result decreases smoothly from $0.2441$ to $0.2280$ and shows numerical convergence toward a finite extrapolated value. This demonstrates that to meaningfully isolate the shell response, one needs to perform the subtraction with respect to the same background, not with respect to any fiducial reference (i.e Schwarzschild in this case).

When the value is at $\ell/\ell_{\rm ext}=0.70$ and $M\omega=5\times10^{-4}$, moving the
direct projection radius through $18M$, $24M$, and $30M$ changes the result by
less than $3.0\times10^{-8}$ relatively for Bardeen and below
$7\times10^{-9}$ for Hayward and Fan--Wang. If one moves the basis
initialization radius through $100M$, $120M$, and $140M$, it changes the real
response by $4.71\times10^{-5}$, $1.59\times10^{-5}$, and
$6.42\times10^{-6}$ relatively, respectively. On the other hand, if one restricts the shell fit to
$R\geq16M$ or $R\geq20M$, the extrapolated finite part changes only at the
level expected from the first omitted inverse-radius terms, and consequently the fit residuals
remain below $1.6\times10^{-5}$ at the representative point \cite{BhattacharyyaKumarKumar_DynamicTLN_RBH_Code}.

We independently check the junction implementation by differentiating Eq.~(\ref{psi_in}),
$\psi_{\rm in}=\widetilde A rj_2(\widetilde\omega\widetilde r)$ numerically \cite{BhattacharyyaKumarKumar_DynamicTLN_RBH_Code}.
The maximum relative derivative error for all three geometries is found to be
$4.8\times10^{-11}$. It is important to note that the agreement between the direct and shell determinations is a nontrivial EFT matching test. It shows that the finite-frequency scalar response can be encoded in a renormalized Wilson coefficient, while the shell-radius dependence is separated into universal running and removable local terms. In this controlled probe sector, the dynamical response is therefore a well-defined, gauge-invariant observable even though an isolated finite coefficient remains scheme dependent. It is equally important to recall that this scalar matching remains separate from both the tensor test field given in Appendix~\ref{app:tensor_probe} and the coupled gravitational Love numbers discussed next.
\endgroup

\paperpart{PART II: Full coupled gravitational-electromagnetic response}
This computes the physical polar and axial response of the coupled Einstein--NLED system. Readers interested in a comparison between genuine coupled dynamics from a spin-2 probe tensor field dynamics are encouraged to look at Appendix~\ref{app:tensor_probe}.

\section{Method for computing the full coupled dynamical Love numbers}
\label{sec:methods}

\textcolor{black}{We obtain the gravitational response from the coupled Einstein--NLED perturbation equations with the appropriate wave boundary condition, namely a purely ingoing solution at the future horizon. The calculation first determines the coupled master-field solutions with the appropriate horizon boundary condition, then separates the source and response branches, and finally extracts the tidal response from the corresponding physical metric variables.}
\subsection{Direct integration and numerical methods}
\label{subsec:methods_wave}
\begingroup
\textcolor{black}{The perturbation equations for both parity sectors can be written in terms of gauge-invariant variables in the matrix form}
\begin{equation}
 \frac{d^2\mathbf U_{\pm}}{dr_*^2}
 +\left[\omega^2\mathbf I-f(r)\mathbf V_{\pm}(r)\right]\mathbf U_{\pm}=0,
 \qquad \frac{dr_*}{dr}=\frac{1}{f(r)},
 \label{eq:master_matrix_main}
\end{equation}
where $+$ and $-$ denote polar and axial parity. Each vector $\mathbf U_\pm$
contains one gravitational and one electromagnetic master amplitude. The
potential matrices are given in Appendix~\ref{app:dynLove}\,. The same equations apply to Bardeen, Hayward, and Fan--Wang after inserting their corresponding $m(r)$, $\mathcal L(r)$, and $\mathcal L_F(r)$.


\textcolor{black}{For the time dependence $e^{-i\omega t}$, the physical solution at the future horizon is purely ingoing and is given by}
\begin{equation}
 \mathbf U_\pm=e^{-i\omega r_*}
 \left[\mathbf a_0+\mathbf a_1(r-r_h)+\cdots\right],
 \qquad
 \frac{d\mathbf U_\pm}{dr_*}=-i\omega\mathbf U_\pm+\mathcal O(r-r_h).
 \label{eq:ingoing_bc}
\end{equation}
Thus $d\mathbf U/dr=-i\omega\mathbf U/f+\mathcal O(1)$ near the horizon. 

We integrate the two-component (refer to Eq.~(\ref{eq:app_master})) ingoing basis from $r_h+\epsilon$ to a common
matching radius. Independently, reference source and response bases are
integrated inward from the asymptotic region. At the matching radius, the horizon ingoing basis is expressed as a linear combination of the asymptotic source and response bases, with $\mathbf A$ and $\mathbf B$ denoting the corresponding coefficient matrices
\begin{equation}
 \mathcal R_\pm(\omega)=\mathbf B\mathbf A^{-1}.
 \label{eq:response_matrix}
\end{equation}
This response matrix is a useful scattering diagnostic, but it is not yet the Love number.

The Love number is obtained only after reconstructing the physical metric. We reconstruct the physical metric
variables and define
\begin{equation}
 Z_+(r,\omega)=-f(r)H_0(r,\omega),\qquad
 Z_-(r,\omega)=h_0(r,\omega).
\end{equation}
For the quadrupole, their near-zone source-response decompositions are (refer to Appendix~(\ref{app:dynLove}) for more details),
\begin{align}
 Z_+&=C_{+,\mathrm{s}}(\omega)r^2
      +C_{+,\mathrm{r}}(\omega)r^{-3}+\cdots,
 \\
 Z_-&=C_{-,\mathrm{s}}(\omega)r^3
      +C_{-,\mathrm{r}}(\omega)r^{-2}+\cdots .
 \label{eq:metric_nearzone_main}
\end{align}
Using the normalization of Ref.~\cite{Coviello:2025pla},
\begin{equation}
 k_{20}^{\mathrm{polar}}
 =-\frac{1}{M^5}\frac{C_{+,\mathrm{r}}}{C_{+,\mathrm{s}}},
 \qquad
 k_{20}^{\mathrm{axial}}
 =\frac{1}{M^5}\frac{C_{-,\mathrm{r}}}{C_{-,\mathrm{s}}}.
 \label{eq:metric_love_conversion_main}
\end{equation}
Appendix~\ref{app:dynLove} derives these factors and gives the exact
master variable to metric component reconstruction formulas. The explicit numerical overlap expansion used to separate the finite-frequency source and response branches is given in Appendix~\ref{app:numerical_asymptotics}. In this overlap region, the four near-zone basis solutions are normalized according to their leading physical gravitational or electromagnetic source and response amplitudes.

At each frequency, we construct four coupled near-zone solutions: gravitational
source, electromagnetic source, gravitational response, and electromagnetic
response. We invert the physical source and response matrices. This step is essential for Fan--Wang, where the two master channels have the same leading asymptotic index. The physical solution is selected by
\begin{equation}
 (C_{g,\mathrm{s}},C_{e,\mathrm{s}})=(1,0),
 \qquad
 (C_{g,\mathrm{r}},C_{e,\mathrm{r}})=(c_{g,\pm},c_{e,\pm}).
 \label{eq:pure_grav_source_main}
\end{equation}
The induced electromagnetic response $c_{e,\pm}$ is retained, but there is no
independent electromagnetic tide. {\color{black}Note that here, $C_{\pm,s}$ and $C_{\pm,r}$ in Eqs.~(\ref{eq:metric_nearzone_main})--(\ref{eq:metric_love_conversion_main}) denote the source and response coefficients of the physical metric variables $Z_{\pm}$, whereas $C_{g/e,s/r}$ in Eq.~(\ref{eq:pure_grav_source_main}) denote the amplitudes of the gravitational/electromagnetic source and response branches in the normalized near-zone basis constructed in Appendix~\ref{app:numerical_asymptotics}. The lowercase coefficients $c_{g,\pm}$ and $c_{e,\pm}$ are the corresponding response amplitudes obtained by matching the horizon-ingoing solution onto this basis.}

\subsubsection{Independent static calculation in all sectors}
{\color{black}As established above, in the general source–response setup, we first determine the $\omega=0$ response, which provides the baseline and is further used to define the contribution coming from the leading frequency correction.} With this, we now determine all six static coefficients. For each coupled master system, we set $\omega=0$ and construct four Frobenius solutions at large radius, corresponding to the gravitational and electromagnetic source and response branches. The gravitational source amplitude is normalized to unity, whereas the independent electromagnetic source is set to zero, allowing to retain only the induced electromagnetic response. Then the resulting solutions are matched to a two-dimensional horizon-regular basis at $r/M=10,12,14,16$.

After reconstructing the metric, a logarithmic term may appear at the same
radial order as the response. We use $\text{Log}(r/2M)$ consistently in all six
calculations and separate this logarithmic running from the finite response.
The source branch may also contain a finite local contribution at the response
order. This contribution is included before forming the physical metric
source--response ratio. Thus
\begin{equation}
 k_{2,+}^{X,\mathrm{static}}=-\frac{c_{\mathrm{hor}}+c_{\mathrm{loc}}}{M^5},
 \qquad
 k_{2,-}^{X,\mathrm{static}}=+\frac{c_{\mathrm{hor}}+c_{\mathrm{loc}}}{M^5},
 \label{eq:all_static_direct_definition}
\end{equation}
where $X\in\{B,H,FW\}$. The sign difference follows directly from the source
sign in the metric asymptotics of Eqs.~\eqref{eq:metric_nearzone_main}-\eqref{eq:metric_love_conversion_main}.

Denoting $x_X=\ell_X/M$ (with $X \in \{B, H, FW)\})$, the numerical continuations over the full sampled
regularization range are
\begin{align}
 k_{2,+}^{B}&=0.08023x_B^2+10.76112x_B^4-0.14944x_B^6-0.18091x_B^8,\nonumber\\
 k_{2,-}^{B}&=-5.76385x_B^4-1.00956x_B^6-0.01640x_B^8-0.45960x_B^{10},\nonumber\\
 k_{2,+}^{H}&=12.80654x_H^2-0.50802x_H^4-0.05286x_H^6-0.07816x_H^8,\nonumber\\
 k_{2,-}^{H}&=1.79976x_H^4+0.10162x_H^6+0.00991x_H^8+0.04609x_H^{10},\nonumber\\
 k_{2,+}^{FW}&=215.47137x_{FW}^3+95.23700x_{FW}^4+1.46516x_{FW}^5-48.49016x_{FW}^6,\nonumber\\
 k_{2,-}^{FW}&=95.33770x_{FW}^3+1.80514x_{FW}^4-12.97444x_{FW}^5+12.73972x_{FW}^6.
 \label{eq:all_static_continuation_fits}
\end{align}
The root mean square (rms) deviations of these fits are between $1.0\times10^{-7}$ and
$5.8\times10^{-6}$. Small-regularization fits independently recover the
leading scalings: $k_{2,+}^{B}\sim0.0798x_B^2+10.7598x_B^4$,
$k_{2,-}^{B}\sim-5.7657x_B^4$, $k_{2,+}^{H}\sim12.8061x_H^2$,
$k_{2,-}^{H}\sim1.7999x_H^4$, $k_{2,+}^{FW}\sim215.37x_{FW}^3$, and
$k_{2,-}^{FW}\sim95.41x_{FW}^3$. These reproduce the powers and
the coefficients reported in Ref.~\cite{Coviello:2025pla}. The least stable
individual coefficient is the subdominant Bardeen-polar quadratic term; its
quartic coefficient and the complete static response are numerically stable.

Table~\ref{tab:static_scaling_comparison} separates the behavior
from the terms required to continue the result toward extremality. The second
column is obtained by fitting only the points with
$\ell_X/\ell_{\rm ext}\leq0.3$. The last column contains the additional terms
in the fit to the complete sampled range. These terms are suppressed near the
Schwarzschild limit, but can have considerable magnitude and affect
the response when $x_X$ is not very small.

\begin{table}[H]
\centering
\caption{Small-regularization scaling of the independently calculated static
Love numbers compared with Ref.~\cite{Coviello:2025pla}. Here
$x_X=\ell_X/M$ ($X \in \{B, H, FW)\}$). The last column lists the higher powers retained in our
full-range continuation. All coefficients, including those of the leading
powers, are refitted over the complete sampled range.}
\label{tab:static_scaling_comparison}
\resizebox{\textwidth}{!}{
\begin{tabular}{lccc}
\hline
Sector & Our small-$x_X$ fit & Ref.~\cite{Coviello:2025pla} & Higher powers in our full-range fit\\
\hline
Bardeen polar & $0.0798x_B^2+10.7598x_B^4$
 & $0.03x_B^2+10x_B^4$ & $-0.1494x_B^6-0.1809x_B^8$\\
Bardeen axial & $-5.7657x_B^4-0.9893x_B^6$
 & $-5.8x_B^4$ & $-0.0164x_B^8-0.4596x_B^{10}$\\
Hayward polar & $12.8061x_H^2-0.5036x_H^4$
 & $13x_H^2$ & $-0.0529x_H^6-0.0782x_H^8$\\
Hayward axial & $1.7999x_H^4+0.1005x_H^6$
 & $1.8x_H^4$ & $0.0099x_H^8+0.0461x_H^{10}$\\
Fan--Wang polar & $215.37x_{FW}^3+96.10x_{FW}^4$
 & $220x_{FW}^3$ & $1.465x_{FW}^5-48.49x_{FW}^6$\\
Fan--Wang axial & $95.407x_{FW}^3-0.0155x_{FW}^4$
 & $95x_{FW}^3$ & $-12.974x_{FW}^5+12.740x_{FW}^6$\\
\hline
\end{tabular}
}
\end{table}

The leading powers agree in every sector. The corresponding coefficients also
agree at the few-percent level, except for the Bardeen-polar $x_B^2$
coefficient. In that sector the quadratic contribution is much smaller than
the quartic contribution over most of the calculated range, so its fitted
coefficient is sensitive to the fitting window. The quartic coefficient is
stable and agrees with the value reported in Ref.~\cite{Coviello:2025pla}.

\paragraph{Relation to the numerical method of Ref.~\cite{Coviello:2025pla}.}
Both calculations solve the zero-frequency coupled gravitational--matter
system with a horizon-regular inner solution, a unit gravitational tide, no
independent matter tide, and the same $\log(r/2M)$ convention. Ref.~\cite{Coviello:2025pla} constructs fifth-order expansions at the horizon and at infinity. They then integrate them from both boundaries and match the perturbations and their derivatives at a common radius. The leading scaling is then determined by fitting the Love number curve over smaller intervals around $X=0$.

Our implementation uses a different numerical basis. We first construct four coupled Frobenius modes at infinity: gravitational and electromagnetic source
and response modes. We then impose the physical source condition by inverting the source and response matrices. This inversion is required for Fan--Wang,
where the leading master indices are degenerate. We then match two horizon-regular modes to the asymptotic basis at $r/M=10,12,14,16$ and extrapolate the response with respect to the matching radius. For the final numerical calculations, we retain terms up to 
\textcolor{black}{For the final numerical calculations, the large-$r$ Frobenius basis is retained through inverse-radius index 22 and logarithmic power $[\ln(r/2M)]^6$.}   
Table~\ref{tab:all_static_direct} gives the convergence test when the radial order is increased to 24. Finally, we reconstruct  $-fH_0$ or $h_0$, keeping the finite response-power term coming from the source branch. Thus, the agreement in the leading small-$x_X$ scaling is a cross-check between two numerical implementations, while our full-range fit makes the
higher-order contributions explicit.
\begin{table}[h]
\centering
\small
\caption{Independent static results at $\ell_X/\ell_{\rm ext}=0.5$ ($X \in \{B, H, FW)\}$). The
third column is the small-regularization expression of
Ref.~\cite{Coviello:2025pla} evaluated at the same point. The last column is
the fractional change when the radial series order is increased from 22 to
24.}
\label{tab:all_static_direct}
\begin{tabular}{lccc}
\hline
Sector & Direct $k_2$ & Ref.~\cite{Coviello:2025pla} small-$\ell$ truncation & Order change\\
\hline
Bardeen polar & $0.247504$ & $0.223923$ & $1.30\times10^{-5}$\\
Bardeen axial & $-0.129825$ & $-0.127298$ & $1.05\times10^{-6}$\\
Hayward polar & $1.885909$ & $1.925926$ & $2.37\times10^{-6}$\\
Hayward axial & $0.039839$ & $0.039506$ & $4.91\times10^{-5}$\\
Fan--Wang polar & $0.746087$ & $0.715338$ & $2.64\times10^{-5}$\\
Fan--Wang axial & $0.310071$ & $0.308896$ & $7.16\times10^{-6}$\\
\hline
\end{tabular}
\end{table}

\begin{figure}[tbh]
\centering
\includegraphics[width=0.98\textwidth]{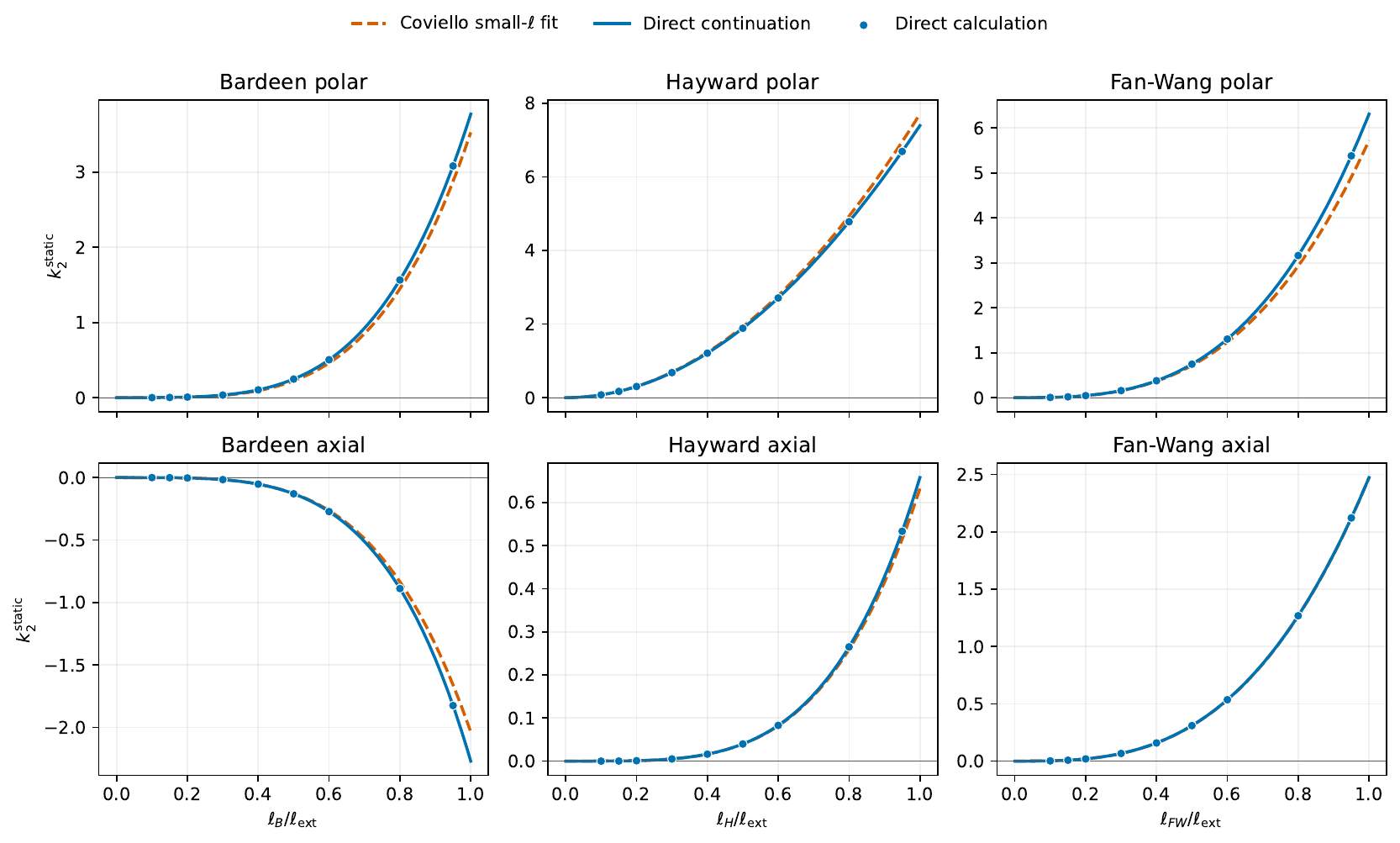}
\caption{Independent quadrupolar static responses in both parities. Points
are obtained directly from the zero-frequency coupled systems. The solid curves
are the full-range continuations in
Eq.~\eqref{eq:all_static_continuation_fits}, and dashed curves are the
small-regularization expressions of Ref.~\cite{Coviello:2025pla}. Their
separation toward extremality measures higher-order regularization effects.}
\label{fig:all_static_direct}
\end{figure}
Figure~\ref{fig:all_static_direct} summarizes these independently calculated static responses over the full regularization range. It also shows where the small-regularization expressions remain accurate and where higher-order regularization corrections become important toward extremality. {\color{black}With the independent static response established and numerically validated, we now use it as the zero-frequency anchor for the dynamical calculation. The finite-frequency problem then determines the change of the response relative to this static limit.}

The raw metric coefficient ratio $c_{g,\pm}^{X}(\omega)$ is retained in the
numerical tables. At finite frequency, we define the dynamical Love number
from the reconstructed metric source--response ratio. The independently
calculated static Love number fixes the zero-frequency value, while the
frequency dependence is determined by the change of the response away from
$\omega=0$:
\begin{equation}
 \mathcal K_{2,\pm}^{X}(\omega)
 =
 k_{2,\pm}^{X,\mathrm{static}}
 +
 \gamma_\pm
 \left[
 c_{g,\pm}^{X}(\omega)-c_{g,\pm}^{X}(0)
 \right],
 \label{eq:metric_dynamic_definition}
\end{equation}
with
\begin{equation}
 \gamma_+=-\frac{1}{M^5},
 \qquad
 \gamma_-=\frac{1}{M^5}.
\end{equation}

Here $c_{g,\pm}^{X}(0)$ is the zero-frequency source--response ratio for the
same model $X$ and parity. Since the static contribution is already included
through the independently calculated $k_{2,\pm}^{X,\mathrm{static}}$,
subtracting $c_{g,\pm}^{X}(0)$ prevents the static response from being counted
twice. Hence,
$c_{g,\pm}^{X}(\omega)-c_{g,\pm}^{X}(0)$ contains only the
frequency-dependent change of the response. Consequently,
\begin{equation}
 \lim_{\omega\to0}\mathcal K_{2,\pm}^{X}(\omega)
 =
 k_{2,\pm}^{X,\mathrm{static}}.
\end{equation}

For every model and parity, the static value in
Eq.~\eqref{eq:metric_dynamic_definition} is the corresponding independently
computed result in Eq.~\eqref{eq:all_static_continuation_fits}. The
conservative response is
\begin{equation}
 k_{2,\pm}(\omega)
 =
 \operatorname{Re}\mathcal K_{2,\pm}(\omega),
\end{equation}
while $\operatorname{Im}\mathcal K_{2,\pm}(\omega)$ gives the dissipative
response. Thus, the static TLN and its small-regularization scaling are first
obtained independently, and the finite-frequency calculation then determines
their dynamical continuation.


\textcolor{black}{The near-zone basis is kept through $O(\omega^4)$ and matched at
$r/M=10,12,14,16$. Quantitative comparisons are restricted to
$M\omega\leq0.004$. The $M\omega=0.006$ point is shown only to illustrate the increased matching sensitivity as the overlap condition $\omega r\ll1$ becomes less accurate. Wider-frequency scans are nevertheless appropriate for the resonance test because a resonance is a feature of the global scattering solution rather than of the intermediate-zone source--response expansion. The resonance analysis therefore uses the full master-field solution and does not rely on the low-frequency overlap condition required for the dynamical-TLN extraction.}

\subsubsection{Numerical strategy for evaluating dynamical TLNs and resonance features}
\label{sec:numerics}
\begingroup
The low-frequency metric calculation uses adaptive complex ODE integration
starting at $r_h+2\times10^{-6}M$, with relative and absolute tolerances of
$5\times10^{-10}$ and $5\times10^{-12}$, respectively.
\textcolor{black}{For the polar sector, the large-$r$ Frobenius basis is kept through inverse-radius index 18 and logarithmic power $[\ln(r/2M)]^5$. In the axial sector, the inverse-radius index is 22; the logarithmic power is 5 for Bardeen and 6 for Hayward and Fan--Wang. These numbers specify the truncation of the asymptotic series used in the numerical analysis.} In all cases, we retain frequency-dependent terms through $O(\omega^4)$.

The metric response is evaluated at four matching radii and extrapolated by
including the omitted leading $(\omega r)^6$ contribution together with
inverse-radius truncation terms. As consistency checks, we monitor the
normalization errors in Eq.~\eqref{eq:pure_grav_source_main}, verify the
frequency-reversal relation
$\mathcal K(-\omega)=\mathcal K(\omega)^*$, and test the stability of the
result under changes in the matching window and asymptotic order.

The resonance analysis is performed separately from the low-frequency metric
extraction. We scan the master-field response for broad maxima along the real
frequency axis and independently fit the time-domain signal of the same coupled
system to a damped sinusoid. For the axial sector, this procedure gives
\begin{equation}
M\omega_{\mathrm{Schw}}=0.372878-0.089066i,
\end{equation}
with a relative complex error of $2.1\times10^{-3}$ compared to the standard
quadrupolar Schwarzschild QNM.

Throughout the analysis, we use the following criterion. If the fitted
ringdown mode is
\begin{equation}
\omega_{\rm QNM}=\omega_R-i|\omega_I|,
\end{equation}
we define the normalized frequency offset by
\begin{equation}
\Delta_{\rm QNM}\equiv
\frac{|\omega_{\rm peak}-\omega_R|}{|\omega_I|}.
\label{eq:qnm_alignment_criterion}
\end{equation}
Using $|\omega_I|$ as the relevant frequency width follows from the standard interpretation of QNMs as resonant states. Near an isolated
quasinormal-mode (QNM) pole, the frequency-domain response takes the
Breit--Wigner form
\begin{equation}
\mathcal R(\omega)\sim
\mathcal R_{\rm bg}(\omega)
+\frac{A}{\omega-\omega_R+i|\omega_I|},
\end{equation}
so that $\omega_R$ sets the characteristic resonance frequency, while
$|\omega_I|$ gives the half-width of the squared resonant amplitude when the non-resonant background varies slowly. Breit--Wigner methods have been used explicitly to extract black hole QNMs from resonant structures along the real-frequency axis \cite{Berti:2009wx}.
Dynamical-TLN studies of Kerr-like compact objects likewise exhibit
oscillatory and resonant features associated with the corresponding QNM
frequencies \cite{Chakraborty:2023zed}.

In the present coupled system, the relevant modes are sufficiently damped that the associated real-frequency features can be broad, and interference with a smooth scattering background can shift the maximum away from $\omega_R$. We therefore classify a numerically stable maximum as \emph{QNM aligned} when
\begin{equation}
\Delta_{\rm QNM}\leq 1,
\end{equation}
or equivalently,
\begin{equation}
|\omega_{\rm peak}-\omega_R|\leq |\omega_I|.
\end{equation}
Thus, $|\omega_I|$ is used as the characteristic frequency width of the mode rather than requiring the exact condition
$\omega_{\rm peak}=\omega_R$. The classification is based on the nominal
values of the extracted frequencies. The ringdown and peak extraction are then repeated under variations of the numerical controls to check that the classification remains unchanged.

This criterion tests whether a resolved real-frequency feature is
spectrally compatible with an independently determined QNM. It is an
operational alignment diagnostic and does not constitute either a necessary or sufficient condition for QNM excitation or a direct identification of a pole in the complex-frequency plane.\footnote{We do not locate poles by analytically continuing the frequency-domain response into the complex $\omega$ plane. Instead, the complex QNM frequency is obtained independently from the time-domain ringdown and compared with the real-frequency maximum using $|\omega_I|$ as the characteristic damping width.}
\begin{table}[htb!]
\centering
\scriptsize
\caption{Explicit QNM-alignment test for the broad real-frequency maxima. Here
$\omega_{\rm QNM}=\omega_R-i|\omega_I|$ is obtained from the independent
time-domain ringdown fit and
$\Delta_{\rm QNM}=|\omega_{\rm peak}-\omega_R|/|\omega_I|$. The last
column is ``Yes'' precisely when $\Delta_{\rm QNM}\leq1$. The classification
is unchanged under the numerical control variations used in the resonance test.}
\label{tab:qnm_alignment}
\resizebox{\textwidth}{!}{
\begin{tabular}{llcccccc}
\hline
Geometry & Parity & $\ell/\ell_{\rm ext}$ & $M\omega_{\rm peak}$ & $M\omega_R$ & $M|\omega_I|$ & $\Delta_{\rm QNM}$ & QNM aligned? \\
\hline
Bardeen & polar & 0.60 & 0.45000 & 0.37977 & 0.08697 & 0.807 & Yes \\
        & polar & 0.90 & 0.45000 & 0.38848 & 0.08328 & 0.739 & Yes \\
        & polar & 0.97 & 0.43750 & 0.38926 & 0.08198 & 0.588 & Yes \\
Hayward & polar & 0.60 & 0.45000 & 0.37362 & 0.08586 & 0.890 & Yes \\
        & polar & 0.90 & 0.45000 & 0.37634 & 0.08166 & 0.902 & Yes \\
        & polar & 0.97 & 0.45000 & 0.37699 & 0.08039 & 0.908 & Yes \\
Fan--Wang & polar & 0.60 & 0.46250 & 0.43992 & 0.09056 & 0.249 & Yes \\
          & polar & 0.90 & 0.76250 & 0.42725 & 0.09272 & 3.616 & No \\
          & polar & 0.97 & 0.78750 & 0.44328 & 0.07407 & 4.647 & No \\
\hline
Bardeen & axial & 0.60 & 0.57500 & 0.36769 & 0.08046 & 2.577 & No \\
        & axial & 0.90 & 0.63750 & 0.39335 & 0.07693 & 3.173 & No \\
        & axial & 0.97 & 0.66250 & 0.39086 & 0.08350 & 3.253 & No \\
Hayward & axial & 0.60 & 0.48750 & 0.36462 & 0.08365 & 1.469 & No \\
        & axial & 0.90 & 0.60625 & 0.37444 & 0.07505 & 3.089 & No \\
        & axial & 0.97 & 0.61250 & 0.37842 & 0.07471 & 3.133 & No \\
Fan--Wang & axial & 0.60 & 0.50625 & 0.43901 & 0.09521 & 0.706 & Yes \\
          & axial & 0.90 & 0.53750 & 0.48622 & 0.08579 & 0.598 & Yes \\
          & axial & 0.97 & 0.55000 & 0.51679 & 0.07675 & 0.433 & Yes \\
\hline
\end{tabular}
}
\end{table}

Table~\ref{tab:qnm_alignment} lists the quantities entering this test for each
sampled geometry, parity, and regularization ratio. Maxima that do not satisfy
the alignment criterion are interpreted as scattering features rather than
QNM-related resonances. When the criterion is satisfied, the agreement supports
the interpretation of the real-frequency maximum as a feature associated with
the corresponding QNM. 
\textcolor{black}{The quality factor,
\begin{equation}
    Q\equiv \frac{\omega_R}{2|\omega_I|},
\end{equation}
then indicates the degree of damping: values of $Q$ of order unity correspond to broad, strongly damped black hole resonances, whereas much larger values are characteristic of narrower, longer-lived resonances, such as those that can arise from reflective cavities
\cite{Chakraborty:2023zed,Berti:2009kk,Kokkotas:1999bd}.}

We independently verify the background identities, the symmetry of both
potential matrices, and the Regge--Wheeler and Zerilli limits
\cite{Regge,Zerelli}. The maximum relative error in
$m'=r^2\mathcal L$ is $1.7\times10^{-8}$, the potential matrices are
symmetric to machine precision, and the largest relative error in the
Schwarzschild potential limits is $5.6\times10^{-8}$. For the most sensitive case, Fan--Wang polar perturbations at $\ell/\ell_{\rm ext}=0.9$, shifting the
matching window changes the real finite-frequency correction by $0.23\%$,
$2.34\%$, and $11.2\%$ at $M\omega=0.002$, $0.004$, and $0.006$,
respectively. We therefore restrict quantitative comparisons to
$M\omega\leq0.004$ and retain the $M\omega=0.006$ points only as an
indication of the growing overlap-window uncertainty.  

Table~\ref{tab:dynamic_tln_convergence} gives the matching-radius
extrapolation check for the direct metric response at
$\ell/\ell_{\rm ext}=0.9$. The residual is normalized by the direct
finite-frequency correction,
\begin{equation}
\Delta k_{2,\pm}
=
\operatorname{Re}
\left[
\mathcal K_{2,\pm}(\omega)-\mathcal K_{2,\pm}(0)
\right].
\end{equation}
At $M\omega=0.004$, the largest relative residual is $2.33\%$ for
Fan--Wang axial perturbations. In this case, however, the direct correction is
only $3.85\times10^{-6}$, corresponding to an absolute matching residual of
$8.99\times10^{-8}$. At the edge point $M\omega=0.006$, the largest
relative residual is $8.11\%$ for Bardeen axial perturbations, again because
the direct correction is small; the corresponding absolute residual is
$1.09\times10^{-6}$. These checks support the use of
$M\omega\leq0.004$ for quantitative comparisons, while the
$M\omega=0.006$ points serve as diagnostics of the limits of the overlap
expansion. 


\begin{table}[htb!]
\centering
\small
\caption{Convergence of the low-frequency metric-level dynamical TLNs at $\ell/\ell_{\rm ext}=0.9$. The third column is the direct real finite-frequency correction $\Delta k_{2,\pm}=\operatorname{Re}[\mathcal K_{2,\pm}(\omega)-\mathcal K_{2,\pm}(0)]$. The last column is the largest residual of the matching-radius extrapolation divided by $|\Delta k_{2,\pm}|$. The absolute residual remains small even when the relative percentage is enhanced by a very small dynamical correction.}
\label{tab:dynamic_tln_convergence}
\begin{tabular}{llccc}
\hline\hline
Geometry & Parity & $M\omega$ & $\operatorname{Re}\Delta k_{2,\pm}$ & Relative residual \\
\hline
Bardeen & polar & 0.002 & $-3.908395\times10^{-5}$ & 0.012\% \\
 &  & 0.004 & $-1.551436\times10^{-4}$ & 0.057\% \\
 &  & 0.006 & $-3.374516\times10^{-4}$ & 0.258\% \\
Hayward & polar & 0.002 & $1.229655\times10^{-3}$ & 0.001\% \\
 &  & 0.004 & $1.499847\times10^{-3}$ & 0.002\% \\
 &  & 0.006 & $1.961391\times10^{-3}$ & 0.038\% \\
Fan--Wang & polar & 0.002 & $3.047325\times10^{-3}$ & $<0.001$\% \\
 &  & 0.004 & $3.187228\times10^{-3}$ & 0.002\% \\
 &  & 0.006 & $3.430919\times10^{-3}$ & 0.023\% \\
Bardeen & axial & 0.002 & $-3.555130\times10^{-6}$ & 0.042\% \\
 &  & 0.004 & $-1.268133\times10^{-5}$ & 0.760\% \\
 &  & 0.006 & $-1.349665\times10^{-5}$ & 8.105\% \\
Hayward & axial & 0.002 & $-2.821972\times10^{-5}$ & 0.007\% \\
 &  & 0.004 & $-1.113535\times10^{-4}$ & 0.087\% \\
 &  & 0.006 & $-2.355823\times10^{-4}$ & 0.464\% \\
Fan--Wang & axial & 0.002 & $6.073789\times10^{-7}$ & 0.200\% \\
 &  & 0.004 & $3.854474\times10^{-6}$ & 2.333\% \\
 &  & 0.006 & $2.258361\times10^{-5}$ & 4.498\% \\
\hline\hline
\end{tabular}
\end{table}
The broad-response convergence data are retained only as a numerical check of the
real-frequency maxima used in the resonance diagnostic. They are independent of the
low-frequency metric-level Love number extraction discussed below and should not be
interpreted as a convergence test of the dynamical TLN itself.

Table~\ref{tab:strict_convergence} reports the largest maximum found in each broad
polar master-field scan and the relative peak-response change when the numerical
controls are tightened simultaneously. The last column is defined as
\begin{equation}
\frac{|\mathcal R_{\rm strict}-\mathcal R_{\rm standard}|}
{|\mathcal R_{\rm strict}|}\label{eq:peak_response_change}
\end{equation}
and is evaluated at the reported maximum. It measures the relative change in
the peak response when the numerical controls are tightened.
The purpose of the table is therefore to establish
that these candidate maxima are resolved features of the scattering calculation rather
than artifacts of the frequency grid, integration range, tolerances, or matching choices.
To stress again, numerical stability alone does not establish a resonance. That interpretation is made only
after the maximum is compared with an independently extracted, control-stable ringdown
frequency. For the polar response, all listed relative peak-response changes are below $1.4\%$. The denser frequency grid resolves the Bardeen maxima at $M\omega=0.4500$, $0.4500$, and $0.4375$, the Hayward maxima at $M\omega=0.4500$, and the near-extremal Fan--Wang maximum at $M\omega=0.7875$.
\begin{table}[tbh]
\centering
\caption{Convergence of the broad polar master-field response used in the resonance diagnostic. For each geometry and regularization ratio, the third column gives the real frequency at which the largest maximum is found, while the last column gives the relative peak-response change defined by Eq.(\ref{eq:peak_response_change}) when the frequency grid, integration range, tolerances, and matching controls are tightened simultaneously. This table tests only the numerical robustness of the candidate scattering maxima.} 
\label{tab:strict_convergence}
\begin{tabular}{lccc}
\hline
Geometry & $\ell/\ell_{\rm ext}$ & $M\omega$ at maximum & relative peak-response change\\
\hline
Bardeen  & 0.60 & 0.4500 & 0.23\% \\
         & 0.90 & 0.4500 & 0.26\% \\
         & 0.97 & 0.4375 & 0.30\% \\
Hayward  & 0.60 & 0.4500 & 0.22\% \\
         & 0.90 & 0.4500 & 0.22\% \\
         & 0.97 & 0.4500 & 0.22\% \\
Fan--Wang & 0.60 & 0.4625 & 0.33\% \\
          & 0.90 & 0.7625 & 0.61\% \\
          & 0.97 & 0.7875 & 1.37\% \\
\hline
\end{tabular}
\end{table}
For completeness, we apply the same broad-response robustness test to the axial master-field scans. Table~\ref{tab:strict_convergence_axial} lists the largest axial maximum for the same regularization ratios and the relative peak-response change under the strict numerical controls. For the axial response, the changes remain below $0.23\%$ in all sampled cases, showing that the axial maxima are likewise resolved features of the scattering calculation. As in the polar sector, this stability test does not determine whether a maximum is a resonance. That question is addressed separately by the ringdown comparison in Section~\ref{sec:results_axial}. {\color{black}With the low-frequency metric extraction and the separate resonance diagnostic thus defined and numerically controlled, we now turn to the polar and axial response of the three regular black hole geometries.}
\endgroup
\section{Numerical results on dynamical TLN: polar sector}
\label{sec:results_polar}
\begingroup
\textcolor{black}{Having completed in Section~\ref{sec:methods} the numerical-stability checks for both the
polar and axial sectors, we now present the polar dynamical Love numbers.
The polar Love number is extracted from the reconstructed metric variable
$-fH_0$ using Eqs.~\eqref{eq:metric_nearzone_main}--%
\eqref{eq:metric_dynamic_definition}. Figs.~(\ref{fig:alpha_vs_chi}), (\ref{fig:hayward_alpha_vs_ell}), and (\ref{fig:fanwang_alpha_vs_ell}) each contain two
complementary metric-level panels for the Bardeen, Hayward, and Fan--Wang
geometries, respectively. The left panel compares the independently
calculated static response with the finite-frequency responses at
$M\omega=0.002$, $0.004$, and $0.006$ as
$\ell_X/\ell_{\rm ext}$ is varied from the Schwarzschild limit toward
extremality, whereas the right panel scans the same low-frequency window
at selected fixed values of $\ell_X/\ell_{\rm ext}$. All curves use
$M\omega\leq0.006$, and quantitative comparisons are restricted to
$M\omega\leq0.004$, in addition, the $M\omega=0.006$ points are retained \textcolor{black}{only to indicate the onset of increased matching sensitivity.}. The plotted quantity is the real part of the reconstructed
gravitational response for a unit gravitational source and no independent
electromagnetic source, with the electromagnetically induced response
retained. The insets show
$10^4[\mathcal K_{2,+}(\omega)-\mathcal K_{2,+}(0)]$ to resolve the leading
frequency correction. Representative complex source--response coefficients
are provided in Appendix~\ref{app:numerical_asymptotics}. The
broad-frequency results in Figs.~(\ref{fig:bardeen_polar_resonance}), (\ref{fig:hayward_polar_resonance}), and (\ref{fig:fanwang_polar_resonance}) are separate scattering
diagnostics and do not extend the metric-level dynamical-TLN extraction
beyond the controlled low-frequency regime. The explicit results for the three geometries are discussed below.
}

\subsection{Bardeen black hole}
\label{sec:numerical_results_bardeen}
\begin{figure}[htb!]
\centering
\begin{minipage}{0.49\textwidth}\centering
\includegraphics[width=\linewidth]{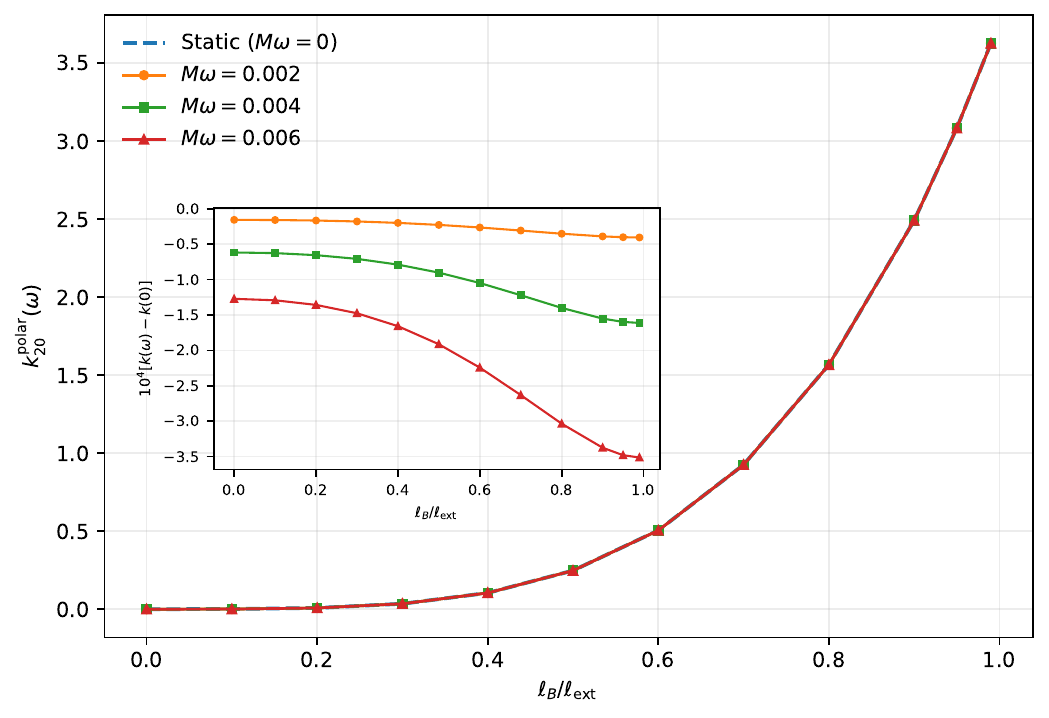}\\[-1mm]
{\small (a) Variation with $\ell_B/\ell_{\rm ext}$.}
\end{minipage}\hfill
\begin{minipage}{0.49\textwidth}\centering
\includegraphics[width=\linewidth]{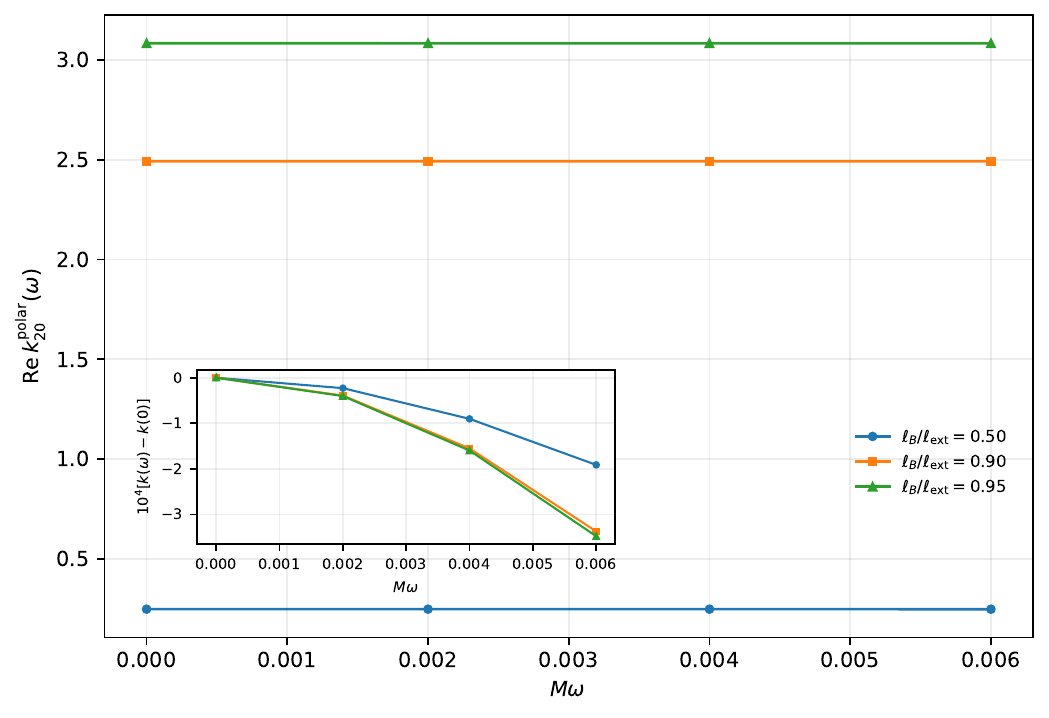}\\[-1mm]
{\small (b) Frequency scans at fixed regularization.}
\end{minipage}
\caption{Bardeen polar dynamical Love number extracted from the reconstructed
metric amplitude $-fH_0$. In panel (a), the dashed curve is the independently
recovered static response and the colored curves show $M\omega=0.002$, $0.004$,
and $0.006$ as $\ell_B/\ell_{\rm ext}$ is increased. The response grows
rapidly toward extremality, while the finite-frequency curves remain below the
static curve. Panel (b) scans the low-frequency dependence at fixed regularization for
$\ell_B/\ell_{\rm ext}=0.50$, $0.90$, and $0.95$, in the same form used for
the Hayward and Fan--Wang polar results. The inset magnifies the difference from the
static value and shows the negative conservative correction beginning at
$O(\omega^2)$. The source is a unit gravitational tide with no independent
electromagnetic tide; the induced electromagnetic response is included in the
metric reconstruction.}
\label{fig:alpha_vs_chi}
\end{figure}

\textcolor{black}{As shown in Figure~\ref{fig:alpha_vs_chi}, the static Bardeen polar response is positive and grows rapidly towards extremality. The finite-frequency curves also maintain this regularization dependence but reside slightly below the static result. The inset resolves the corresponding smooth negative conservative correction that begins at $O(\omega^2)$ and increases in magnitude towards extremality; this remains small throughout the controlled near-zone window. Further, the broad-frequency scan in Figure~\ref{fig:bardeen_polar_resonance} indicates a separate master-field scattering diagnostic and does not extend the metric-level dynamical-TLN extraction to $M\omega\sim0.4$. Across the sampled regularization range $0.60\leq\ell_B/\ell_{\rm ext}\leq0.97$, the broad maxima occur in the region $M\omega\simeq0.437$--$0.45$ and lie within one damping width of the independently extracted QNMs listed in Table~\ref{tab:qnm_alignment}. Thus, the Bardeen polar peaks are QNM aligned throughout the sampled regularization range. Their numerical robustness under the tightened broad-scan controls is confirmed in Table~\ref{tab:strict_convergence}, and also summarized in the caption of Figure \ref{fig:bardeen_polar_resonance}.}
\begin{figure}[htb!]
\centering
\includegraphics[width=0.96\textwidth]{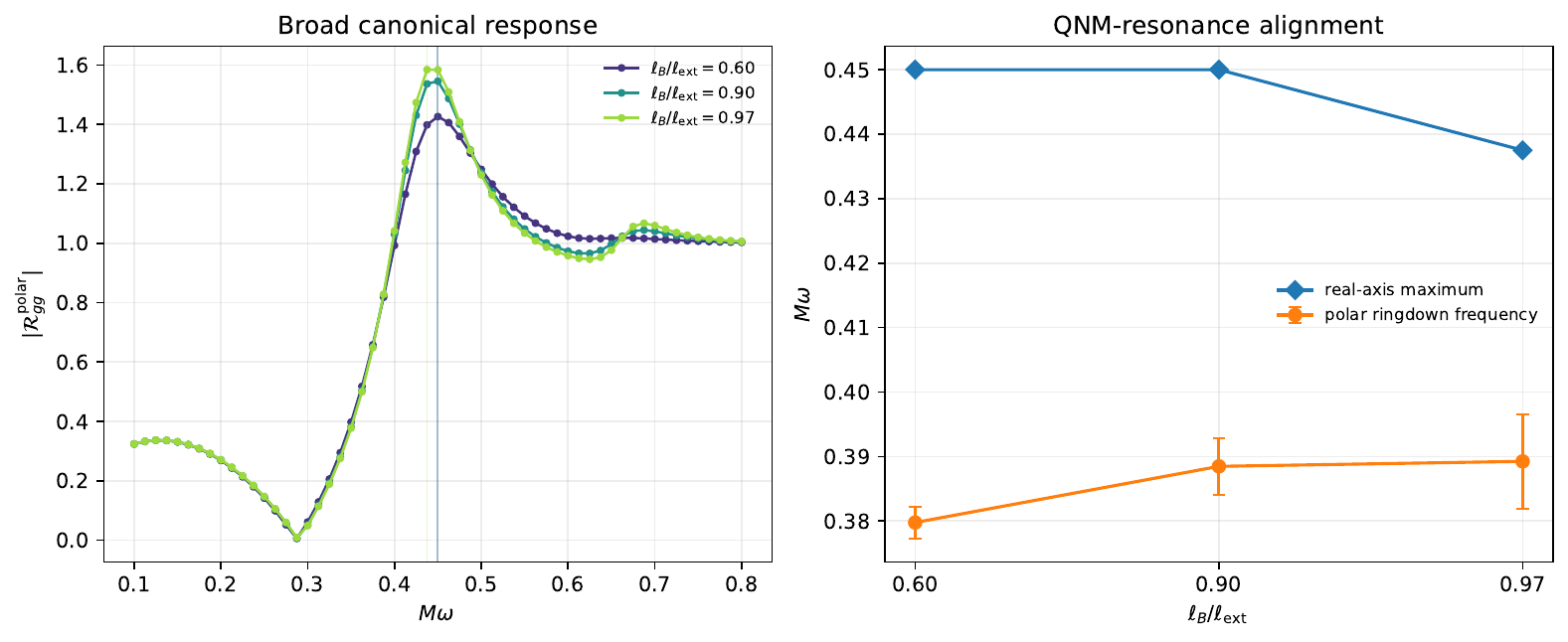}
\caption{Bardeen polar resonance diagnostic. The left panel shows the absolute
value of the broad real-frequency canonical response for three regularization
ratios, with vertical lines marking the largest sampled maxima. The right panel
compares the peak frequencies with the real parts of the independently
extracted time-domain ringdown frequencies. The error bars indicate the largest
frequency shifts obtained under variations of the ringdown analysis. The
response peaks at $M\omega=0.4500$ for
$\ell_B/\ell_{\rm ext}=0.60$ and $0.90$, and at $M\omega=0.4375$ for
$\ell_B/\ell_{\rm ext}=0.97$, while the corresponding ringdown frequencies
lie near $M\omega_R\simeq0.38$--$0.39$. For all three regularization ratios,
the frequency offset is smaller than the corresponding damping width, so the
maxima satisfy the QNM-alignment criterion.}
\label{fig:bardeen_polar_resonance}
\end{figure}
\subsection{Hayward black hole}
\label{subsec:hayward_results}
\begin{figure}[h]
\centering
\begin{minipage}{0.49\textwidth}\centering
\includegraphics[width=\linewidth]{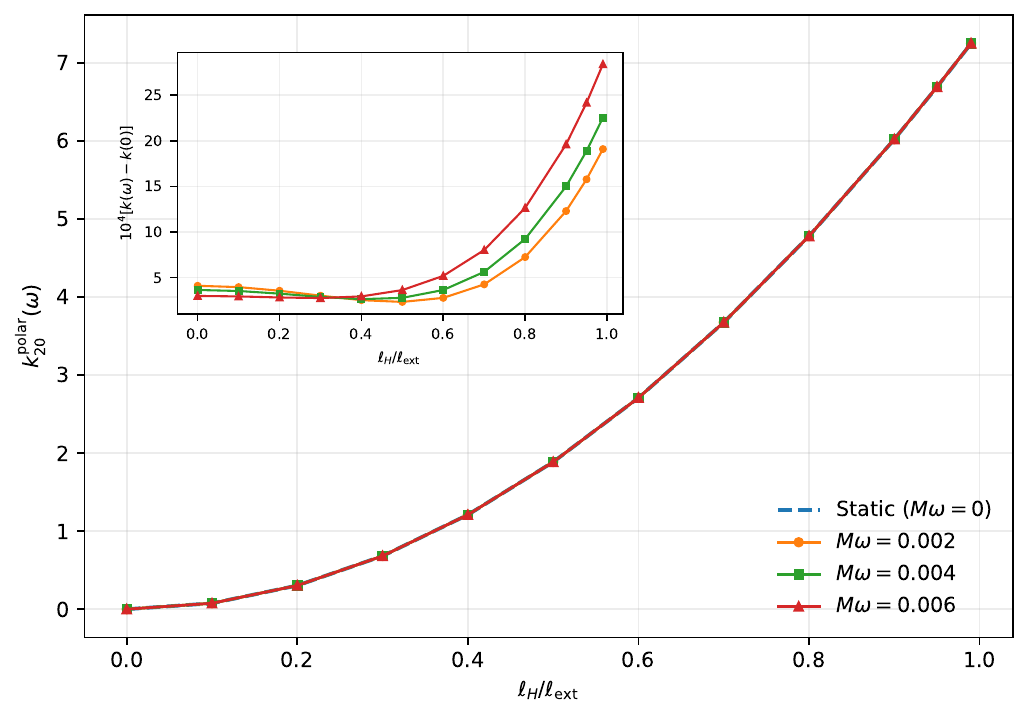}\\[-1mm]
{\small (a) Variation with $\ell_H/\ell_{\rm ext}$.}
\end{minipage}\hfill
\begin{minipage}{0.49\textwidth}\centering
\includegraphics[width=\linewidth]{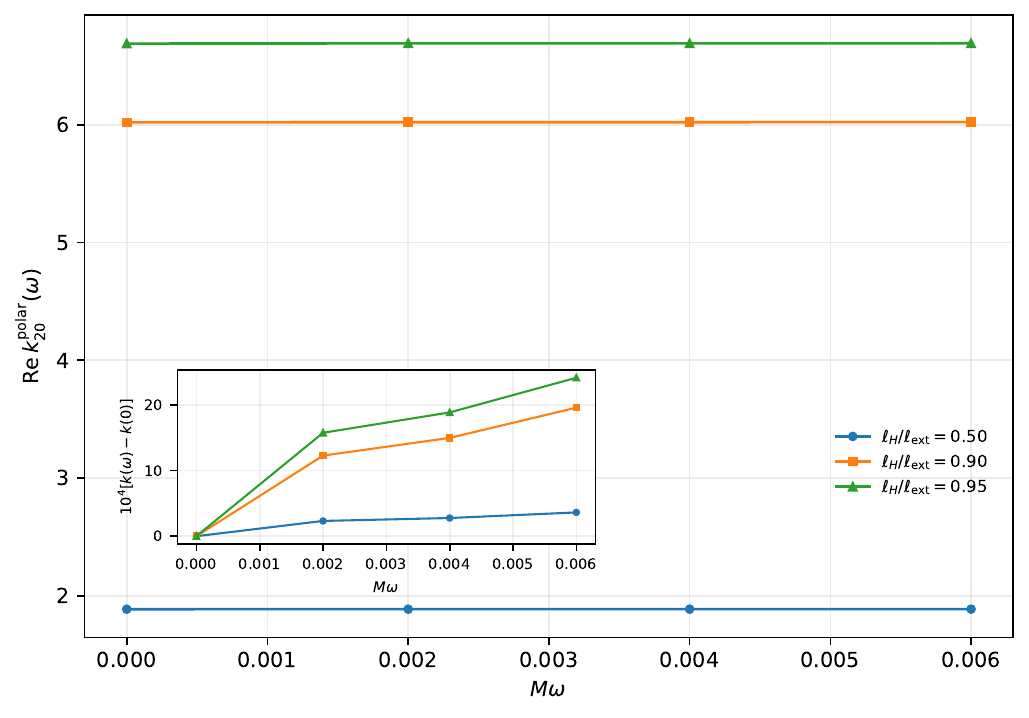}\\[-1mm]
{\small (b) Frequency scans at fixed regularization.}
\end{minipage}
\caption{Hayward polar dynamical Love number from the reconstructed metric
perturbation. Panel (a) varies $\ell_H/\ell_{\rm ext}$ and compares the
independent static curve with the three finite-frequency curves. Panel (b)
shows the low-frequency scan at fixed regularization for several values of
$\ell_H/\ell_{\rm ext}$. The response is positive and smooth across the
regularization range. The inset shows the small difference from the static value. The correction is positive and starts at $\mathcal{O}(\omega^2)$. For all curves, the gravitational source is set to unity, the independent electromagnetic source is set to zero and the induced electromagnetic response is retained in the reconstructed metric.}
\label{fig:hayward_alpha_vs_ell}
\end{figure}
The curves corresponding to the Hayward polar correction have the opposite sign to those of Bardeen: they lie above the static response as shown in Figure~\ref{fig:hayward_alpha_vs_ell}. The inset illustrates the small positive shift of order $O(\omega^2)$, whereas the main panel indicates that this correction does not affect the monotonic increase of the response with regularization.
The Hayward static response increases smoothly toward extremality. In contrast to Bardeen, the finite-frequency curves are above the static curve. The leading conservative correction therefore increases the response. The effect is small on the scale of the main plot and is most clearly seen in the inset. The near-zone extrapolation remains stable, with a maximum residual of order $10^{-6}$ at the upper frequency edge, which is why the $M\omega \leq 0.004$ points are used for quantitative comparisons and the $M\omega = 0.006$ points are retained \textcolor{black}{only to indicate the onset of increased matching sensitivity.}
\begin{figure}[t]
\centering
\includegraphics[width=0.92\textwidth]{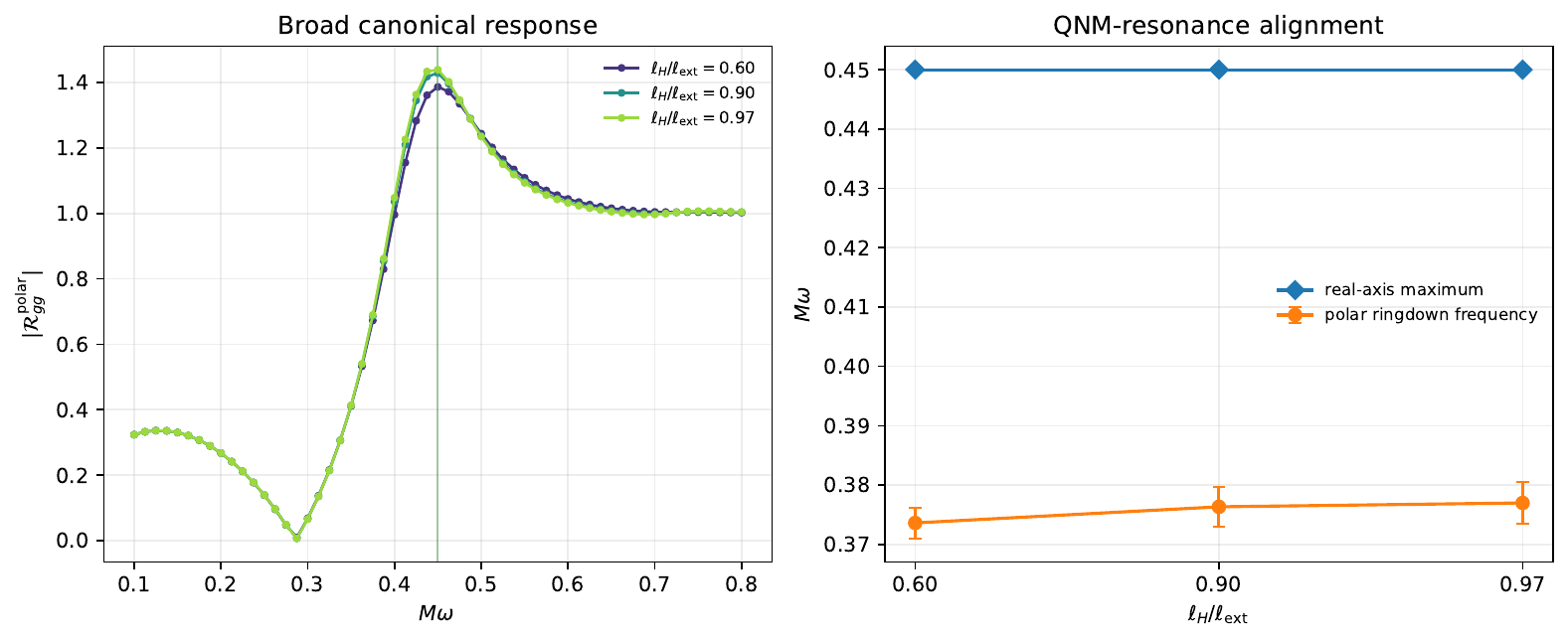}
\caption{Hayward polar resonance diagnostic. The left panel shows the absolute
value of the broad real-frequency canonical response for three regularization
ratios; the vertical lines mark the largest sampled maxima. The right panel compares the
broad real-frequency master-field maximum with the real part of the independently extracted
ringdown frequency. For the polar response, the largest maximum occurs at $M\omega=0.4500$ for all three sampled regularization ratios. The frequency offsets are $\Delta_{\rm QNM}=0.890$, $0.902$, and $0.908$, respectively, and therefore remain within one damping width. The sampled Hayward polar maxima are consequently classified as broad QNM-aligned response features.}
\label{fig:hayward_polar_resonance}
\end{figure}
As can be seen in Figure~\ref{fig:hayward_polar_resonance}, the highest value among the sampled maxima remains close to $M\omega=0.45$ when the regularization is changed. Even though this feature is broad, the offsets listed earlier in Table~\ref{tab:qnm_alignment} are all less than one damping width for each of the three samples. Furthermore, in view of the stability test given in Table~\ref{tab:strict_convergence}, this indicates that the feature can be classified as a broad QNM-aligned response feature rather than a numerical artifact.

\subsection{Fan--Wang black hole}
\label{subsec:fanwang_results}
\begin{figure}[h]
\centering
\begin{minipage}{0.49\textwidth}\centering
\includegraphics[width=\linewidth]{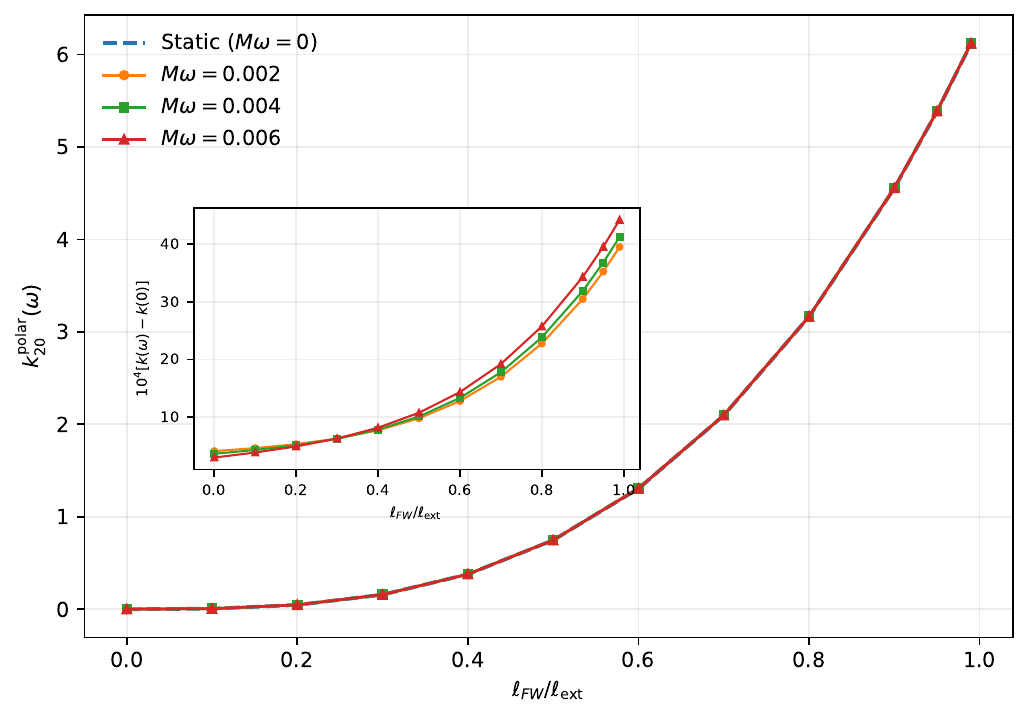}\\[-1mm]
{\small (a) Variation with $\ell_{FW}/\ell_{\rm ext}$.}
\end{minipage}\hfill
\begin{minipage}{0.49\textwidth}\centering
\includegraphics[width=\linewidth]{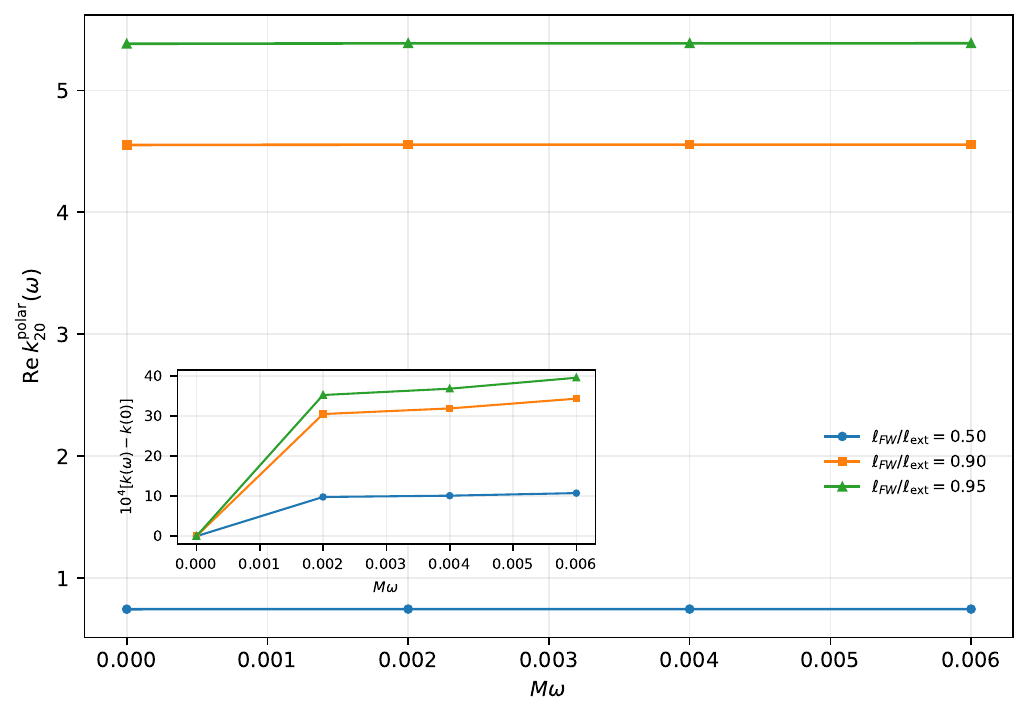}\\[-1mm]
{\small (b) Frequency scans at fixed regularization.}
\end{minipage}
\caption{Fan--Wang polar dynamical Love number. Panel (a) shows the
regularization scan for the static system and for $M\omega=0.002$, $0.004$,
and $0.006$. Panel (b) shows the low-frequency response at fixed values of
$\ell_{FW}/\ell_{\rm ext}$; the inset magnifies the frequency-dependent change.
The static response rises steeply and begins at cubic order in
$\ell_{FW}/M$, while the controlled conservative correction is positive and
starts at $O(\omega^2)$. The leading gravitational and electromagnetic master
indices are degenerate in this model, so the complete physical source matrix is
inverted before imposing a unit gravitational source and zero independent
electromagnetic source. The induced electromagnetic response is retained.}
\label{fig:fanwang_alpha_vs_ell}
\end{figure}
Figure~\ref{fig:fanwang_alpha_vs_ell} shows the steep growth of the Fan–Wang static polar response together with the smaller positive finite-frequency correction. As in the inset, the dynamical effect is separated from the much larger static background and a smooth increase of order $O(\omega^2)$ is observed over the entire low-frequency range under control.
The steep rise in the left panel reflects the cubic leading static scaling and
the higher-order terms that become important close to extremality. The finite
frequency curves remain close to the static curve because the calculation is
restricted to the near-zone expansion. The positive shift is nevertheless
resolved by the inset and grows monotonically with frequency. This behavior is
qualitatively different from the negative Bardeen-polar correction and the
positive but smaller Hayward-polar shift, showing that the sign of the
conservative response depends on the regular black hole geometry.

\begin{figure}[t]
\centering
\includegraphics[width=0.92\textwidth]{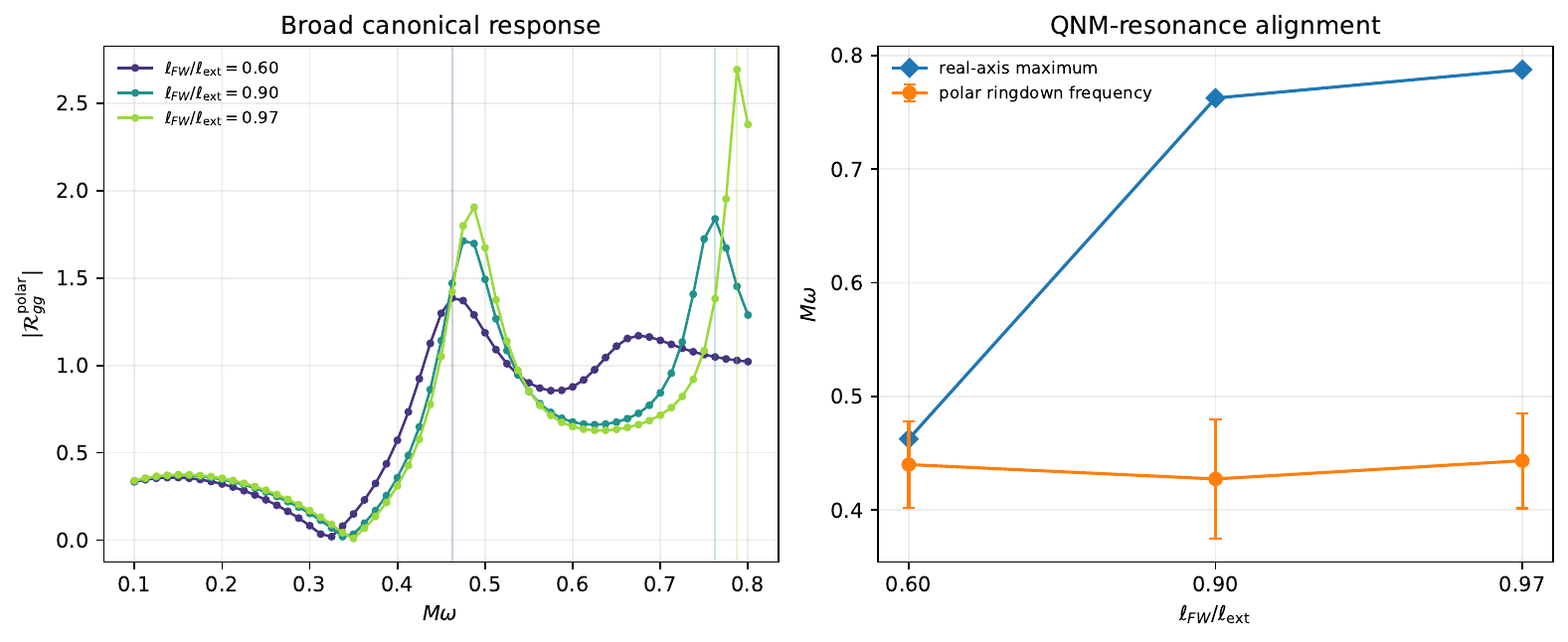}
\caption{Fan--Wang polar resonance diagnostic. The left panel shows the
absolute value of the broad real-frequency canonical response for three
regularization ratios, with vertical lines marking the largest sampled maxima.
The right panel compares these maxima with the independently extracted
time-domain ringdown frequencies. The polar response peaks at
$M\omega=0.4625$, $0.7625$, and $0.7875$ for
$\ell_{FW}/\ell_{\rm ext}=0.60$, $0.90$, and $0.97$, respectively.
The maximum for $\ell_{FW}/\ell_{\rm ext}=0.60$ satisfies the damping-width
alignment criterion, whereas those for $0.90$ and $0.97$ do not. Thus, only
the lowest sampled regularization case is QNM aligned in the Fan--Wang polar
scan, while the higher-regularization maxima are interpreted as broad
scattering features.}
\label{fig:fanwang_polar_resonance}
\end{figure}
The broad-frequency behavior in Figure~\ref{fig:fanwang_polar_resonance} differs
qualitatively from the Bardeen and Hayward polar cases. As shown again earlier in
Table~\ref{tab:qnm_alignment}, only the maximum at
$\ell_{FW}/\ell_{\rm ext}=0.60$ lies within one damping width of the
corresponding QNM and is therefore classified as QNM aligned. The maxima at
$\ell_{FW}/\ell_{\rm ext}=0.90$ and $0.97$ lie well outside the alignment
window and are instead interpreted as broad scattering features. Their
numerical stability is nevertheless confirmed by
Table~\ref{tab:strict_convergence}.

These higher-frequency maxima remain useful for characterizing the scattering
response of the coupled master system, but they lie outside the controlled
metric-level near-zone regime used to extract the dynamical TLNs. They are
therefore not included in the polar TLN curves. {\color{black}These results demonstrate that both the low-frequency conservative correction and the broad-frequency response are sensitive to the regular black hole geometry. We now turn to the corresponding axial sector.}
\endgroup

\section{Numerical results on dynamical TLN: axial sector}
\label{sec:results_axial}
\begingroup
The axial Love number is extracted from the physical Regge-Wheeler-gauge
metric amplitude $h_0$. Appendix~\ref{app:dynLove} gives the
master-to-metric reconstruction and the equivalent axial metric system, while
Appendix~\ref{app:numerical_asymptotics} presents the
intermediate-near-zone source and response solutions used in the numerical
extraction.

The axial figures follow the same two-panel format as the polar figures shown earlier. In the
left panel, we vary $\ell_X/\ell_{\rm ext}$ ($X \in \{B, H, FW)\}$) and compare the static response
with the results at $M\omega=0.002$, $0.004$, and $0.006$. The right panel
shows the frequency dependence over the same low-frequency range at fixed
regularization. The axial response is obtained from the $r^3$ source
coefficient and the $r^{-2}$ response coefficient of $h_0$, using the
normalization given in Eq.~\eqref{eq:metric_love_conversion_main}. The plotted curves
are the real conservative response. The insets show the rescaled change from
the static value. Because the axial static response can be either negative or
positive, the sign of the inset directly indicates whether finite frequency
makes the response more or less positive. As in the polar sector, the source
is purely gravitational, the induced electromagnetic response is retained,
and only $M\omega\leq0.004$ is used for quantitative comparisons. The separate
resonance panels compare broad master-field maxima with time-domain ringdown
frequencies and are not extensions of the metric-level TLN calculation.

Just as in the polar case, the low-frequency curves and the broad-frequency resonance test are each controlled separately. \textcolor{black}{Table~\ref{tab:dynamic_tln_convergence} tests the numerical stability of the axial Love number extraction. Table~\ref{tab:qnm_alignment} performs the resonance test by comparing each broad maximum with the independently extracted QNM frequency and determining whether their separation is within one damping width. Table~\ref{tab:strict_convergence_axial} separately verifies the numerical robustness of these maxima using a denser frequency grid and stricter integration-range, numerical-tolerance, and matching controls. The following figures show how these tests apply to the axial responses of the three regular black hole geometries under consideration.}

\begin{table}[htb!]
\centering
\caption{Convergence of the broad axial master-field response used in the resonance diagnostic. For each geometry and regularization ratio, the third column gives the real frequency of the largest maximum found in the broad scan, and the last column gives the relative peak-response change defined by Eq.~(\ref{eq:peak_response_change}). The table tests the numerical robustness of the candidate axial scattering maxima only.}
\label{tab:strict_convergence_axial}
\begin{tabular}{lccc}
\hline
Geometry & $\ell/\ell_{\rm ext}$ & $M\omega$ at maximum & relative peak-response change \\
\hline
Bardeen  & 0.60 & 0.5750  & 0.08\% \\
         & 0.90 & 0.6375  & 0.15\% \\
         & 0.97 & 0.6625  & 0.20\% \\
Hayward  & 0.60 & 0.4875  & 0.11\% \\
         & 0.90 & 0.60625 & 0.07\% \\
         & 0.97 & 0.6125  & 0.09\% \\
Fan--Wang & 0.60 & 0.50625 & 0.14\% \\
          & 0.90 & 0.5375  & 0.19\% \\
          & 0.97 & 0.5500  & 0.23\% \\
\hline
\end{tabular}
\end{table}

\subsection{Bardeen black hole}
\begin{figure}[htb!]
\centering
\begin{minipage}{0.49\textwidth}\centering
\includegraphics[width=\linewidth]{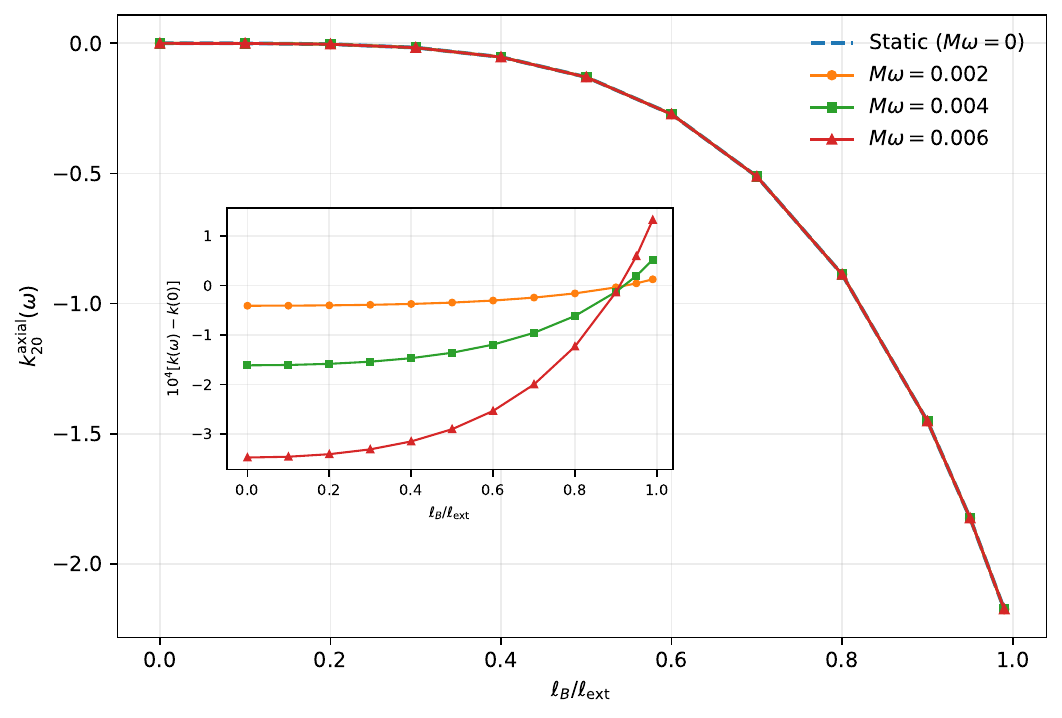}\\[-1mm]
{\small (a) Variation with $\ell_B/\ell_{\rm ext}$.}
\end{minipage}\hfill
\begin{minipage}{0.49\textwidth}\centering
\includegraphics[width=\linewidth]{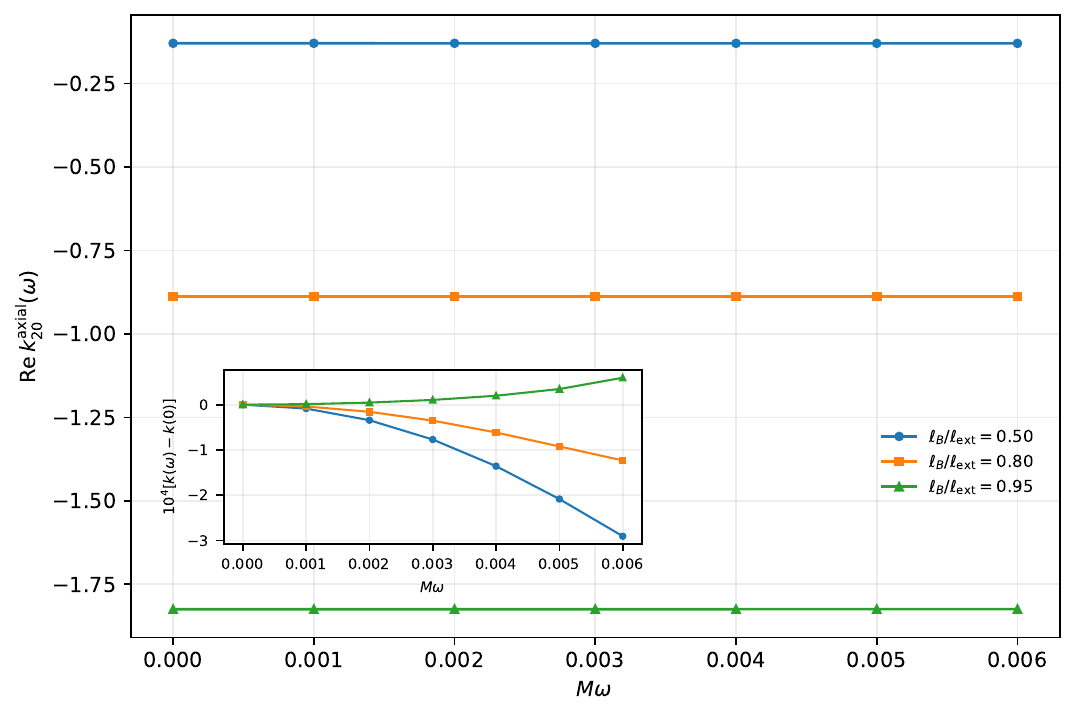}\\[-1mm]
{\small (b) Frequency scans at fixed regularization.}
\end{minipage}
\caption{Bardeen axial dynamical Love number.
Panel (a) varies $\ell_B/\ell_{\rm ext}$ and shows the independently recovered
static curve together with the three controlled finite-frequency curves. Panel
(b) scans $M\omega$ at fixed regularization for several values of
$\ell_B/\ell_{\rm ext}$. The static response is negative. The direct finite-frequency correction is negative through most of the sampled regularization range, making the response slightly more negative, but it changes sign close to extremality (between the $0.90$ and $0.95$ samples). The correction begins at $O(\omega^2)$. The
inset magnifies this small shift. }
\label{fig:axial_dynamic}
\end{figure}
The negative Bardeen axial static response is shown in Figure~\ref{fig:axial_dynamic}. The finite-frequency correction is negative throughout most of the sampled range, but the inset reveals a sign change near extremality. The dynamical correction, therefore, alters its qualitative effect depending on the regularization, even though it remains small in comparison to the static contribution.

\begin{figure}[t]
\centering
\includegraphics[width=0.92\textwidth]{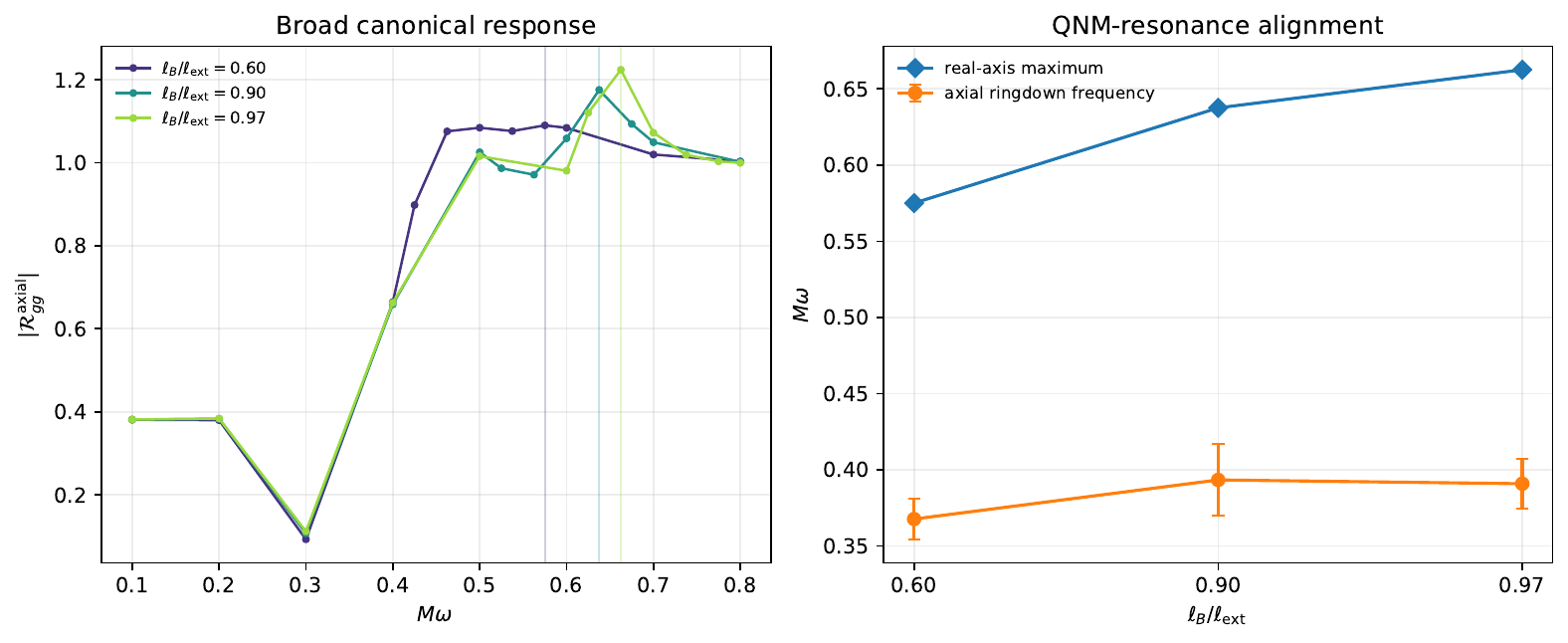}
\caption{Bardeen axial resonance diagnostics. The left panel shows the absolute
value of the broad real-frequency canonical response for three regularization
ratios. The vertical lines mark the largest sampled maxima. The right panel shows the largest maxima from the broad real-frequency scans of the axial master field and compares them with the corresponding ringdown frequencies. For the axial response, the maxima occur at $M\omega=0.5750$, $0.6375$, and $0.6625$ for $\ell_B/\ell_{\rm ext}=0.60$, $0.90$, and $0.97$, respectively, whereas the ringdown frequencies remain near $0.37$--$0.39$. The corresponding normalized offsets are $2.577$, $3.173$, and $3.253$, all greater than one damping width, so the maxima remain classified as broad scattering features rather than QNM-related resonances.}
\label{fig:bardeen_axial_resonance}
\end{figure}
Looking at the wider scan, Figure~\ref{fig:bardeen_axial_resonance} reveals maxima whose real frequencies shift to much higher values as the regularization increases. In all three cases, the deviation from the QNMs determined independently is greater than one damping width (as shown in earlier in Table~\ref{tab:qnm_alignment}), and therefore the features are regarded as broad scattering maxima and not as resonances related to the QNMs. Table~\ref{tab:strict_convergence_axial} confirms that this conclusion is not due to poor numerical resolution of the maxima.

\subsection{Hayward black hole}
\begin{figure}[t]
\centering
\begin{minipage}{0.49\textwidth}\centering
\includegraphics[width=\linewidth]{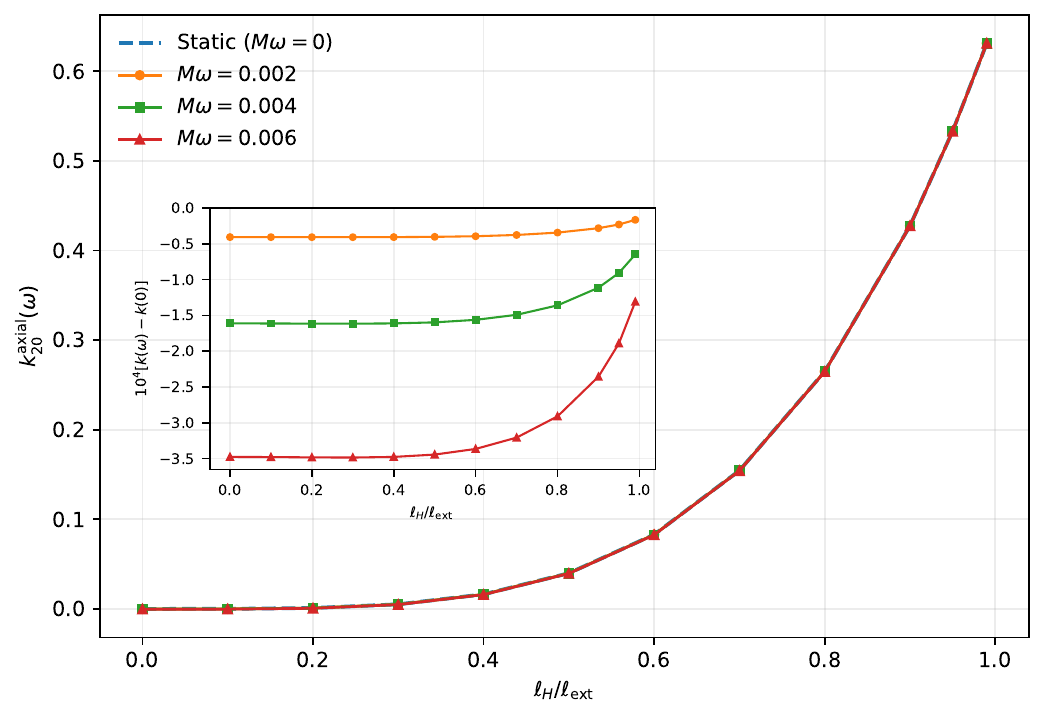}\\[-1mm]
{\small (a) Variation with $\ell_H/\ell_{\rm ext}$.}
\end{minipage}\hfill
\begin{minipage}{0.49\textwidth}\centering
\includegraphics[width=\linewidth]{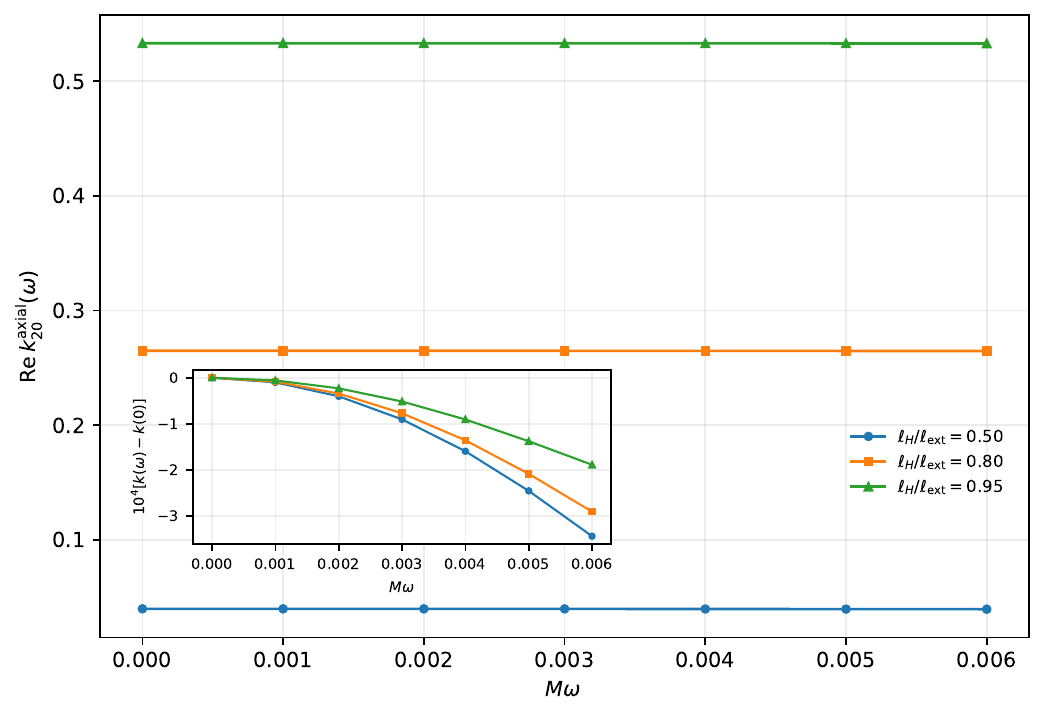}\\[-1mm]
{\small (b) Frequency scans at fixed regularization.}
\end{minipage}
\caption{Hayward axial dynamical Love number. Panel (a) gives the dependence on
$\ell_H/\ell_{\rm ext}$ for the static and three finite-frequency responses,
while panel (b) gives the low-frequency scans at fixed regularization. The static response is positive, whereas the direct conservative finite-frequency correction is negative over the sampled regularization range. The finite-frequency curves therefore lie slightly below the static curve, with the $O(\omega^2)$ shift visible mainly in the inset because it is much smaller than the static response. The source is a unit gravitational tide, and the induced
electromagnetic response is included in the reconstructed $h_0$ coefficient.}
\label{fig:hayward_axial_metric}
\end{figure}
As shown in Figure~\ref{fig:hayward_axial_metric}, the Hayward axial static response is positive, whereas the finite-frequency correction is negative over the entire sampled range of regularisation. Because of this, the curves for the finite frequency lie just below those of the static case, and it is for this reason that the inset is required in order to display the smooth $O(\omega^2)$ decrease.

\begin{figure}[t]
\centering
\includegraphics[width=\textwidth]{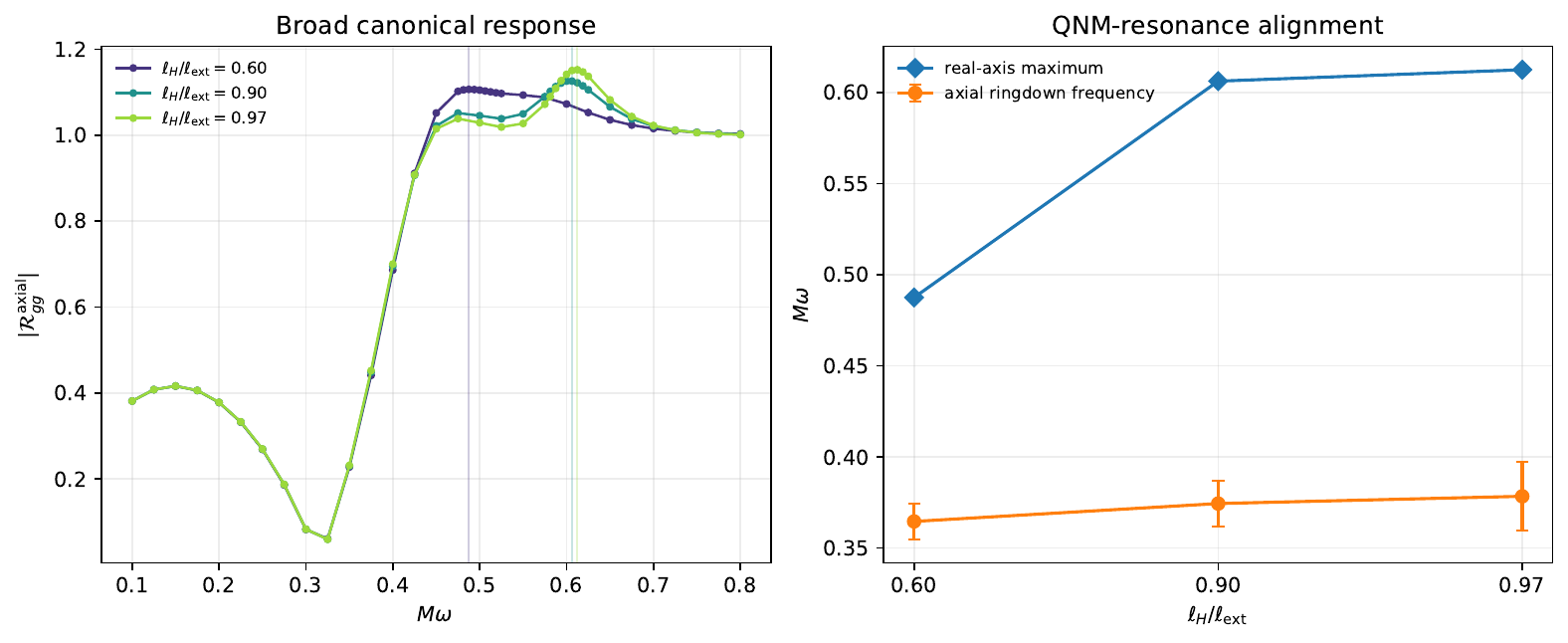}
\caption{Hayward axial resonance diagnostic. The left panel shows the absolute
value of the broad real-frequency canonical response for three regularization
ratios; the vertical lines mark the largest sampled maxima. The right panel
compares these maxima with the real parts of the independently extracted
time-domain ringdown frequencies. The error bars give the largest frequency
shift under the ringdown control variations. For the axial response, the maxima occur at $M\omega=0.4875$, $0.60625$, and $0.6125$ for $\ell_H/\ell_{\rm ext}=0.60$, $0.90$, and $0.97$, respectively. Their normalized frequency offsets are $1.469$, $3.089$, and $3.133$, so none satisfies the peak--mode alignment criterion. They are therefore classified as broad scattering features rather than QNM-related resonances.}
\label{fig:hayward_axial_resonance}
\end{figure}
The broad maxima shown in Figure~\ref{fig:hayward_axial_resonance} do not follow the Hayward axial QNM frequencies closely enough to meet the damping-width criterion. The normalized offsets ($\Delta_{\rm QNM}$) given in Table~\ref{tab:qnm_alignment} are all greater than one, and for this reason they are regarded as scattering features. The fact that these features change only slightly in response to the controls listed in Table~\ref{tab:strict_convergence_axial} indicates that the lack of alignment with the QNMs is a physical property within the diagnostic and not due to an unstable peak position.
\subsection{Fan--Wang black hole}
\begin{figure}[t]
\centering
\begin{minipage}{0.49\textwidth}\centering
\includegraphics[width=\linewidth]{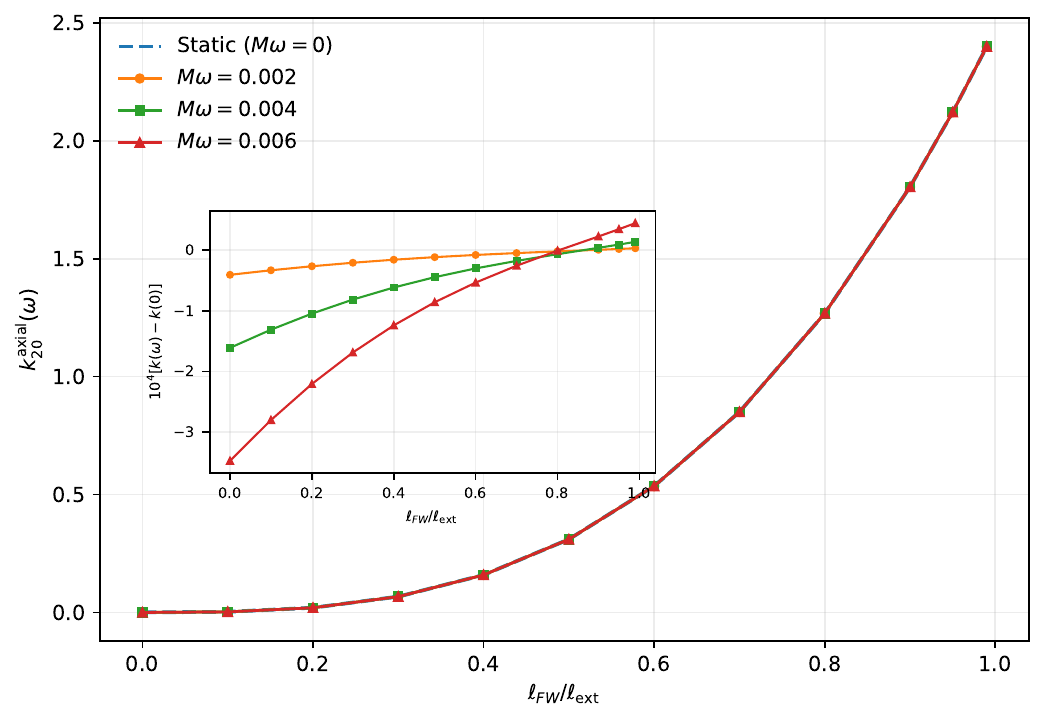}\\[-1mm]
{\small (a) Variation with $\ell_{FW}/\ell_{\rm ext}$.}
\end{minipage}\hfill
\begin{minipage}{0.49\textwidth}\centering
\includegraphics[width=\linewidth]{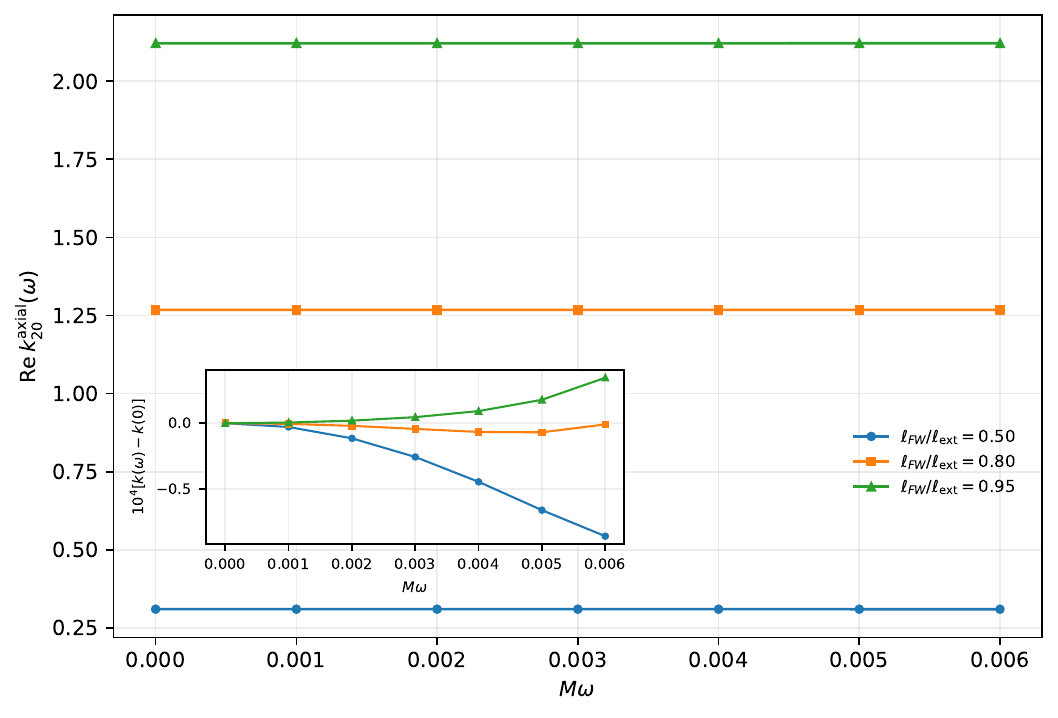}\\[-1mm]
{\small (b) Frequency scans at fixed regularization.}
\end{minipage}
\caption{Fan-Wang axial dynamical Love number. Panel (a) shows the
regularization dependence of the static and finite-frequency responses, and
Panel (b) shows the frequency scans at fixed regularization. The static response is positive. The direct conservative correction is negative at low and intermediate regularization, and changes sign toward large regularization, becoming positive for the high-regularization samples. It begins at $O(\omega^2)$, and the inset resolves this small change. As in the polar Fan--Wang calculation, the leading gravitational and electromagnetic master indices are degenerate, so the full physical source matrix is inverted before selecting a unit gravitational source and a zero independent electromagnetic source.}
\label{fig:fanwang_axial_metric}
\end{figure}
The sign of the axial dynamical correction depends most strongly on the regularization, as shown in Figure \ref{fig:fanwang_axial_metric}. It is negative at low and intermediate regularization, but turns positive for the high-regularization samples. The inset shows this crossing without covering the much larger positive static response.

\begin{figure}[t]
\centering
\includegraphics[width=\textwidth]{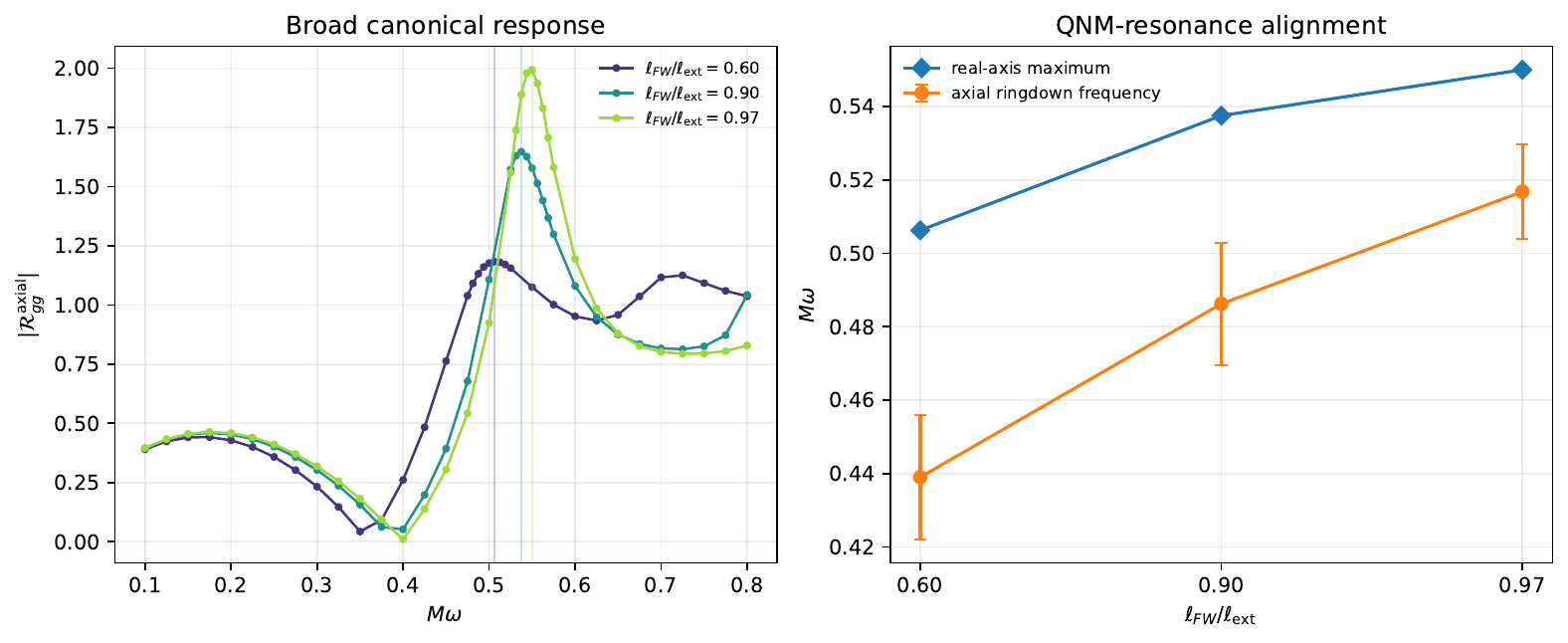}
\caption{Fan--Wang axial resonance diagnostic. The left panel shows the
absolute value of the broad real-frequency canonical response for three
regularization ratios; the vertical lines mark the largest sampled maxima. The
right panel compares these maxima with the real parts of the independently
extracted time-domain ringdown frequencies, with error bars showing the largest
frequency shifts under the ringdown control variations. For the axial response, the maxima occur at $M\omega=0.50625$, $0.5375$, and $0.5500$ for $\ell_{FW}/\ell_{\rm ext}=0.60$, $0.90$, and $0.97$, respectively. For the near-extremal case, $M\omega_{\rm peak}=0.5500$ lies within the damping width of the stable mode with $M\omega_R\simeq0.5168$. The other two sampled maxima also lie within their corresponding damping widths. Across the three cases the quality factor ranges from $Q\simeq2.31$ to $3.37$, so these are broad damped black hole resonance features rather than narrow reflective-cavity modes.}
\label{fig:fanwang_axial_resonance}
\end{figure}
The clearest case of an axial QNM-type is shown in Figure~\ref{fig:fanwang_axial_resonance}. All three maxima fall within one damping width of the independently extracted QNMs listed in Table~\ref{tab:qnm_alignment}, and their peak heights and positions are stable under the  controls described in Table~\ref{tab:strict_convergence_axial}. The near-extremal point supports the interpretation of a broad damped QNM-related resonance feature, aside from the low-frequency metric Love number extraction.
All three sampled Fan--Wang axial maxima satisfy the damping-width alignment criterion. For the near-extremal case, $M\omega_{\rm peak}=0.5500$ and $M\omega_R\simeq0.5168$, with $Q\simeq3.37\,.$ The lower-regularization cases have $Q\simeq2.31$ and $2.83$. We therefore identify the sampled Fan--Wang axial maxima as broad QNM-related resonance features. The resonance is consequently a property of the wider-frequency coupled
scattering problem, not a pole inferred from the low-frequency metric fit. The
axial TLN curves themselves remain restricted to the overlap region where the
near-zone source--response split is controlled. {\color{black}Overall, the axial sector exhibits a stronger dependence of the sign of the dynamical correction on regularization, while its broad-frequency response shows a different pattern of QNM alignment from that found in the polar sector.}

\textcolor{black}{Therefore, based on Secs. (\ref{sec:results_polar}) and (\ref{sec:results_axial}), we can summarize the diagnostics of polar and axial resonance, considering the quantitative alignment test in Table~\ref{tab:qnm_alignment}, as follows. In the polar sector, the peaks are QNM aligned for all sampled Bardeen and Hayward configurations and for the Fan--Wang configuration at $\ell_{FW}/\ell_{\rm ext}=0.60$, whereas the Fan--Wang peaks at $\ell_{FW}/\ell_{\rm ext}=0.90$ and $0.97$ are not aligned. In the axial sector, all sampled Fan--Wang peaks are QNM aligned, while none of the Bardeen or Hayward peaks satisfies the alignment criterion. Thus, broad QNM-related resonance features occur only in the aligned cases; the remaining numerically resolved maxima are classified as broad scattering features. This comparison shows that QNM alignment depends on both the regular black hole geometry and the perturbation parity.}

\section{Conclusions and Discussion}\label{sec:conclusion}
We studied the quadrupolar frequency-dependent tidal response of the Bardeen, Hayward, and Fan--Wang regular black holes using the full coupled Einstein--nonlinear-electrodynamics perturbation equations and a complementary shell EFT construction in a controlled scalar probe sector. The main conclusions are as follows:

\begin{itemize}
\item Although for numerical convenience and stability, we solved the perturbation equation in terms of master variables, the gravitational Love numbers are extracted from the reconstructed physical metric amplitudes, $-fH_0$ in the polar sector and $h_0$ in the axial sector. A full physical source matrix imposes a unit gravitational tide and no independent electromagnetic tide while retaining the induced electromagnetic response. This provides a consistent metric-level definition of the coupled dynamical response.

\item For each of the two parities associated with all three models, we separately obtain the static TLN together with its small-regularization scaling from the zero-frequency coupled system, and then continue it dynamically. The leading-order powers and coefficients agree with those found in the existing small-regularization studies Ref.~\cite{Coviello:2025pla}, whereas the full dynamical continuations yield the higher-order corrections in the neighbourhood of extremality. The direct dynamical corrections start smoothly at $\mathcal{O}(\omega^2)$. The polar correction is negative in the case of Bardeen and positive for Hayward and Fan--Wang. In the axial sector, Hayward exhibits a negative correction over the range sampled, while Bardeen and Fan--Wang show a sign change that depends on the regularization and which tends towards the near-extremal regime. Quantitative comparisons are limited to $M\omega\leq0.004$, although the points with $M\omega=0.006$ indicate the onset of \textcolor{black}{of increased matching sensitivity as the overlap condition ($\omega r\ll1)$ becomes less accurate}. 

\item \textcolor{black}{The shell EFT calculation using the scalar probe gives an explicit EFT description of the dynamical tidal response. When the same-background source subtraction is applied, the response extracted from the derivative jump across the shell agrees with the direct scalar source--response ratio for all three geometries. The response can thus be expressed in terms of renormalized Wilson coefficients in the probe sector without requiring a particular gauge choice.} The analytic low-frequency calculation determines the asymptotic
source-response structure and the associated logarithmic running, while the finite part of the response is fixed by the global ingoing solution and the shell matching. It should therefore not be viewed as an independent asymptotic determination of the complete dynamical Love coefficient. Although the isolated finite coefficients remain scheme dependent, the full matched response is independent of the arbitrary shell radius. The result demonstrates that the frequency-dependent tidal response can be described in terms of well-defined EFT observables. It should, however, be noted that the scalar coefficient is not equivalent to the full gravitational Love number obtained by solving the coupled perturbation equation. Extension of the shell-EFT analysis for the full gravito-electromagnetic sector is left for future work.

\item We also looked for any possible resonances by comparing broad real-frequency scans of the master field with separate time-domain ringdown fits. With the damping-width criterion, the Bardeen and Hayward polar maxima are QNM aligned, together with the polar maximum of the Fan-Wang at $\ell_{FW}/\ell_{\rm ext}=0.60\,.$ The Fan--Wang polar maxima at 0.90 and 0.97 are not aligned. The Bardeen and Hayward axial maxima are not aligned, while all three of the sampled Fan--Wang axial maxima are. In the case of the near-extremal Fan--Wang axial mode, $M\omega_{\rm peak}=0.5500$ and $M\omega_R\simeq0.5168$, with $Q\simeq3.37$. The aligned cases consist of broad QNM-related response features and not of narrow cavity modes. The calculation of the metric-level Love number is not extrapolated into this wider-frequency range.

\end{itemize}

Taken together, the results show that the dynamical tidal response is a controlled low-frequency observable of regular black holes and serves as a useful means of differentiating between their geometries and \textcolor{black}{coupled matter dynamics.}  The analysis based on the \textcolor{black}{shell EFT calculation organizes the response into universal running and finite matching terms}, while the comparison with the QNMs provides \textcolor{black}{a quantitative} criterion for distinguishing genuine resonance-related features from ordinary scattering maxima. {\color{black} A natural next step is to extend the shell EFT construction to the fully coupled gravitational system and to incorporate the resulting response functions into compact-binary dynamics.} This approach would further help in gaining a better understanding of the ambiguities and technical difficulties connected with the classical method of extracting gravitational Love numbers.

\acknowledgments
The research of S.K. is funded by the National Post-Doctoral Fellowship (Grant No. PDF/2023/000369) from the ANRF (formerly SERB), Department of Science and Technology (DST), Government of India. S.K. also acknowledges the support and research facilities provided by the Department of Physics, IIT Kharagpur. A.B. is supported by the Core Research Grant (CRG/2023/005112) by the Department of Science and Technology, Science and Anusandhan National Research Foundation (formerly SERB), Government of India. A.B. also acknowledges the associateship program of the Indian Academy of Science, Bengaluru. The authors thank the speakers of ``Testing Aspects of General Relativity-IV'' for the illuminating discussion. We also acknowledge the use of ChatGPT-Codex 5.6 Sol (OpenAI) for language improvement, grammatical corrections, and debugging some parts of our numerical codes.

\section*{Code Availability Statement}
The numerical codes used for all the results presented in this paper are publicly available in Ref.~\cite{BhattacharyyaKumarKumar_DynamicTLN_RBH_Code}.

\resumetoc

\appendix
\section{Dynamical perturbations and extraction of frequency--dependent Love numbers}
\label{app:dynLove}

\begingroup
This appendix contains both the coupled master equations and the details of conversion to the metric variables in the Regge-Wheeler gauge, from which we can find the dynamical Love
numbers. {\color{black}Briefly, we first summarize the coupled master equations, then give the axial and polar master variables to metric relations. We also verify their finite-frequency Schwarzschild limit, and finally relate the asymptotic metric coefficients to the Love number normalization used in the main text.} Define
\begin{equation}
 m(r)=\frac{r}{2}[1-f(r)],\qquad
 \lambda=(\ell-1)(\ell+2),\qquad
 \kappa=1+\frac{2F\mathcal L_{FF}}{\mathcal L_F},
\end{equation}
and
\begin{equation}
 D_- =\sqrt{\mathcal L_F}\,\frac{d}{dr}
 \left[f\frac{d}{dr}(\mathcal L_F^{-1/2})\right],\qquad
 D_+ =\mathcal L_F^{-1/2}\frac{d}{dr}
 \left[f\frac{d}{dr}(\sqrt{\mathcal L_F})\right].
\end{equation}
The constrained gauge-invariant Einstein--NLED system
\cite{Moreno:2002gg,Chaverra:2016ttw,Nomura:2020tpc} is \textcolor{black}{written in terms of the gauge-invariant gravitational master amplitude $\Psi_\pm$ and electromagnetic master amplitude $\Phi_\pm$, where $+$ and $-$ denote polar and axial parity:}
\begin{equation}
 \frac{d^2\mathbf U_\pm}{dr_*^2}
 +\left[\omega^2\mathbf I-f\mathbf V_\pm\right]\mathbf U_\pm=0,
 \qquad \mathbf U_\pm=(\Psi_\pm,\Phi_\pm)^T,
 \label{eq:app_master}
\end{equation}
where $dr_*/dr=f^{-1}$. For axial parity,
\begin{align}
 V^-_{11}&=\frac{\ell(\ell+1)}{r^2}-\frac{6m}{r^3}+2\mathcal L,
 \\
 V^-_{12}=V^-_{21}&=-\frac{q}{r^3}\sqrt{4\lambda\mathcal L_F},
 \\
 V^-_{22}&=\frac{\ell(\ell+1)}{r^2}+D_-
 +\frac{4q^2\mathcal L_F}{r^4}.
 \label{eq:app_axial_matrix}
\end{align}
For polar parity, let
\begin{equation}
 a=\frac{6m}{r}-2r^2\mathcal L,
 \qquad b=\lambda+\frac{4q^2\mathcal L_F}{r^2}.
\end{equation}
Then
\begin{align}
 V^+_{11}={}&
 \frac{\ell(\ell+1)\lambda-2f\lambda+a(a-4m/r)}{r^2(a+\lambda)}
 +\frac{2f\lambda b}{r^2(a+\lambda)^2},
 \\
 V^+_{22}={}&\frac{\kappa\ell(\ell+1)}{r^2}+D_+
 +\frac{4q^2\mathcal L_F[\lambda+1-f+2r^2\mathcal L+4f\kappa]}
 {r^4(a+\lambda)}
 +\frac{8fq^2\mathcal L_F b}{r^4(a+\lambda)^2},
 \\
 V^+_{12}=V^+_{21}={}&-\frac{q}{r^3}\sqrt{4\lambda\mathcal L_F}
 \left[
 \frac{\lambda+1-f+2r^2\mathcal L+2f\kappa}{a+\lambda}
 +\frac{2fb}{(a+\lambda)^2}
 \right].
 \label{eq:app_polar_matrix}
\end{align}
These expressions apply to all three models after inserting
\begin{align}
 m_B&=\frac{Mr^3}{(r^2+\ell_B^2)^{3/2}}, &
 \mathcal L_B&=\frac{3M\ell_B^2}{(r^2+\ell_B^2)^{5/2}}, &
 \mathcal L_{F,B}&=\frac{15Mr^6}{2(r^2+\ell_B^2)^{7/2}},
 \\
 m_H&=\frac{Mr^3}{r^3+2M\ell_H^2}, &
 \mathcal L_H&=\frac{6M^2\ell_H^2}{(r^3+2M\ell_H^2)^2}, &
 \mathcal L_{F,H}&=\frac{18M^2r^7}{(r^3+2M\ell_H^2)^3},
 \\
 m_{FW}&=\frac{Mr^3}{(r+\ell_{FW})^3}, &
 \mathcal L_{FW}&=\frac{3M\ell_{FW}}{(r+\ell_{FW})^4}, &
 \mathcal L_{F,FW}&=\frac{6Mr^5}{\ell_{FW}(r+\ell_{FW})^5}.
 \label{eq:app_model_functions}
\end{align}
Here $q=\ell_B,\ell_H,\ell_{FW}$ in the corresponding row.

\subsection{Axial master-to-metric conversion}
For odd parity,
in Regge--Wheeler gauge and with time dependence $e^{-i\omega t}$, this gives
\begin{equation}
 h_0=\frac{f}{\sqrt\lambda}\frac{d}{dr}(r\Psi_-),
 \qquad
 h_1=-\frac{i\omega r}{f\sqrt\lambda}\Psi_-.
 \label{eq:app_axial_metric_map}
\end{equation}
The overall orientation sign of the two-dimensional Hodge dual multiplies both
source and response and therefore cancels from their ratio. Combining
Eq.~\eqref{eq:app_axial_metric_map} with the first component of
Eq.~\eqref{eq:app_master} yields an equivalent metric first-order
system for $\omega\neq0$,
\begin{align}
 (fh_1)'&=-\frac{i\omega}{f}h_0,
 \\
 h_0'&=\frac{2}{r}h_0
 +\frac{if}{\omega r}
 \left(f'-\frac{2f}{r}-\frac{r\omega^2}{f}+rV^-_{11}\right)h_1
 +\frac{rV^-_{12}}{\sqrt\lambda}\Phi_-.
 \label{eq:app_axial_metric_system}
\end{align}
The electromagnetic master equation \textcolor{black}{which is the second component of the coupled master system in Eq.(\ref{eq:app_master})} supplies the remaining coupled equation.
The static limit is obtained directly from
Eqs.~\eqref{eq:app_master} and~\eqref{eq:app_axial_metric_map}.
The first-order system in Eq.~\eqref{eq:app_axial_metric_system} is valid for $\omega\neq0$ and contains explicit factors of $1/\omega$, so it is not used to take the static limit.

\subsection{Polar master-to-metric conversion}
For even parity, define the intermediate gauge-invariant variables
\begin{equation}
 \zeta=\frac{a+\lambda}{\sqrt\lambda}\Psi_+,
 \qquad
 \varphi=\frac{\Phi_+}{2\sqrt{\mathcal L_F}}
 +\frac{q\Psi_+}{r\sqrt\lambda}.
 \label{eq:app_polar_intermediate}
\end{equation}
The polar constraint gives
\begin{equation}
 K=\frac{8fq\mathcal L_F\varphi/r-2rf\zeta'
 -\ell(\ell+1)\zeta}{r(a+\lambda)},
 \label{eq:app_polar_K}
\end{equation}
and the temporal component of the metric perturbation in polar sector is
\begin{equation}
 H_0=-\zeta'-rK'+\frac{4q\mathcal L_F}{r^2}\varphi.
 \label{eq:app_polar_H0}
\end{equation}
Equations~\eqref{eq:app_master},
\eqref{eq:app_polar_intermediate}--\eqref{eq:app_polar_H0} are the conversion
implemented in the polar code. The electromagnetic master
field contributes to the reconstructed metric and is not discarded after the pure gravitational source condition is imposed. {\color{black}As a consistency check on these coupled master equations and metric relations, we next verify that the gravitational sector reduces exactly to the standard Regge–Wheeler and Zerilli systems for the Schwarzwald case when the regularization scale vanishes.}

\subsection{Explicit Schwarzschild reduction}
\label{app:schwarzschild_reduction}
The Schwarzschild limit is obtained by sending the regularization scale of
any of the three models to zero while keeping $M$, $\omega$, and the
multipole number fixed.  Denoting the corresponding scale by
$\ell_{\rm reg}$, Eqs.~\eqref{eq:app_model_functions} give
\begin{equation}
 \ell_{\rm reg}\to0:
 \qquad m(r)\to M,\qquad
 f(r)\to f_S(r)=1-\frac{2M}{r},\qquad
 \mathcal L\to0,
 \label{eq:app_schw_background_limit}
\end{equation}
and the combinations that mix the gravitational and electromagnetic sectors
satisfy
\begin{equation}
 q\sqrt{\mathcal L_F}\to0,
 \qquad q^2\mathcal L_F\to0.
 \label{eq:app_schw_mixing_limit}
\end{equation}
Consequently $V^{\pm}_{12}=V^{\pm}_{21}=0$, and the first component of the
coupled system becomes a closed gravitational equation.  This statement does
not require taking $\omega\to0$; it holds for the complete frequency-domain
problem.

\paragraph{Axial parity.}
Using $m=M$ and $\mathcal L=0$ in Eq.~\eqref{eq:app_axial_matrix} gives
\begin{equation}
 V^-_{11}\big|_S
 =\frac{\ell(\ell+1)}{r^2}-\frac{6M}{r^3}.
 \label{eq:app_schw_axial_v11}
\end{equation}
The gravitational master equation therefore becomes
\begin{equation}
 \frac{d^2\Psi_-}{dr_*^2}
 +\left[\omega^2-V_{\rm RW}(r)\right]\Psi_-=0,
 \qquad
 V_{\rm RW}(r)=f_S(r)
 \left[\frac{\ell(\ell+1)}{r^2}-\frac{6M}{r^3}\right],
 \label{eq:app_regge_wheeler_limit}
\end{equation}
which is the Regge--Wheeler equation.  The metric reconstruction makes the
same reduction explicit.  Define the conventional Regge--Wheeler metric
variable
\begin{equation}
 Q_{\rm RW}=\frac{f_S h_1}{r}
 =-\frac{i\omega}{\sqrt\lambda}\Psi_-.
 \label{eq:app_Qrw_definition}
\end{equation}
Because $Q_{\rm RW}$ differs from $\Psi_-$ only by an $r$-independent
normalization, Eq.~\eqref{eq:app_regge_wheeler_limit} is equivalently
\begin{equation}
 \frac{d^2Q_{\rm RW}}{dr_*^2}
 +\left[\omega^2-V_{\rm RW}(r)\right]Q_{\rm RW}=0.
 \label{eq:app_Qrw_equation}
\end{equation}
The remaining axial metric amplitude follows from
\begin{equation}
 h_0=\frac{i f_S}{\omega}\frac{d}{dr}\left(rQ_{\rm RW}\right),
 \label{eq:app_schw_h0_from_Qrw}
\end{equation}
for the $e^{-i\omega t}$ convention.  Equivalently, the metric first-order
system in Eq.~\eqref{eq:app_axial_metric_system} reduces to
\begin{align}
 (f_S h_1)'&=-\frac{i\omega}{f_S}h_0,
 \label{eq:app_schw_axial_metric_1}
 \\
 h_0'&=\frac{2}{r}h_0
 +\frac{i f_S}{\omega r}
 \left[
 f_S'-\frac{2f_S}{r}-\frac{r\omega^2}{f_S}
 +\frac{\ell(\ell+1)}{r}-\frac{6M}{r^2}
 \right]h_1.
 \label{eq:app_schw_axial_metric_2}
\end{align}
Eliminating $h_0$ from these two equations and setting
$Q_{\rm RW}=f_Sh_1/r$ reproduces Eq.~\eqref{eq:app_Qrw_equation} exactly.
Thus the axial metric equations and the axial master equation have the same
Schwarzschild limit.

\paragraph{Polar parity.}
In the same limit,
\begin{equation}
 a\to\frac{6M}{r},\qquad b\to\lambda,
 \qquad \ell(\ell+1)=\lambda+2.
 \label{eq:app_schw_polar_ab}
\end{equation}
Direct substitution in the polar gravitational entry $V^+_{11}$ above yields
\begin{equation}
 V^+_{11}\big|_S
 =\frac{
 \lambda^2(\lambda+2)r^3+6\lambda^2Mr^2
 +36\lambda M^2r+72M^3}
 {r^3(\lambda r+6M)^2}.
 \label{eq:app_schw_polar_v11}
\end{equation}
The potential appearing in the wave equation is therefore
\begin{equation}
 V_Z(r)=f_S(r)V^+_{11}\big|_S.
 \label{eq:app_zerilli_potential_lambda}
\end{equation}
Introducing the conventional Zerilli parameter
$n=(\ell-1)(\ell+2)/2=\lambda/2$, this becomes
\begin{equation}
 V_Z(r)=
 \frac{2f_S(r)}{r^3(nr+3M)^2}
 \left[
 n^2(n+1)r^3+3n^2Mr^2+9nM^2r+9M^3
 \right].
 \label{eq:app_zerilli_potential_standard}
\end{equation}
Hence, with $Z\equiv\Psi_+$ up to an irrelevant constant normalization, the
polar gravitational equation is
\begin{equation}
 \frac{d^2Z}{dr_*^2}
 +\left[\omega^2-V_Z(r)\right]Z=0,
 \label{eq:app_zerilli_equation}
\end{equation}
which is the Zerilli equation.

The polar metric reconstruction also reduces directly to the standard
Regge--Wheeler-gauge reconstruction.  Equations
\eqref{eq:app_polar_intermediate}--\eqref{eq:app_polar_H0} give
\begin{align}
 \zeta_S&=\frac{\lambda+6M/r}{\sqrt\lambda}\,Z,
 \label{eq:app_schw_zeta}
 \\
 K_S&=-\frac{2f_S}{\sqrt\lambda}\frac{dZ}{dr}
 -\frac{\lambda(\lambda+2)r^2+6\lambda Mr+24M^2}
 {\sqrt\lambda\,r^2(\lambda r+6M)}Z,
 \label{eq:app_schw_K_from_Z}
 \\
 H_{0,S}&=-\frac{d\zeta_S}{dr}-r\frac{dK_S}{dr}.
 \label{eq:app_schw_H0_from_Z}
\end{align}
The $tr$ Einstein equation gives $H_{1,S}$ algebraically in terms of $Z$. Substituting Eqs.~\eqref{eq:app_schw_K_from_Z} and
\eqref{eq:app_schw_H0_from_Z} into the even-parity metric equations reduces all remaining constraints to the Zerilli equation, Eq.~\eqref{eq:app_zerilli_equation}. The same result follows if the metric constraint equations are eliminated first. Thus the polar metric equations are equivalent to the Zerilli master equation, with no additional frequency-dependent term.

For the quadrupole used in the numerical analysis, $\lambda=4$ and $n=2$, so
the two Schwarzschild potentials are explicitly
\begin{align}
 V_{\rm RW}^{\ell=2}
 &=f_S\left(\frac{6}{r^2}-\frac{6M}{r^3}\right),
 \\
 V_Z^{\ell=2}
 &=\frac{6f_S
 \left(4r^3+4Mr^2+6M^2r+3M^3\right)}
 {r^3(2r+3M)^2}.
 \label{eq:app_quadrupole_schw_potentials}
\end{align}
Equations~\eqref{eq:app_regge_wheeler_limit} and
\eqref{eq:app_zerilli_equation} are therefore the exact finite-frequency
Schwarzschild limits of the axial and polar gravitational equations used in this work. {\color{black}With metric variables and their Schwarzschild limit, we now specify the asymptotic source–response normalization used to extract the physical Love numbers.}

\subsection{Metric source--response coefficients and Love normalization}
For the quadrupole, the metric asymptotics of Ref.~\cite{Coviello:2025pla} are
\begin{align}
 -fH_0&=-\mathcal E_{20}r^2
 +2\sqrt{\frac{4\pi}{5}}\,\mathcal M_{20}r^{-3}+\cdots,
 \\
 h_0&=\frac{1}{3}\mathcal B_{20}r^3
 +2\sqrt{\frac{\pi}{5}}\,\mathcal S_{20}r^{-2}+\cdots.
 \label{eq:app_metric_asymptotics}
\end{align}
Writing the coefficients of the displayed source and response powers as
$C_{+,\mathrm{s}},C_{+,\mathrm{r}}$ and
$C_{-,\mathrm{s}},C_{-,\mathrm{r}}$, respectively, one obtains
\begin{equation}
 k_{20}^{\mathrm{polar}}=-\frac{1}{M^5}
 \frac{C_{+,\mathrm{r}}}{C_{+,\mathrm{s}}},\qquad
 k_{20}^{\mathrm{axial}}=\frac{1}{M^5}
 \frac{C_{-,\mathrm{r}}}{C_{-,\mathrm{s}}}
 \label{eq:app_metric_love_factors}
\end{equation}
This polar normalization is included in the numerical scripts \cite{BhattacharyyaKumarKumar_DynamicTLN_RBH_Code}. Logarithmic
and noninteger-power descendants are included in the
near-zone basis, but do not replace the two powers used in
Eq.~\eqref{eq:app_metric_asymptotics}.

For a Schwarzschild normalization check, the quadrupolar Riccati basis gives
\begin{align}
 H_0[\widehat j_2]&=\frac{2}{5}\omega^3r^2+\cdots,
 &H_0[\widehat y_2]&=-\frac{3}{\omega^2r^3}+\cdots,
 \\
 h_0[\widehat j_2]&=\frac{2}{15}\omega^3r^3+\cdots,
 &h_0[\widehat y_2]&=\frac{3}{2\omega^2r^2}+\cdots.
\end{align}
Therefore, the leading master-to-metric coefficient ratios are
\begin{equation}
 \frac{C_{+,\mathrm{r}}}{C_{+,\mathrm{s}}}
 =-\frac{15}{2\omega^5}\mathcal R_{gg,+},
 \qquad
 \frac{C_{-,\mathrm{r}}}{C_{-,\mathrm{s}}}
 =\frac{45}{4\omega^5}\mathcal R_{gg,-}.
 \label{eq:app_leading_master_metric}
\end{equation}
Here $\mathcal R_{gg,\pm}$ denotes the gravitational-to-gravitational component of the coupled response matrix $\mathcal R_\pm=BA^{-1}$. We impose a unit gravitational source with no independent electromagnetic source, while retaining the induced electromagnetic response. The numerical extraction does not rely on these leading formulas alone; it builds
frequency-corrected coupled near-zone modes and performs the metric projection
at finite matching radii. Finally, the physical basis is chosen so that
\begin{equation}
 (C_{g,\mathrm{s}},C_{e,\mathrm{s}})=(1,0).
\end{equation}
For Fan--Wang, where the leading indices are degenerate, the complete physical
$2\times2$ source and response matrices are inverted. The finite-frequency
quantity plotted in the paper is Eq.~\eqref{eq:metric_dynamic_definition},
with $\gamma_+=-1/M^5$ and $\gamma_-=1/M^5$.

The finite-frequency Schwarzschild reduction is displayed explicitly in
Eqs.~\eqref{eq:app_regge_wheeler_limit}--\eqref{eq:app_quadrupole_schw_potentials}. At $\omega=0$, the reconstruction reduces to the horizon-regular metric
system used in Ref.~\cite{Coviello:2025pla}.
\endgroup

\section{Numerical intermediate-near-zone asymptotic solutions}
\label{app:numerical_asymptotics}

At a nonzero frequency, the fixed-frequency limit $r\to\infty$ is a radiative region, where the solutions are oscillatory ingoing/outgoing waves rather than the static tidal power laws. The source-response expansion relevant for computing the dynamical Love number numerically must therefore be constructed in the overlap region to control the filling errors. 
\begin{equation}
 M\ll r\ll \omega^{-1}.
 \label{eq:app_overlap_region}
\end{equation}
In this sense, the ``asymptotic'' solutions below are numerical intermediate-near-zone asymptotic solutions. They are the finite-frequency counterpart of the static large-radius expansions used in Ref.~\cite{Coviello:2025pla}, but they are not meant to be continued as power series to fixed-frequency null infinity.

It is useful to state explicitly how the numerical basis used in Part II is constructed. Define
\begin{equation}
 x=\left(\frac{M}{r}\right)^{1/2},\qquad \widehat\omega=M\omega .
\end{equation}
For each parity and for each of the four physical branches $a\in\{g{\rm s},e{\rm s},g{\rm r},e{\rm r}\}$, the two-component master field is generated recursively in the form
\begin{equation}
 \mathbf U_{a,\pm}(r,\omega)
 =\sum_{k=0}^{2}\widehat\omega^{2k}
 x^{\sigma_{a,\pm}-4k}
 \sum_{n=0}^{N_\pm}\sum_{p=0}^{P_\pm}
 \mathbf c^{(a,\pm)}_{knp}\,x^n(\log x)^p
 +O(\widehat\omega^6),
 \label{eq:app_dynamic_nearzone_series}
\end{equation}
where $g/e$ labels gravitational/electromagnetic character and ${\rm s}/{\rm r}$ labels source/response character. The indicial exponent $\sigma_{a,\pm}$ and the coefficient vectors $\mathbf c^{(a,\pm)}_{knp}$ are obtained from the large-radius expansion of the same coupled potential matrix used in the numerical integration. Whenever the radial recurrence becomes resonant, the logarithmic coefficients are solved simultaneously rather than discarding the logarithmic branch. This is why the basis can include logarithmic and, through the indicial roots, noninteger-power descendants.

The truncations used in the calculation are summarized in Table~\ref{tab:nearzone_basis_orders}. They are the same truncations used for the metric-level results in Secs.~\ref{sec:results_polar} and~\ref{sec:results_axial}.
\begin{table}[H]
\centering
\caption{Numerical intermediate-near-zone basis used for the coupled gravitational response. The frequency order $k=0,1,2$ corresponds to terms through $O(\omega^4)$. The different axial logarithmic orders are model-specific numerical choices of the recurrence solver.}
\label{tab:nearzone_basis_orders}
\begin{tabular}{lcccc}
\hline
Sector & models & $N_\pm$ & $P_\pm$ & frequency order\\
\hline
Polar & Bardeen, Hayward, Fan--Wang & 18 & 5 & $O(\omega^4)$\\
Axial & Bardeen & 22 & 5 & $O(\omega^4)$\\
Axial & Hayward, Fan--Wang & 22 & 6 & $O(\omega^4)$\\
\hline
\end{tabular}
\end{table}

After the master fields are reconstructed into the physical metric and electromagnetic variables, the basis is normalized by its leading physical source and response coefficients. In particular,
\begin{align}
 Z_{+,g{\rm s}}&=r^2\left[1+O\!\left(\frac{M}{r}\right)+O((\omega r)^2)\right],
 &Z_{+,g{\rm r}}&=r^{-3}\left[1+O\!\left(\frac{M}{r}\right)+O((\omega r)^2)\right],\\
 Z_{-,g{\rm s}}&=r^3\left[1+O\!\left(\frac{M}{r}\right)+O((\omega r)^2)\right],
 &Z_{-,g{\rm r}}&=r^{-2}\left[1+O\!\left(\frac{M}{r}\right)+O((\omega r)^2)\right],
 \label{eq:app_metric_basis_leading}
\end{align}
where $Z_+=-fH_0$ and $Z_-=h_0$. The omitted terms are not discarded in the calculation: they are generated by Eq.~\eqref{eq:app_dynamic_nearzone_series} and retained through the orders in Table~\ref{tab:nearzone_basis_orders}.

For the physical horizon-ingoing solution, the unit gravitational source and vanishing independent electromagnetic source imply
\begin{equation}
 \mathbf U^{\rm phys}_\pm
 =\mathbf U_{g{\rm s},\pm}
 +c_{g,\pm}(\omega)\,\mathbf U_{g{\rm r},\pm}
 +c_{e,\pm}(\omega)\,\mathbf U_{e{\rm r},\pm}.
 \label{eq:app_physical_nearzone_solution}
\end{equation}
Thus $c_{g,\pm}$ is the gravitational response amplitude in this numerical near-zone basis, while $c_{e,\pm}$ is the electromagnetic response induced by the gravitational tide. There is no $\mathbf U_{e{\rm s},\pm}$ term in Eq.~\eqref{eq:app_physical_nearzone_solution}. The electromagnetic response is nevertheless retained because the metric reconstruction depends on the coupled solution.

The coefficients are determined at the four matching radii $r/M=10,12,14,16$. To remove the first omitted frequency and inverse-radius terms, the match-radius dependence of either complex amplitude $c_a$ is extrapolated as
\begin{equation}
 c_a(r_m)=c_a^{(0)}+A_a(\omega r_m)^6+\frac{B_a}{r_m^2},
 \qquad a\in\{g,e\},
 \label{eq:app_nearzone_extrapolation}
\end{equation}
which is the same extrapolation used in the metric-level Love number calculation. Tables~\ref{tab:polar_nearzone_amplitudes} and
\ref{tab:axial_nearzone_amplitudes} list representative extrapolated
amplitudes at two frequencies within the controlled range and for three
regularization ratios. These quantities are the raw complex amplitudes appearing
in Eq.~\eqref{eq:app_physical_nearzone_solution} and should not be identified
directly with the physical Love numbers shown in the main text. The latter are
obtained from $c_{g,\pm}$ only after applying
Eq.~\eqref{eq:metric_dynamic_definition}, with the zero-frequency value fixed
by the independently recovered static Love number.
\begin{table}[t]
\centering
\small
\setlength{\tabcolsep}{4pt}
\caption{Polar numerical intermediate-near-zone amplitudes. The physical solution has unit gravitational source and zero independent electromagnetic source. $c_{g,+}$ multiplies the gravitational response branch and $c_{e,+}$ the induced electromagnetic response branch in Eq.~\eqref{eq:app_physical_nearzone_solution}. The imaginary parts encode the dissipative phase generated by the ingoing horizon condition.}
\label{tab:polar_nearzone_amplitudes}
\begin{tabular}{lcccccc}
\hline
Geometry & $\ell/\ell_{\rm ext}$ & $M\omega$ & $\Re c_{g,+}$ & $\Im c_{g,+}$ & $\Re c_{e,+}$ & $\Im c_{e,+}$\\
\hline
Bardeen & 0.50 & 0.002 & $-0.610049$ & $0.002696$ & $-0.230324$ & $0.001634$ \\
Bardeen & 0.50 & 0.004 & $-0.609982$ & $0.005393$ & $-0.230437$ & $0.003269$ \\
Bardeen & 0.80 & 0.002 & $-0.234143$ & $9.173037\times10^{-4}$ & $-1.995275$ & $9.106719\times10^{-4}$ \\
Bardeen & 0.80 & 0.004 & $-0.234039$ & $0.001836$ & $-1.995481$ & $0.001822$ \\
Bardeen & 0.95 & 0.002 & $-0.113139$ & $1.794619\times10^{-4}$ & $-4.50606$ & $2.153965\times10^{-4}$ \\
Bardeen & 0.95 & 0.004 & $-0.113019$ & $3.593208\times10^{-4}$ & $-4.506514$ & $4.312538\times10^{-4}$ \\
Hayward & 0.50 & 0.002 & $-0.378849$ & $0.002956$ & $-0.36339$ & $-3.504280\times10^{-4}$ \\
Hayward & 0.50 & 0.004 & $-0.378894$ & $0.005913$ & $-0.363396$ & $-7.010972\times10^{-4}$ \\
Hayward & 0.80 & 0.002 & $0.614144$ & $0.001229$ & $-1.496604$ & $-2.366728\times10^{-4}$ \\
Hayward & 0.80 & 0.004 & $0.613946$ & $0.002459$ & $-1.496672$ & $-4.736037\times10^{-4}$ \\
Hayward & 0.95 & 0.002 & $1.319672$ & $3.128598\times10^{-4}$ & $-2.517376$ & $-7.247714\times10^{-5}$ \\
Hayward & 0.95 & 0.004 & $1.319361$ & $6.265247\times10^{-4}$ & $-2.5175$ & $-1.451412\times10^{-4}$ \\
Fan--Wang & 0.50 & 0.002 & $-2.375006$ & $0.001001$ & $0.107487$ & $-4.483200\times10^{-5}$ \\
Fan--Wang & 0.50 & 0.004 & $-2.375038$ & $0.002003$ & $0.107489$ & $-8.969359\times10^{-5}$ \\
Fan--Wang & 0.80 & 0.002 & $-3.470063$ & $1.915699\times10^{-4}$ & $0.245881$ & $-1.353121\times10^{-5}$ \\
Fan--Wang & 0.80 & 0.004 & $-3.47017$ & $3.833058\times10^{-4}$ & $0.245889$ & $-2.707416\times10^{-5}$ \\
Fan--Wang & 0.95 & 0.002 & $-4.09589$ & $2.504192\times10^{-5}$ & $0.342217$ & $-2.090269\times10^{-6}$\\
Fan--Wang & 0.95 & 0.004 & $-4.096048$ & $5.012632\times10^{-5}$& $0.342231$ & $-4.184113\times10^{-6}$ \\
\hline
\end{tabular}
\end{table}

\begin{table}[t]
\centering
\small
\setlength{\tabcolsep}{4pt}
\caption{Axial numerical intermediate-near-zone amplitudes, with the same normalization as Table~\ref{tab:polar_nearzone_amplitudes}. The nonzero $c_{e,-}$ values show explicitly that a purely gravitational external tide still induces an electromagnetic response through the coupled perturbation equations.}
\label{tab:axial_nearzone_amplitudes}
\begin{tabular}{lcccccc}
\hline
Geometry & $\ell/\ell_{\rm ext}$ & $M\omega$ & $\Re c_{g,-}$ & $\Im c_{g,-}$ & $\Re c_{e,-}$ & $\Im c_{e,-}$\\
\hline
Bardeen & 0.50 & 0.002 & $-0.02816$ & $-0.004678$ & $0.207737$ & $0.002324$ \\
Bardeen & 0.50 & 0.004 & $-0.028261$ & $-0.009358$ & $0.207695$ & $0.00465$ \\
Bardeen & 0.80 & 0.002 & $-0.221591$ & $-0.001788$ & $1.169273$ & $0.001219$ \\
Bardeen & 0.80 & 0.004 & $-0.221637$ & $-0.003579$ & $1.169173$ & $0.002438$ \\
Bardeen & 0.95 & 0.002 & $-0.499126$ & $-2.736724\times10^{-4}$ & $2.331202$ & $2.027829\times10^{-4}$\\
Bardeen & 0.95 & 0.004 & $-0.499111$ & $-5.479532\times10^{-4}$ & $2.331019$ & $4.060214\times10^{-4}$ \\
Hayward & 0.50 & 0.002 & $0.039812$ & $-0.005265$ & $0.043404$ & $1.279071\times10^{-4}$ \\
Hayward & 0.50 & 0.004 & $0.039693$ & $-0.010534$ & $0.043409$ & $2.559038\times10^{-4}$ \\
Hayward & 0.80 & 0.002 & $0.265032$ & $-0.002792$ & $0.177186$ & $9.533026\times10^{-5}$ \\
Hayward & 0.80 & 0.004 & $0.26493$ & $-0.005588$ & $0.177189$ & $1.907630\times10^{-4}$ \\
Hayward & 0.95 & 0.002 & $0.533287$ & $-7.553341\times10^{-4}$ & $0.295969$ & $2.834646\times10^{-5}$ \\
Hayward & 0.95 & 0.004 & $0.53322$ & $-0.001513$ & $0.295967$ & $5.676626\times10^{-5}$\\
Fan--Wang & 0.50 & 0.002 & $0.282999$ & $-0.001159$ & $0.048588$ & $4.047084\times10^{-5}$ \\
Fan--Wang & 0.50 & 0.004 & $0.282966$ & $-0.002319$ & $0.048589$ & $8.096777\times10^{-5}$ \\
Fan--Wang & 0.80 & 0.002 & $1.202982$ & $-1.071413\times10^{-4}$ & $0.235022$ & $5.271341\times10^{-6}$ \\
Fan--Wang & 0.80 & 0.004 & $1.202978$ & $-2.143753\times10^{-4}$ & $0.235023$ & $1.054719\times10^{-5}$ \\
Fan--Wang & 0.95 & 0.002 & $2.050533$ & $-3.346234\times10^{-6}$ & $0.422659$ & $1.851009\times10^{-7}$ \\
Fan--Wang & 0.95 & 0.004 & $2.05054$ & $-6.698443\times10^{-6}$ & $0.422661$ & $3.699025\times10^{-7}$ \\
\hline
\end{tabular}
\end{table}

The tables also make two features of the numerical construction clear. First,
at sufficiently small $M\omega$, the imaginary parts vary approximately
linearly with frequency, as expected for the leading dissipative contribution
of the ingoing solution. By contrast, the conservative correction to the
physical dynamical Love number begins at even order in frequency. Second,
$c_{e,\pm}$ remains nonzero even though the independent electromagnetic source
is set to zero. This quantity represents the induced matter response retained
in the metric reconstruction, rather than an externally applied
electromagnetic tidal field.


\section{Fixed-background tensor test-field comparison}
\label{app:tensor_probe}

This appendix gives a separate spin-2 probe field computation. It is included only to show how much of the finite-frequency behavior can arise from propagation on the fixed regular black hole geometry before gravitational--electromagnetic backreaction is included. It is not part of the scalar shell-EFT matching in Part I and is not used to define the gravitational Love numbers in Part II.

For $\phi=0$, the test field obeys the Regge--Wheeler-type equation
\begin{align}
 \frac{d^2\psi_T}{dr_*^2}+[\omega^2-V_T(r)]\psi_T&=0,\\
 V_T(r)&=f(r)\left[\frac{\ell(\ell+1)}{r^2}
 -\frac{6m(r)}{r^3}+\frac{2m'(r)}{r^2}\right].
 \label{eq:tensor_probe_rw}
\end{align}
The field is purely ingoing at the outer horizon. Since it is treated as a test field, it does not perturb the background metric and also does not couple to the nonlinear-electrodynamics perturbations. Therefore, its response describes scattering on a fixed background, which is separate from the coupled gravitational Love number.

We now extract the tensor response in the same intermediate-region spirit used for the full coupled calculation. At $\omega=0$ we solve Eq.~\eqref{eq:tensor_probe_rw} independently, imposing horizon regularity and decomposing the solution into the static source and response basis,
\begin{equation}
 \psi_T(r,0)=C_{T,s}(0)\,\psi_{T,s}(r,0)+C_{T,r}(0)\,\psi_{T,r}(r,0),
 \qquad
 \frac{\alpha_T(0)}{M^5}=\frac{1}{M^5}\operatorname{Re}\!\left[\frac{C_{T,r}(0)}{C_{T,s}(0)}\right].
 \label{eq:tensor_probe_static_direct}
\end{equation}
The large-radius basis is constructed in the same $\log(r/2M)$ convention used in the static analysis of Ref.~\cite{Coviello:2025pla}. This is relevant for Bardeen because its static tensor response contains a logarithmic running term. As a check, a small regularization fit to our independently extracted static solution gives
\begin{equation}
 \alpha_{T,\rm static}^{B}\simeq 8.34\,M\ell_B^4,
 \qquad
 \alpha_{T,\rm static}^{H}\simeq 4.83\,M\ell_H^4,
 \label{eq:tensor_probe_static_scaling}
\end{equation}
in agreement with the corresponding leading coefficients $42/5=8.4$ and $24/5=4.8$ obtained analytically in Ref.~\cite{Coviello:2025pla}. For Fan--Wang, the independently extracted static response has the
small-regularization behavior
\begin{equation}
\alpha_{T,\rm static}^{FW}
\simeq \frac{128}{25}M^3\ell_{FW}^2.
\end{equation}
At the smallest sampled regularization value, the numerical coefficient is
$5.1144$, in close agreement with the analytic value
$128/25=5.12$.

At finite frequency we construct the source and response solutions perturbatively in $\omega^2$ through $\mathcal{O}(\omega^6)$, using radial series order 26, and match the horizon-ingoing numerical solution at $r/M=10,12,14,16$, where $M\ll r\ll\omega^{-1}$ for the frequencies used below. The higher-order basis leaves the static response unchanged while reducing the matching-window sensitivity of the small finite-frequency correction.
Writing
\begin{equation}
 c_T(\omega)=\frac{C_{T,r}(\omega)}{C_{T,s}(\omega)},
\end{equation}
we define the conservative dynamical tensor response by anchoring the finite-frequency change to the independently computed static value,
\begin{equation}
 \frac{\alpha_T(\omega)}{M^5}
 =\frac{\alpha_T(0)}{M^5}
 +\frac{1}{M^5}\operatorname{Re}\!\left[c_T(\omega)-c_T(0)\right].
 \label{eq:tensor_probe_dynamic_direct}
\end{equation}

\begin{figure}[htb!]
\centering
\includegraphics[width=0.96\textwidth]{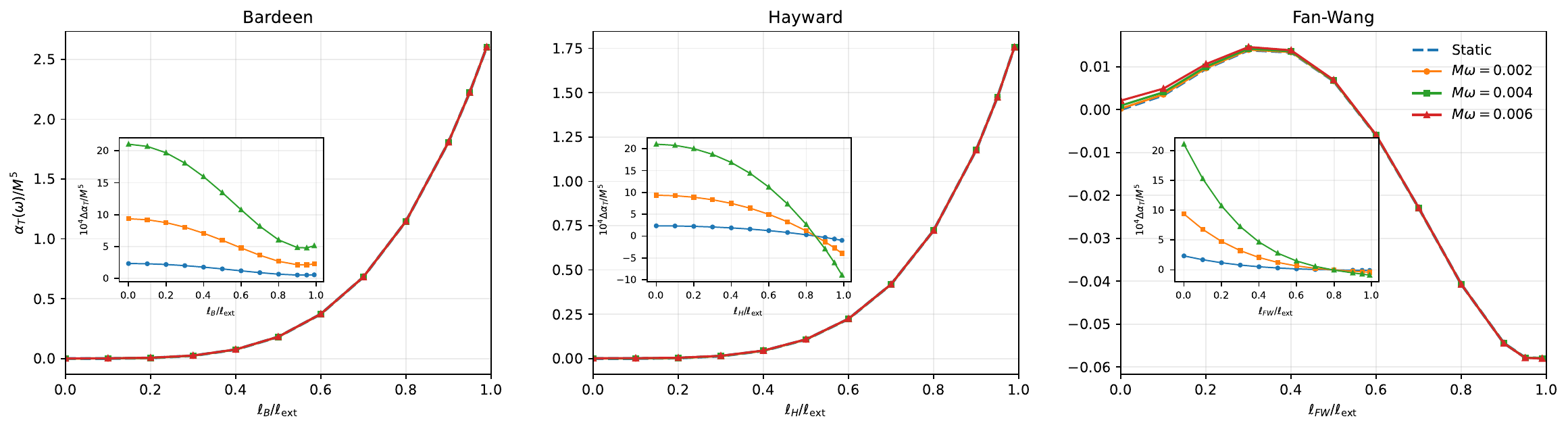}
\caption{Fixed-background tensor test-field response extracted in the
intermediate region. The three panels show the Bardeen, Hayward, and Fan--Wang
cases from left to right. The dashed curve gives the independently computed
static response, while the solid curves show the full dynamical response at
$M\omega=0.002$, $0.004$, and $0.006$. Because the finite-frequency
corrections are much smaller than the static response, the insets display
$10^4\Delta\alpha_T/M^5$, where
$\Delta\alpha_T\equiv\alpha_T(\omega)-\alpha_T(0)$. The Bardeen static
curve reproduces the positive behavior reported in Ref.~\cite{Coviello:2025pla},
while the Hayward static response is also positive and monotonic. For
Fan--Wang, the static response is positive at small regularization, crosses
zero near $\ell_{FW}/\ell_{\rm ext}\simeq0.56$, and becomes negative toward
extremality. The finite-frequency curves are obtained using the
$\mathcal{O}(\omega^6)$, radial-order-26 near-zone basis. The correction
remains positive over the sampled Bardeen range, changes sign toward the
near-extremal Hayward regime, and is negative for near-extremal Fan-Wang.}
\label{fig:tensor_testfield}
\end{figure}

\paragraph{Bardeen background.}
The independently reproduced static response is positive and grows rapidly with $\ell_B/\ell_{\rm ext}$, consistently with the static test-field behavior of Ref.~\cite{Coviello:2025pla}. The full finite-frequency curves lie very close to the static curve, so their separation is most clearly visible in the inset of Figure~\ref{fig:tensor_testfield}. The conservative correction remains positive over the sampled regularization range and decreases in magnitude as the regularization is increased, with a mild upturn close to extremality. At $\ell_B/\ell_{\rm ext}=0.99$,
\begin{equation}
 \frac{\Delta\alpha_T^B}{M^5}
 =5.67\times10^{-5},\quad 2.28\times10^{-4},\quad 5.13\times10^{-4}
\end{equation}
for $M\omega=0.002$, $0.004$, and $0.006$, respectively.

\paragraph{Hayward background.}
The static response is again positive and monotonic, with the small-regularization coefficient in Eq.~\eqref{eq:tensor_probe_static_scaling}. The dynamical correction is positive at low and intermediate regularization but decreases as $\ell_H/\ell_{\rm ext}$ grows and becomes negative in the near-extremal region. At $\ell_H/\ell_{\rm ext}=0.99$,
\begin{equation}
 \frac{\Delta\alpha_T^H}{M^5}
 =-9.91\times10^{-5},\quad -3.97\times10^{-4},\quad -8.94\times10^{-4}
\end{equation}
for the same three frequencies. The main panels, therefore, remain almost indistinguishable from the static curves, while the insets resolve the finite-frequency structure directly.

\paragraph{Fan--Wang background.}
The Fan--Wang static response is positive at small regularization, reaches a
maximum at intermediate values of $\ell_{FW}/\ell_{\rm ext}$, crosses zero
near $\ell_{FW}/\ell_{\rm ext}\simeq0.56$, and becomes negative toward
extremality. At $\ell_{FW}/\ell_{\rm ext}=0.99$, the static response is
\begin{equation}
\frac{\alpha_T^{FW}(0)}{M^5}=-0.0579344,
\end{equation}
while the finite-frequency corrections are
\begin{equation}
\frac{\Delta\alpha_T^{FW}}{M^5}
=-9.69\times10^{-6},\quad
-3.88\times10^{-5},\quad
-8.73\times10^{-5}
\end{equation}
for $M\omega=0.002$, $0.004$, and $0.006$, respectively. These corrections
are much smaller than the static response, so the inset is needed to make the
finite-frequency dependence visible.

\bibliography{Dynamic_TLN}
\bibliographystyle{JHEP}
\end{document}